\documentclass[12pt]{article}

\usepackage[margin=1.1in]{geometry}
\usepackage[T1]{fontenc}
\usepackage{lmodern}
\usepackage{amsmath,amssymb,amsthm,mathtools}
\usepackage{physics}
\usepackage{bm}
\usepackage{graphicx}
\usepackage{tikz}
\usepackage{booktabs}
\usepackage{enumitem}
\usepackage{microtype}
\usepackage{authblk}
\usepackage[colorlinks=true,allcolors=magenta]{hyperref}
\hypersetup{pdftitle={The cone of states over time: exact tests and convergent two-sided hierarchies},pdfauthor={Matt Hoogsteder-Riera}}

\theoremstyle{plain}
\newtheorem{theorem}{Theorem}[section]
\newtheorem{proposition}[theorem]{Proposition}
\newtheorem{lemma}[theorem]{Lemma}
\newtheorem{corollary}[theorem]{Corollary}
\theoremstyle{definition}
\newtheorem{definition}[theorem]{Definition}
\newtheorem{example}[theorem]{Example}

\theoremstyle{remark}
\newtheorem{remark}[theorem]{Remark}

\newcommand{\bQ}{\widehat{\mathcal{Q}}}          
\newcommand{\HA}{\mathcal{H}_A}
\newcommand{\HB}{\mathcal{H}_B}
\newcommand{\cB}{\mathcal{B}}                     
\newcommand{\base}{\mathfrak{B}}                  
\newcommand{\BHS}{B_{\mathrm{HS}}}                
\newcommand{\cE}{\mathcal{E}}
\newcommand{\cF}{\mathcal{F}}
\newcommand{\cJ}{\mathcal{J}}                     
\newcommand{\cS}{\mathcal{S}}                     
\newcommand{\one}{\mathbf{1}}
\DeclareMathOperator{\cone}{cone}
\DeclareMathOperator{\conv}{conv}
\DeclareMathOperator{\diag}{diag}
\DeclareMathOperator{\id}{id}
\DeclareMathOperator{\Ad}{Ad}
\newcommand{\SEP}{\mathrm{SEP}}
\newcommand{\PPT}{\mathrm{PPT}}
\newcommand{\PSD}{\mathrm{PSD}}

\newcommand{\opnorm}[1]{\lVert #1 \rVert_{\mathrm{op}}}
\newcommand{\hsnorm}[1]{\lVert #1 \rVert_{\mathrm{HS}}}
\newcommand{\trnorm}[1]{\lVert #1 \rVert_{1}}
\renewcommand{\ip}[2]{\langle #1, #2\rangle}
\newcommand{\TA}{{T_A}}

\title{\fontsize{20}{24}\selectfont
  \textbf{The cone of states over time: exact tests and convergent two-sided hierarchies}}

\author[1,2]{Matt Hoogsteder-Riera\thanks{\href{mailto:hoogstedermatt@gmail.com}{hoogstedermatt@gmail.com}}}

\affil[1]{\textit{Centre for Quantum Technologies,\protect\\[-1mm] National University of Singapore,\protect\\[-1mm] 3 Science Drive 2, Singapore 117543}}
\affil[2]{\textit{Campus for Research Excellence and Technological Enterprise (CREATE),\protect\\[-1mm] National University of Singapore,\protect\\[-1mm] 1 Create Way, Singapore 13860}}

\date{}

\begin{document}

\maketitle

\vspace*{-1.2cm}

\begin{abstract}
A state over time (stote) pairs a state $\rho$ with a channel $\cE$ as the Hermitian, generally non-positive operator $(\rho\otimes\one)\star J_\cE$. We study the convex cone $\bQ$ generated by these operators. Our main result is a convergent outer semidefinite hierarchy with an explicit trace-norm Hausdorff bound on its trace-one slices. Specializing constrained symmetric-extension theory to the input--channel representation, we show that Bose-extension rounding preserves the channel marginal constraint and gives error at most $2d_A^2(d_A-1)/(k+d_A-1)$ at level $k$. A complementary inner hierarchy retains all channels at each chosen pure-input direction. For $d_A\ge2$, its error is at most $2(d_A-1)\sin^2\delta$, where $\delta$ is the Fubini--Study covering radius; nested sets of $N$ directions give error at most $2\pi^2(d_A-1)N^{-1/(d_A-1)}$, independently of $d_B$. These bounds provide explicit size--accuracy guarantees and certified support-value bounds, without guaranteeing finite termination for exact boundary membership. For binary classical outputs, we obtain an exact SDP membership test, including indefinite output blocks and mixtures that are not single stotes. At faithful marginal the test reduces to a scalar inequality; every trace-one member of this sector with input marginal $\rho$ has a constructive realization using at most $\operatorname{rank}\rho$ pure-input stotes with binary measurement channels. Finally, we compute the exact isotropic boundary and prove $\PPT\subsetneq\bQ\cap\PSD\subsetneq\PSD$ for equal local dimensions at least two.
\end{abstract}


\section{Introduction}\label{sec:1}

\subsection{States over time and cone membership}\label{sec:1.1}

Classical probability has no difficulty treating ``the state of a system at two times'' as a single object: a joint distribution $\pi(x,y)$ with prescribed time-marginals, factorizing as $\pi(x,y) = p(x)\,T(y|x)$ into an initial distribution and a stochastic map. The set of such couplings, entrywise-nonnegative matrices with fixed marginals, is the feasible set of optimal transport \cite{Villani2009}, and its simple convex geometry (a polytope, characterized by nonnegativity) is what makes classical transport a linear program.

Quantum temporal correlations need not admit a density-operator representation, as the pseudo-density-matrix construction of Fitzsimons, Jones and Vedral \cite{FitzsimonsJonesVedral2015} shows. Horsman, Heunen, Pusey, Barrett and Spekkens \cite{HorsmanHeunenPuseyBarrettSpekkens2017} proved a no-go theorem for a binary operation on operators satisfying Hermiticity, compatibility with probabilistic mixing, the correct classical limit, the correct single-time marginals, and associative composition. The no-go result was circumvented later by Fullwood and Parzygnat \cite{FullwoodParzygnat2022} who bypass the obstruction by restricting the assignment to a domain adapted to the state--channel data. Following this work and the causally neutral formulation of quantum Bayesian inference of Leifer and Spekkens \cite{LeiferSpekkens2013}, we use a \emph{state over time} (stote) to encode an initial state and a channel as a Hermitian, generally non-positive operator on $\HA\otimes\HB$. Specifically, we use the Jordan-product prescription
\begin{equation}\label{eq:jordan-stote}
Q \;=\; (\rho\otimes\one_B)\star J_{\cE}, \qquad X\star Y = \tfrac{1}{2}(XY+YX),
\end{equation}
with $J_{\cE}$ the Jamio{\l}kowski operator \cite{Jamiolkowski1972,Choi1975} of the channel $\cE\colon \cB(\HA)\to\cB(\HB)$, proposed in this form by Fullwood and Parzygnat \cite{FullwoodParzygnat2022}, characterized axiomatically by \cite{FullwoodParzygnat2022,ParzygnatFullwood2023}, and shown by Lie and Ng \cite{LieNg2024} to be \emph{the} quantum state over time under natural operational axioms. Its diagonal entries in local measurement bases are the real parts of the corresponding two-time Kirkwood--Dirac quasiprobabilities \cite{Kirkwood1933,Dirac1945,LostaglioEtAl2023}. We study the convex cone generated by all such operators, allowing nonnegative sums rather than only a single state--channel pair:
\begin{equation}\label{eq:cone-def}
\bQ(\HA{:}\HB) \;=\; \cone\{(\rho\otimes\one)\star J:\ \rho \text{ a state},\ J \text{ a channel Jamio{\l}kowski matrix}\},
\end{equation}
and its dual cone $\bQ^*$. Their characterization was posed as an open problem in the quantum-transport framework of Hoogsteder-Riera, Calsamiglia and Winter \cite{HoogstederCalsamigliaWinter2025}. At trace one, cone membership asks whether an operator is a convex mixture of stotes, not necessarily a single stote. The latter compatibility problem already has a spectral characterization \cite{SongParzygnat2025}; the distinction is essential here. Related compatibility questions arise in quantum Bayesian inference and retrodiction \cite{ParzygnatFullwood2023,ParzygnatBuscemi2023}, the statistics of incompatible observables \cite{LostaglioEtAl2023,ArvidssonShukurEtAl2024}, temporal correlations and quantum causal inference \cite{FitzsimonsJonesVedral2015,RiedEtAl2015,Fullwood2025}, and quantum optimal transport \cite{HoogstederCalsamigliaWinter2025,GolseMouhotPaul2016,DePalmaTrevisan2021}.

General membership in $\bQ$ is NP-hard, and in the dimensions covered by the known reduction the cone has no finite semidefinite representation \cite{PaperNPHardness}. These results motivate the questions studied here: which structural descriptions remain exact, which special cases admit complete membership tests, and how can the full cone be approximated from both sides? Our main result is a convergent outer semidefinite hierarchy with an explicit approximation rate, complemented by finite-input inner approximations and exact membership tests in special cases.

\subsection{Main results}\label{sec:1.2}

Throughout, $\base$ denotes the trace-one base of $\bQ$, $h(K) = \max_{Q\in\base}\ip{Q}{K}$ its support function, and $d_A, d_B$ the local dimensions; formal statements appear in the body.

The outer hierarchy is the main result, complemented by the inner hierarchy and the exact structural and optimization results below.
\begin{itemize}
\item \emph{Outer hierarchy (Section~\ref{sec:outer-hierarchy}).} We specialize constrained symmetric-extension methods \cite{DohertyParriloSpedalieri2004,BertaBorderiFawziScholz2022} to construct closed SDP-representable cones $\bQ_k\downarrow\bQ$, using Bose-symmetric extensions with a channel marginal constraint and optional PPT constraints (Theorem~\ref{thm:outer-hierarchy}). The existence and qualitative convergence of such relaxations follow from the general theory; our main contribution here is their explicit stote formulation and the trace-norm Hausdorff bound of Theorem~\ref{thm:outer-rate},
\begin{equation}\label{eq:intro-outer-rate}
d_H\bigl(\bQ_k\cap\{\Tr=1\},\base\bigr)\le\frac{2d_A^2(d_A-1)}{k+d_A-1}.
\end{equation}
The quantitative proof specializes the Bose-extension rounding formula of \cite{NavascuesOwariPlenio2009,Chiribella2011}, preserving the channel marginal constraint pointwise and controlling the error after the stote map. Every nonmember is excluded at a finite outer level; for a trace-one nonmember at a known positive distance from $\base$, the bound gives an explicit sufficient level. Permutation invariance alone also gives convergence, with an $O(k^{-1/2})$ bound at fixed dimensions from constrained de~Finetti theory \cite{BertaBorderiFawziScholz2022}; Bose support gives the displayed $O(1/k)$ bound independent of $d_B$ (Remark~\ref{rem:outer-rate}).

As a secondary result, we correct an unjustified quantifier exchange in the completeness proof of \cite[Section~V, between Eqs.~(29) and~(30)]{DohertyParriloSpedalieri2004}. The compactness and diagonal-extraction argument in Step~2 of the proof of Theorem~\ref{thm:outer-hierarchy} supplies the missing consistent family needed for the infinite de~Finetti theorem. This repairs the proof without changing its conclusion; Remark~\ref{rem:definetti-consistency} explains the gap and its relation to later finite de~Finetti proofs. Example~\ref{ex:dicke-extension-consistency} in Appendix~\ref{app:consistency-example} gives PPT-free certificates with internally consistent finite marginals whose maximal extension levels depend on the candidate, and exhibits their consistent infinite limit. The paragraph following the example shows that separately chosen PPT extensions can also be incompatible.
\item \emph{Inner hierarchy (Section~\ref{sec:inner-hierarchy}).} We discretize only the pure inputs, retaining the full channel set at each chosen direction. The resulting closed SDP-representable cones satisfy $\bQ^{(m)}\subseteq\bQ^{(m+1)}\subseteq\bQ$ and $\overline{\bigcup_m\bQ^{(m)}}=\bQ$ (Corollary~\ref{prop:inner-validity}). For $d_A\ge2$, Theorem~\ref{thm:inner-rate} constructs nested sets of $N_m$ directions with Fubini--Study covering radius $\delta_m$ such that
\begin{equation}\label{eq:intro-inner-rate}
d_H\bigl(\bQ^{(m)}\cap\{\Tr=1\},\base\bigr)\le 2(d_A-1)\sin^2\delta_m\le 2\pi^2(d_A-1)N_m^{-1/(d_A-1)}.
\end{equation}
Thus $N_m\ge\lceil(2\pi^2(d_A-1)/\eta)^{d_A-1}\rceil$ suffices for error $\eta>0$, using $N_m$ PSD blocks of size $d_Ad_B$ (Remark~\ref{rem:covering-numbers}). The error bound and sufficient number of directions are independent of $d_B$. This is a sufficient size, not an optimality claim. Together the two hierarchies give certified support-value bounds and one-sided membership certificates (Remark~\ref{rem:two-sided}); exact inner membership need not occur at a finite level for boundary points (Remark~\ref{rem:boundary-membership}).
\item \emph{Exact isotropic boundary and strict cone inclusions.} For equal local dimensions $d\ge2$, we prove $\PPT \subsetneq \bQ\cap\PSD \subsetneq \PSD$ with explicit witnesses for both strict inclusions (Corollary~\ref{cor:strict-chain}). A general twirling reduction (Proposition~\ref{prop:twirling-reduction}) reduces invariant membership to support values in the commutant. Its isotropic application gives the exact boundary, with maximal singlet fraction $F_{\max}(d) = \tfrac{1}{d}\bigl(1 + \tfrac{d-1}{4}\bigr)$ for $2 \le d \le 5$ and $\sqrt{d-1}/d$ beyond (Theorem~\ref{thm:isotropic-boundary}), and supplies the required witnesses.
\item \emph{Exact membership tests (Section~\ref{sec:membership-tests}).} We recall the known single-stote test at faithful input marginal and the inclusion $\PPT\subseteq\bQ$ \cite{SongParzygnat2025} as Theorem~\ref{thm:spectral-test}. Theorem~\ref{thm:binary-classical-membership} instead characterizes convex-cone membership for binary classical outputs by an exact SDP, including indefinite output blocks and mixtures that are not single stotes. At faithful marginal, its test reduces to a scalar inequality. For every trace-one member with input marginal $\rho$, the proof constructs a realization using at most $\operatorname{rank}\rho$ pure-input stotes with binary measurement channels. Remark~\ref{rem:binary-classical-use} gives a linear-algebra test for recognizing this sector without a prescribed output basis.
\end{itemize}

\subsection{Notation and preliminaries}\label{sec:1.4}

We write $\cS(\mathcal H)$ for the density operators on $\mathcal H$, $\PSD$ for the positive semidefinite cone, and $\SEP$ for the cone generated by positive product operators on $\HA\otimes\HB$.

We follow \cite{HoogstederCalsamigliaWinter2025}. All spaces are finite-dimensional, $d_A = \dim\HA$, $d_B = \dim\HB$, $N = d_A d_B$. The Choi matrix \cite{Choi1975} of a CP map $\cE : \cB(\HA) \to \cB(\HB)$ is $C_{\cE} = \sum_{ij} \ketbra{i}{j} \otimes \cE(\ketbra{i}{j}) \ge 0$, with trace preservation $\Leftrightarrow$ $\Tr_B C_{\cE} = \one_A$; the Jamio{\l}kowski matrix \cite{Jamiolkowski1972} is $J_{\cE} = C_{\cE}^{\TA}$ ($\TA$ = partial transpose on $A$ in the computational basis), and $\cJ = \cJ(\HA \to \HB)$ denotes the (convex, spectrahedral \cite{BlekhermanParriloThomas2013}) set of Jamio{\l}kowski matrices of channels, $\cJ = \{\tilde C^{\TA} : \tilde C \succeq 0,\ \Tr_B[\tilde C] = \one_A\}$. The stote of $(\rho,\cE)$ is $Q = (\rho\otimes\one_B)\star J_{\cE}$; it satisfies $\Tr_B[Q] = \rho$, $\Tr_A[Q] = \cE(\rho)$, $\Tr[Q] = 1$. We write
\begin{equation}\label{eq:cone-base-def}
\bQ = \cone\{(\rho\otimes\one)\star J : \rho\in\cS(\HA),\ J\in\cJ\}, \qquad \base = \conv\{(\rho\otimes\one)\star J\} \cap \{\Tr = 1\},
\end{equation}
$\varphi_\rho(X) := (\rho\otimes\one)\star X$, and use two elementary facts throughout: $\varphi_\rho$ is self-adjoint for the Hilbert--Schmidt inner product, and for faithful $\rho$ it is invertible with
\begin{equation}\label{eq:jordan-inverse-integral}
\varphi_\rho^{-1}(\omega)=2\int_0^\infty(e^{-t\rho}\otimes\one_B)\omega(e^{-t\rho}\otimes\one_B)\,dt.
\end{equation}
This is the inverse of a Lyapunov/Sylvester equation (cf.\ \cite[Ch.~VII]{Bhatia1997}, \cite[Ch.~4]{HornJohnson1991}; eq.~(2.11) of \cite{HoogstederCalsamigliaWinter2025}).

Operators are compared in three norms: the operator norm $\opnorm{\cdot}$, the trace norm $\trnorm{\cdot}$, and the \emph{Hilbert--Schmidt} (Schatten-2, or \emph{Frobenius} in the matrix-analysis literature) norm $\hsnorm{X} := \sqrt{\Tr[X^\dagger X]}$, the norm induced by the Hilbert--Schmidt inner product $\ip{X}{Y} := \Tr[X^\dagger Y]$ \cite[Ch.~IV]{Bhatia1997}, \cite[\S 5.6]{HornJohnson2013}; we write $\norm{\cdot}_2$ for the same norm when it appears alongside its Schatten siblings. The Hilbert--Schmidt metric is the Euclidean geometry of the ambient real vector space of Hermitian operators, and it is the metric in which the convex-geometric statements of Section~\ref{sec:2} are made (the Hausdorff-distance convergence rates of Section~\ref{sec:3} are stated in trace norm, which dominates it): $\BHS(x,r)$ denotes the closed Hilbert--Schmidt ball of radius $r$ centered at $x$, and the Gurvits--Barnum separable ball \cite{GurvitsBarnum2002} is such a ball (Corollary~\ref{cor:sep-ppt-in-cone}). Since three distinct objects would otherwise compete for the letter $B$, we distinguish them typographically: $\cB(\mathcal{H})$ (calligraphic) is the algebra of operators on a Hilbert space $\mathcal{H}$, $\base$ (Fraktur) is the base of $\bQ$ at trace one, and $\BHS$ (italic, subscripted) is a Hilbert--Schmidt ball.

For a Hermitian cost operator $K$, the support function of $\base$ is \cite[\S 13]{Rockafellar1970}
\begin{equation}\label{eq:support-function}
h(K):=\max_{\omega\in\base}\ip{\omega}{K}.
\end{equation}
The maximum is attained by compactness (Proposition~\ref{prop:cone-properties}(i)). Since $\base$ is closed and convex, a trace-one Hermitian operator $\omega$ belongs to $\base$ if and only if $\ip{\omega}{K}\le h(K)$ for every Hermitian $K$ \cite[Thm.~13.1]{Rockafellar1970}.

For nonempty compact sets $E,F$ of Hermitian operators, $d_H$ denotes the \emph{Hausdorff distance} in trace norm (the standard Hausdorff metric; cf.\ \cite[\S 1.8]{Schneider2014}):
\begin{equation*}
d_H(E,F):=\max\left\{\sup_{x\in E}\inf_{y\in F}\trnorm{x-y},\ \sup_{y\in F}\inf_{x\in E}\trnorm{x-y}\right\}.
\end{equation*}
We also write $\operatorname{dist}_1(x,F):=\inf_{y\in F}\trnorm{x-y}$. For nested sets $F\subseteq E$, the Hausdorff distance simplifies to
\begin{equation}\label{eq:hausdorff-nested}
d_H(E,F)=\sup_{x\in E}\operatorname{dist}_1(x,F),\qquad F\subseteq E,
\end{equation}
which measures the largest error in approximating a point of $E$ by a point of $F$. For compact sets, $d_H(E,F)=0$ holds exactly when $E=F$. Replacing the trace norm by the Hilbert--Schmidt norm defines the corresponding Hausdorff distance, which is no larger because $\hsnorm{X}\le\trnorm{X}$.

Fix once and for all an orthonormal (``computational'') basis $\{\ket{m}\}_{m=1}^{d_A}$ of $\HA$. Relative to it, every $X \in \cB(\HA\otimes\HB)$ has a unique \emph{block decomposition}
\begin{equation}\label{eq:block-decomposition}
X = \sum_{m,n} \ketbra{m}{n} \otimes X_{mn}, \qquad X_{mn} := (\bra{m}\otimes\one_B)\, X\, (\ket{n}\otimes\one_B) \in \cB(\HB),
\end{equation}
a $d_A\times d_A$ array whose entries are operators on $\HB$; $X$ is Hermitian iff $X_{nm} = X_{mn}^\dagger$ for all $m,n$. We use the same block notation relative to any other stated orthonormal basis of $\HA$ --- an eigenbasis of $\rho$, or a basis whose first vector is a prescribed unit vector, which we call a basis \emph{adapted} to that vector. $K$ always denotes a Hermitian cost operator, with blocks $K_{mn} \in \cB(\HB)$ as in \eqref{eq:block-decomposition} --- the letter $C$ is reserved for Choi matrices; $\ket{\psi}$ a unit vector of $\HA$, $P_\psi = \ketbra{\psi}{\psi}$. $\PPT$ denotes the cone of positive operators with positive partial transpose \cite{Peres1996,HorodeckiHorodeckiHorodecki1996}.

\section{Basic cone properties and membership tests}\label{sec:2}

We first establish the general properties of $\bQ$ used in Section~\ref{sec:3}, then study membership criteria for individual operators.

\subsection{General cone properties}\label{sec:cone-properties}

Proposition~\ref{prop:cone-properties} establishes compactness of the trace-one base, closedness of the cone, local unitary invariance, and stability under channel post-processing. Lemma~\ref{lem:pure-input} shows that pure input states suffice to generate the cone; this is used in the outer hierarchy, the inner approximation, and the symmetry reductions of Section~\ref{sec:3}.

\begin{proposition}\label{prop:cone-properties}
The following properties hold (they are noted, in large part, already in \cite{HoogstederCalsamigliaWinter2025}; we include self-contained proofs for completeness):
\begin{enumerate}[label=(\roman*)]
\item every generator $(\rho\otimes\one)\star J$ has unit trace, and $\base$ is compact and convex with $0\notin\base$ --- so the name given to $\base$ in Section~\ref{sec:1.4} is correct: it is a base of $\bQ$ \cite[Ch.~1]{AliprantisTourky2007} (a base, by definition, never contains $0$);
\item $\bQ$ is closed, pointed and spanning;
\item $\bQ$ is invariant under $Q\mapsto(U\otimes V)Q(U\otimes V)^\dagger$ for all unitaries $U,V$;
\item the intertwining identity $(\id\otimes\cF)\big((\rho\otimes\one)\star J_{\cE}\big)=(\rho\otimes\one)\star J_{\cF\circ\cE}$ holds for every CP map $\cF$ on $\cB(\HB)$; consequently every trace-preserving CP $\cF$ maps $\base$ into $\base$, i.e.\ $\bQ$ is invariant under post-processing by channels.
\end{enumerate}
\end{proposition}

\begin{proof}
(i) \emph{Unit trace of generators.} This was already recorded (without proof) among the marginal properties in Section~\ref{sec:1.4}, so we only note the one-line verification: by cyclicity $\Tr Q=\Tr[(\rho\otimes\one)J]$, and by trace duality of the partial transpose ($\Tr[XY^{\TA}]=\Tr[X^{\TA}Y]$), $\Tr[(\rho\otimes\one)\tilde C^{\TA}]=\Tr[(\rho^T\otimes\one)\tilde C]=\Tr[\rho^T\,\Tr_B[\tilde C]]=\Tr[\rho^T]=1$.

\emph{Compactness of the generating set, hence of $\base$ and the base property.} The parameter set $\cS(\HA)\times\cJ$ is compact: states form a closed bounded subset of a finite-dimensional space, and $\cJ$ is the image of the closed bounded set $\{\tilde C\succeq 0,\Tr_B[\tilde C]=\one_A\}$ (bounded because $\tilde C\succeq 0$ and $\Tr[\tilde C]=d_A$ force $\opnorm{\tilde C}\le d_A$) under the linear homeomorphism $\TA$. The map $(\rho,J)\mapsto(\rho\otimes\one)\star J$ is continuous (bilinear on finite-dimensional spaces), so the generating set $\mathcal{G}:=\{(\rho\otimes\one)\star J\}$ is compact, and $\base=\conv(\mathcal{G})$ is compact by Carath\'eodory (the convex hull of a compact set in $\mathbb{R}^n$ is compact \cite{Rockafellar1970}). Every element of $\base$ has unit trace (trace is linear and constant $=1$ on $\mathcal{G}$), so $0\notin\base$; since $\bQ=\bigcup_{t\ge 0}t\,\base$ by construction and every nonzero element of $\bQ$ has a unique representation $t\,b$ (with $t=\Tr$ of the element), $\base$ is a base.

(ii) \emph{Closedness, pointedness, spanning.} A cone generated by a compact convex set not containing the origin is closed (cf.\ \cite[\S 9]{Rockafellar1970}, \cite[Ch.~1]{AliprantisTourky2007}): if $t_nb_n\to x$ with $b_n\in\base$, $t_n\ge 0$, then $t_n=\Tr[t_nb_n]\to\Tr[x]=:t\ge 0$; if $t>0$ then $b_n\to x/t\in\base$ by closedness of $\base$ and $x=t(x/t)\in\bQ$, while if $t=0$ then $\norm{t_nb_n}\le t_n\sup_{b\in\base}\norm{b}\to 0$ so $x=0\in\bQ$. Pointedness: $Q,-Q\in\bQ$ with $Q\ne 0$ would give $\Tr[Q]>0$ and $\Tr[-Q]>0$ (every nonzero element has trace $=t\,\Tr[b]>0$), a contradiction.

For spanning, consider the prepare channel $\cE_\sigma(X)=\Tr[X]\sigma$ ($\sigma$ a state). Its Choi matrix is $\tilde C=\one_A\otimes\sigma$ ($\ge 0$, $\Tr_B[\tilde C]=\one_A$, so $\cE_\sigma$ is a channel), and its Jamio{\l}kowski matrix $J=\tilde C^{\TA}=\one_A\otimes\sigma$ is unchanged. For a pure input $P_\psi$,
\begin{equation}\label{eq:prepare-channel-product}
(P_\psi\otimes\one_B)\star(\one_A\otimes\sigma)=\tfrac12\big[(P_\psi\otimes\one)(\one\otimes\sigma)+(\one\otimes\sigma)(P_\psi\otimes\one)\big]=P_\psi\otimes\sigma,
\end{equation}
since the two factors act on different tensor slots and hence commute. So every product state $P_\psi\otimes\sigma$ (with $\sigma$ a state) --- in particular every pure product state $P_\psi\otimes P_\phi$ --- is a generator, and $\base\supseteq\conv\{P_\psi\otimes P_\phi\}=\SEP\cap\{\Tr=1\}$ by taking convex hulls.

These pure product states span the Hermitian operators: rank-one projectors span the Hermitian operators of each factor --- the projectors onto $\ket{i}$, $\ket{j}$, $(\ket{i}+\ket{j})/\sqrt{2}$, $(\ket{i}+i\ket{j})/\sqrt{2}$ already span the $(i,j)$ matrix-unit directions --- and tensor products of spanning sets span the tensor product. Hence $\bQ-\bQ=\cB(\HA\otimes\HB)_{\mathrm{herm}}$.

(iii) \emph{Local unitary invariance.} Write $W:=U\otimes V$ (unitary) and $\Ad_W(X):=WXW^\dagger$. Inserting $W^\dagger W=\one$ between the two factors of each term of the Jordan product,
\begin{equation}\label{eq:lu-covariance}
\Ad_W\big((\rho\otimes\one)\star J\big)=\tfrac12\big[W(\rho\otimes\one)W^\dagger\,WJW^\dagger+WJW^\dagger\,W(\rho\otimes\one)W^\dagger\big]=(U\rho U^\dagger\otimes\one)\star\Ad_W(J),
\end{equation}
using $W(\rho\otimes\one)W^\dagger=U\rho U^\dagger\otimes\one$. It remains to check $\Ad_{U\otimes V}(J)\in\cJ$. In Choi coordinates, $J'=\Ad_{U\otimes V}(\tilde C^{\TA})=\big((\bar U\otimes V)\tilde C(\bar U\otimes V)^\dagger\big)^{\TA}$ (using $(U\otimes V)X^{\TA}(U\otimes V)^\dagger=\big((\bar U\otimes V)X(\bar U\otimes V)^\dagger\big)^{\TA}$, valid since $\TA\circ\Ad_U=\Ad_{\bar U}\circ\TA$ on the $A$ factor); the new Choi matrix $\tilde C'=(\bar U\otimes V)\tilde C(\bar U\otimes V)^\dagger$ is PSD and satisfies $\Tr_B[\tilde C']=\bar U(\Tr_B[\tilde C])\bar U^\dagger=\bar U\one_A\bar U^\dagger=\one_A$. Hence conjugated generators are generators (of the rotated input state), and the invariance of the set $\bQ$ follows by linearity and closure.

(iv) \emph{Post-processing.} Let $\cF:\cB(\HB)\to\cB(\HB)$ be CP. Since $\id_A\otimes\cF$ acts only on the $B$ factor and $\rho\otimes\one_B$ acts as $\rho$ on $A$ tensor a multiple of the identity on $B$, the module property $(\id\otimes\cF)\big((\rho\otimes\one)X\big)=(\rho\otimes\one)(\id\otimes\cF)(X)$ holds (check on product operators $X=a\otimes b$: $(\id\otimes\cF)(\rho a\otimes b)=\rho a\otimes\cF(b)=(\rho\otimes\one)(a\otimes\cF(b))$; extend linearly), and similarly on the right. Hence $\id\otimes\cF$ commutes with $\varphi_\rho$, giving $(\id\otimes\cF)\big((\rho\otimes\one)\star J_{\cE}\big)=(\rho\otimes\one)\star(\id\otimes\cF)(J_{\cE})$. It remains to identify $(\id\otimes\cF)(J_{\cE})=J_{\cF\circ\cE}$: on Choi matrices, $(\id\otimes\cF)(C_{\cE})=\sum_{ij}\ketbra{i}{j}\otimes\cF(\cE(\ketbra{i}{j}))=C_{\cF\circ\cE}$ directly from the definition, and $\id\otimes\cF$ commutes with $\TA$ (they act on different factors), so the same identity holds for Jamio{\l}kowski matrices. If $\cF$ is additionally trace-preserving then $\cF\circ\cE$ is a channel, so the image of a generator is a generator with the same input state, and $(\id\otimes\cF)(\base)\subseteq\base$ follows by convexity.
\end{proof}

\begin{lemma}[exact pure-input generation]\label{lem:pure-input}
\begin{equation}\label{eq:pure-input-generation}
\base = \conv\{(P_\psi \otimes \one_B) \star J : \ket\psi \text{ a unit vector},\ J \in \cJ\}.
\end{equation}
\end{lemma}

\begin{proof}
Every pure-input generator belongs to $\base$, so the right-hand convex hull is contained in $\base$. For the reverse inclusion, fix a generator $(\rho\otimes\one_B)\star J$ and spectrally decompose $\rho = \sum_i \lambda_i P_{\psi_i}$ with $\lambda_i \geq 0$ and $\sum_i \lambda_i = 1$ \cite{HornJohnson2013}. Linearity in the input gives
\begin{equation}\label{eq:pure-input-decomposition}
(\rho \otimes \one) \star J = \sum_i \lambda_i \, (P_{\psi_i} \otimes \one) \star J.
\end{equation}
This is a convex combination of pure-input generators with the same channel $J$. Taking the convex hull over all generators proves the reverse inclusion.
\end{proof}

The Jordan map $\varphi_\rho=(\rho\otimes\one_B)\star(\cdot)$ relates an input state and a channel to their stote. Proposition~\ref{prop:schur-multiplier} describes its inverse at faithful $\rho$.

\begin{proposition}\label{prop:schur-multiplier}
Let $\rho$ be faithful with eigenbasis $\{\ket{i}\}$ and eigenvalues $\{p_i\}$. In this basis, $\varphi_\rho^{-1}$ acts on the block indices as the Hadamard (Schur) multiplier with symbol the Cauchy-type matrix $[2/(p_i+p_j)]_{ij}$ --- that is, as the map multiplying the $(i,j)$-th block entrywise by $2/(p_i+p_j)$; ``symbol'' here is the standard term for the coefficient matrix $M$ of a Hadamard multiplier $X\mapsto M\circ X$, see \cite[Ch.~5]{HornJohnson1991}. This statement is eq.~(2.3) of \cite{HoogstederCalsamigliaWinter2025}, and is an instance of the classical solution theory of the Sylvester equation $\tfrac12(\rho X+X\rho)=\omega$ \cite[Ch.~4]{HornJohnson1991}. Consequently:
\begin{enumerate}[label=(\roman*)]
\item $\varphi_\rho^{-1}$ is completely positive: $[1/(p_i+p_j)]$ is a Gram matrix, hence PSD, hence its Schur multiplier preserves positivity (Schur product theorem \cite{Schur1911}). Here $\varphi_\rho^{-1}$ is the inverse Jordan map on $A$ tensored with $\id_B$, the identity map on $\cB(\HB)$.
\item With $\TA$ taken in the eigenbasis of $\rho$, $\varphi_\rho^{-1}\circ\TA=\TA\circ\varphi_\rho^{-1}$ (the symbol is symmetric).
\item The symbol $[(p_i+p_j)/2]$ of $\varphi_\rho$ has rank $\le 2$. If $\rho$ is not proportional to $\one_A$, this symbol has mixed signature and $\varphi_\rho$ is not a positive map. If $\rho$ is proportional to $\one_A$, then $\varphi_\rho$ is a positive scalar multiple of the identity map.
\end{enumerate}
\end{proposition}

\begin{proof}
Write any $X\in\cB(\HA\otimes\HB)$ in $\{\ket{i}\}$-block form, $X=\sum_{ij}\ketbra{i}{j}\otimes X_{ij}$ with $X_{ij}\in\cB(\HB)$. Since $\rho\otimes\one_B=\sum_i p_i\ketbra{i}{i}\otimes\one_B$ is diagonal in this basis,
\begin{equation}\label{eq:rho-left-right-blocks}
(\rho\otimes\one_B)X=\sum_{ij}p_i\ketbra{i}{j}\otimes X_{ij},\qquad X(\rho\otimes\one_B)=\sum_{ij}p_j\ketbra{i}{j}\otimes X_{ij},
\end{equation}
so $\varphi_\rho(X)_{ij}=\tfrac12(p_i+p_j)X_{ij}$: a Hadamard multiplier with symbol $[(p_i+p_j)/2]_{ij}$, block-by-block. Since every $p_i>0$ ($\rho$ faithful), each scalar $(p_i+p_j)/2>0$, so this multiplier is invertible block-by-block with inverse symbol $[2/(p_i+p_j)]_{ij}$, proving the displayed formula for $\varphi_\rho^{-1}$.

(i) Let $c_{ij}:=1/(p_i+p_j)=\int_0^\infty e^{-tp_i}e^{-tp_j}\,dt$; the integral representation exhibits $[c_{ij}]$ as a Gram-type limit of PSD matrices $[e^{-tp_i}e^{-tp_j}]_{ij}=u(t)u(t)^\dagger$ ($u(t)_i:=e^{-tp_i}$) integrated against Lebesgue measure, hence $[c_{ij}]\succeq 0$. Diagonalize $[c_{ij}]=\sum_k v_kv_k^\dagger$ (finite sum, $v_k\in\mathbb{C}^{d_A}$) and set $D_k:=\sqrt{2}\,\diag(v_k)\otimes\one_B$ (diagonal in the block index). Then
\begin{equation}\label{eq:phi-inv-kraus}
\sum_k D_kXD_k^\dagger=\sum_{ij}\Big(2\sum_k v_k(i)\overline{v_k(j)}\Big)\ketbra{i}{j}\otimes X_{ij}=\sum_{ij}2c_{ij}\,\ketbra{i}{j}\otimes X_{ij}=\varphi_\rho^{-1}(X),
\end{equation}
using $c_{ij}=c_{ji}$ real. This exhibits $\varphi_\rho^{-1}$ as a Kraus sum, hence completely positive \cite{Kraus1983,Choi1975}; tensoring $\one_B$ with a further ancilla changes nothing in the argument, since $D_k\otimes\one$ is again a valid Kraus operator.

(ii) In the eigenbasis of $\rho$, $\TA$ acts on block matrices by $(X^{\TA})_{ij}=X_{ji}$ (it permutes block position without touching block content, since $\TA$ only transposes the $\HA$ factor, $\ketbra{i}{j}\mapsto\ketbra{j}{i}$). Hence $(\varphi_\rho^{-1}(\TA(X)))_{ij}=\tfrac{2}{p_i+p_j}X_{ji}$ while $(\TA(\varphi_\rho^{-1}(X)))_{ij}=(\varphi_\rho^{-1}(X))_{ji}=\tfrac{2}{p_j+p_i}X_{ji}$: identical, since $p_i+p_j=p_j+p_i$.

(iii) The symbol matrix $M:=[(p_i+p_j)/2]_{ij}=\tfrac12(p\mathbf{e}^T+\mathbf{e}p^T)$, with $p=(p_1,\dots,p_{d_A})^T$ and $\mathbf{e}$ the all-ones vector, is a sum of two rank-one matrices, so $\rank M\le 2$. If $\rho$ has two distinct eigenvalues $p_i\ne p_j$ (true for any faithful $\rho$ not proportional to $\one_A$), the principal $2\times 2$ minor on indices $i,j$ has determinant $p_ip_j-\left(\tfrac{p_i+p_j}{2}\right)^2=-\tfrac14(p_i-p_j)^2<0$, so $M$ restricted to this $2\times 2$ block is indefinite; by the Cauchy interlacing theorem \cite{Cauchy1829} (see \cite[Ch.~III]{Bhatia1997} for a modern treatment) $M$ itself has a negative eigenvalue. Since $\Tr[M]=\sum_i p_i=1>0$, $M$ also has a positive eigenvalue: $M$ has mixed signature.

To see that mixed signature forces non-positivity of $\varphi_\rho$, fix any nonzero $\sigma\ge 0$ on $\HB$ and consider the test operator
\begin{equation}\label{eq:test-operator-Xsigma}
X_\sigma:=(\mathbf{e}\mathbf{e}^T)\otimes\sigma,
\end{equation}
where $\mathbf{e}\mathbf{e}^T$ is the (real, rank-one, PSD) all-ones $d_A\times d_A$ matrix in the $\{\ket{i}\}$ basis; $X_\sigma\ge 0$ as a tensor product of PSD operators, and every block of $X_\sigma$ equals $\sigma$: $(X_\sigma)_{ij}=\sigma$ for all $i,j$. By the block formula, $\varphi_\rho(X_\sigma)_{ij}=\tfrac{p_i+p_j}{2}\,\sigma=M_{ij}\,\sigma$, i.e.\ $\varphi_\rho(X_\sigma)=M\otimes\sigma$. The eigenvalues of $M\otimes\sigma$ are the products $\lambda_k(M)\mu_l(\sigma)$; since $M$ has a negative eigenvalue $\lambda<0$ and $\sigma$ has some positive eigenvalue $\mu>0$ (as $\sigma\ge 0$, $\sigma\ne 0$), $\lambda\mu<0$ is an eigenvalue of $\varphi_\rho(X_\sigma)$, so $\varphi_\rho(X_\sigma)\not\ge 0$ although $X_\sigma\ge 0$. Thus the Jordan map can create negativity from positive operators. Stote negativity can also come from the Jamio{\l}kowski matrix itself: for the identity channel and $\rho=\one_A/d_A$, the stote is the swap operator divided by $d_A$, which is not positive when $d_A\ge 2$.
\end{proof}

\subsection{Membership tests}\label{sec:membership-tests}

Theorem~\ref{thm:spectral-test} restates the single-stote compatibility criterion and PPT inclusion of Song and Parzygnat \cite[Thm.~III.11, Cor.~III.12 and App.~C]{SongParzygnat2025}; we include the proof to make the later arguments self-contained. Corollary~\ref{cor:sep-ppt-in-cone} gives a quantitative inner ball, and its inclusion $\PPT\subseteq\bQ$ is used in Corollary~\ref{cor:strict-chain}. We then turn to classical outputs, culminating in the exact convex-cone membership criterion of Theorem~\ref{thm:binary-classical-membership}, which also covers indefinite output blocks and mixtures that are not single stotes.

\begin{theorem}[fixed-marginal membership is a spectral test; $\PPT\subseteq\bQ$; Song and Parzygnat {\cite[Thm.~III.11, Cor.~III.12 and App.~C]{SongParzygnat2025}}]\label{thm:spectral-test}
Let $\omega=\omega^\dagger$ with $\Tr\omega=1$ and $\rho:=\Tr_B[\omega]$ positive definite (i.e.\ $\rho>0$: a faithful state). Then $\omega$ is a stote of some channel with input $\rho$ if and only if $\varphi_\rho^{-1}(\omega^{\TA})\ge 0$, with $\TA$ in an eigenbasis of $\rho$. In particular every PPT state with faithful marginal is a \emph{single} stote of its own marginal, and $\PPT\subseteq\bQ$.
\end{theorem}

\begin{proof}
\emph{Faithful case.} Since $\rho$ is faithful, $\varphi_\rho$ is a bijection on Hermitian operators (Proposition~\ref{prop:schur-multiplier}; cf.\ the Lyapunov-equation inverse, eq.~(2.11) of \cite{HoogstederCalsamigliaWinter2025}: its symbol $(p_i+p_j)/2$ is strictly positive in every entry, hence nonzero, hence the blockwise multiplier is invertible; the same statement with $d_B=1$ shows $x\mapsto\rho\star x$ is invertible on Hermitian operators on $\HA$ alone). Define $J:=\varphi_\rho^{-1}(\omega)$; this is the \emph{unique} Hermitian operator with $\varphi_\rho(J)=\omega$, so ``$\omega$ is a stote of some channel with input $\rho$'' reduces to the single question of whether this specific $J$ is the Jamio{\l}kowski matrix of a trace-preserving CP map --- there is no further existential choice to make. $J$ is Hermitian because $\varphi_\rho^{-1}$ preserves Hermiticity: in the eigenbasis of $\rho$, $X=X^\dagger$ means $X_{ji}=X_{ij}^\dagger$, and the real symmetric scalar symbol $2/(p_i+p_j)=2/(p_j+p_i)$ then gives $(\varphi_\rho^{-1}(X))_{ji}=\big((\varphi_\rho^{-1}(X))_{ij}\big)^\dagger$.

\emph{Trace preservation is automatic.} For any Hermitian $X$ on $\HA\otimes\HB$, the identities $\Tr_B[(\rho\otimes\one_B)X]=\rho\,\Tr_B X$ and $\Tr_B[X(\rho\otimes\one_B)]=(\Tr_B X)\,\rho$ (operators acting on $A$ can be taken outside the partial trace over $B$) give $\Tr_B[\varphi_\rho(X)]=\rho\star\Tr_B[X]$. Applying this to $X=J$ and using $\varphi_\rho(J)=\omega$ together with the hypothesis $\Tr_B[\omega]=\rho=\rho\star\one_A$ (the latter since $\rho\star\one_A=\tfrac12(\rho\cdot\one_A+\one_A\cdot\rho)=\rho$) yields $\rho\star\Tr_B[J]=\rho\star\one_A$; invertibility of $x\mapsto\rho\star x$ then forces $\Tr_B[J]=\one_A$, i.e.\ $J$ is automatically the Jamio{\l}kowski matrix of a \emph{trace-preserving} map, regardless of complete positivity.

\emph{Complete positivity is the spectral test.} $J$ is the Jamio{\l}kowski matrix of a valid channel iff, by definition ($J=\tilde C^{\TA}$, $\TA$ an involution), the Choi matrix $\tilde C=J^{\TA}$ satisfies $\tilde C\ge 0$ (Choi's theorem \cite{Choi1975}). By the commutation of Proposition~\ref{prop:schur-multiplier}(ii),
\begin{equation}\label{eq:phi-inv-commutes-TA}
J^{\TA}=\big(\varphi_\rho^{-1}(\omega)\big)^{\TA}=\varphi_\rho^{-1}\big(\omega^{\TA}\big)
\end{equation}
(with $\TA$ taken in the eigenbasis of $\rho$ throughout), so the CP condition \emph{is} the stated spectral test, verbatim. This completes the proof of the equivalence.

\emph{PPT states pass the test.} If moreover $\omega\in\PPT$, then $\omega^{\TA}\ge 0$ by definition, and now --- and only now --- complete positivity of $\varphi_\rho^{-1}$ (Proposition~\ref{prop:schur-multiplier}(i), with the Schur multiplier on $A$ tensored with $\id_B$) applies: $\varphi_\rho^{-1}(\omega^{\TA})\ge 0$. Hence every PPT state with faithful marginal is a single stote of its own marginal.

\emph{Non-faithful marginals.} The zero operator is already in $\bQ$. By positive rescaling, it suffices to consider $\omega\in\PPT$ with $\Tr\omega=1$ and possibly singular $\rho:=\Tr_B[\omega]$. For $\epsilon\in(0,1]$ set
\begin{equation}\label{eq:omega-epsilon}
\omega_\epsilon:=(1-\epsilon)\,\omega+\epsilon\,\frac{\one_A}{d_A}\otimes\frac{\one_B}{d_B}.
\end{equation}
Then $\Tr_B[\omega_\epsilon]=(1-\epsilon)\rho+\epsilon\,\one_A/d_A$ is faithful for every $\epsilon>0$; $\omega_\epsilon\ge 0$ and $\omega_\epsilon^{\TA}=(1-\epsilon)\omega^{\TA}+\epsilon\,\one_A/d_A\otimes\one_B/d_B\ge 0$, as convex combinations of PSD operators (the maximally mixed state is invariant under partial transpose in any product basis). So $\omega_\epsilon\in\PPT$ with faithful marginal, hence $\omega_\epsilon\in\bQ$ by the faithful case just proved. Since $\bQ$ is closed (Proposition~\ref{prop:cone-properties}) and $\omega_\epsilon\to\omega$ as $\epsilon\to 0^+$ (in any norm on the finite-dimensional space of operators), $\omega=\lim_{\epsilon\to 0^+}\omega_\epsilon\in\bQ$. This proves $\PPT\subseteq\bQ$ unconditionally (the ``single stote of its own marginal'' statement itself is only asserted, and only needed, in the faithful case).
\end{proof}

\begin{corollary}\label{cor:sep-ppt-in-cone}
$\SEP\subseteq\PPT\subseteq\bQ$. For $N\ge 2$, the Gurvits--Barnum ball \cite{GurvitsBarnum2002} lies inside $\base$:
\begin{equation}\label{eq:gurvits-barnum-ball}
\BHS\bigl(\one/N,\ 1/\sqrt{N(N-1)}\bigr)\cap\{\Tr=1\}\ \subseteq\ \SEP\cap\{\Tr=1\}\ \subseteq\ \base.
\end{equation}
\end{corollary}

\begin{proof}
The first inclusion is the Peres criterion \cite{Peres1996}: if $\omega=\sum_k\lambda_k\,\rho_k\otimes\sigma_k$ with $\lambda_k\ge 0$, $\rho_k,\sigma_k\ge 0$, then $\omega^{\TA}=\sum_k\lambda_k\,\rho_k^T\otimes\sigma_k\ge 0$ since transposition preserves positivity of each $\rho_k$. The second inclusion is Theorem~\ref{thm:spectral-test}.

Finally, for $N\ge 2$ the Gurvits--Barnum theorem \cite{GurvitsBarnum2002} states that every \emph{Hermitian} $\omega$ with $\Tr[\omega]=1$ and $\hsnorm{\omega-\one/N}\le 1/\sqrt{N(N-1)}$ is separable. This proves \eqref{eq:gurvits-barnum-ball}. For $N=1$, all three cones are $\mathbb{R}_+$ and $\base=\{1\}$. For $N\ge2$, the ball gives a full-dimensional neighbourhood of $\one/N$ within the affine space $\{\Tr=1\}$.
\end{proof}

The following corollary shows that every positive classical-output state can be realized by one input state and one measurement, and gives the measurement explicitly.

\begin{corollary}[measurement reconstruction]\label{cor:measurement-reconstruction}
Let $\omega_y\succeq0$ with $\rho:=\sum_y\omega_y$ a state. There is a POVM $\{M_y\}_y$ satisfying $\omega_y=\rho\star M_y$, so $\sum_y\omega_y\otimes\ketbra{y}{y}$ is a single stote. If $\rho$ is faithful, the POVM is uniquely given by $M_y=\varphi_\rho^{-1}(\omega_y)$, where here $\varphi_\rho(X)=\rho\star X$ acts on $\HA$. If $\rho$ is singular, this formula applies on $\operatorname{supp}\rho$, with any POVM on $\ker\rho$ completing the operators by direct sum.
\end{corollary}

\begin{proof}
For faithful $\rho$, the block-diagonal state $\sum_y\omega_y\otimes\ketbra{y}{y}$ is PPT, so the reconstruction in Theorem~\ref{thm:spectral-test} gives $J=\sum_y M_y\otimes\ketbra{y}{y}$ with $M_y=\varphi_\rho^{-1}(\omega_y)$. Positivity of the inverse gives $M_y\succeq0$, and linearity gives $\sum_y M_y=\varphi_\rho^{-1}(\rho)=\one_A$. Thus these operators form a POVM and satisfy $\rho\star M_y=\omega_y$; invertibility gives uniqueness. If $\rho$ is singular, $0\preceq\omega_y\preceq\rho$ implies $\operatorname{supp}\omega_y\subseteq\operatorname{supp}\rho$. Apply the faithful construction on that support and complete by an arbitrary POVM on the kernel, where $\rho\star M_y=0$.
\end{proof}

\begin{lemma}[classical-output operators]\label{lem:classical-output}
Fix an output basis and let $\omega=\sum_y\omega_y\otimes\ketbra{y}{y}$ be Hermitian with unit trace. By post-processing with output dephasing (Proposition~\ref{prop:cone-properties}(iv)), $\omega\in\base$ if and only if $\omega_y=\sum_i\rho_i\star M_{i,y}$, where $\rho_i\succeq0$, $\sum_i\Tr\rho_i=1$, and each $\{M_{i,y}\}_y$ is a POVM. This is the Jordan-product analogue of an assemblage in EPR steering \cite{WisemanJonesDoherty2007}. The positive-block case is covered by Corollary~\ref{cor:measurement-reconstruction}; positivity is not necessary for such a representation, since Jordan products of positive operators can have negative eigenvalues. For two output outcomes, this representation gives the complete membership test below, including operators with indefinite blocks.
\end{lemma}

\begin{theorem}[binary classical-output membership]\label{thm:binary-classical-membership}
Let $\omega=\omega_0\otimes\ketbra{0}{0}+\omega_1\otimes\ketbra{1}{1}$, where $\omega_0,\omega_1$ are Hermitian operators on $\HA$, and set $\rho=\omega_0+\omega_1$ and $X=\omega_0-\omega_1$. Then $\omega\in\base$ if and only if $\Tr\rho=1$ and the following semidefinite feasibility problem has a solution:
\begin{equation}\label{eq:binary-classical-sdp}
\begin{pmatrix}\rho&Y\\Y^\dagger&T\end{pmatrix}\succeq0,\qquad \frac{Y+Y^\dagger}{2}=X,\qquad \Tr T\le1,
\end{equation}
where $Y$ is an operator on $\HA$ and $T=T^\dagger$. If $\rho\succ0$ and $\Tr\rho=1$, this is equivalently the scalar test
\begin{equation}\label{eq:binary-classical-spectral}
\Tr\bigl[X\varphi_\rho^{-1}(X)\bigr]=\sum_{i,j}\frac{2\abs{X_{ij}}^2}{p_i+p_j}\le1,
\end{equation}
where $\varphi_\rho(L)=\rho\star L$ acts on $\HA$, and $X_{ij}$ are the entries in an eigenbasis of $\rho$ with eigenvalues $p_i$. Every member has a realization as a convex combination of at most $\operatorname{rank}\rho$ pure-input stotes with binary measurement channels.
\end{theorem}

\begin{proof}
\emph{Necessity.} Suppose $\omega\in\base$. By Lemma~\ref{lem:classical-output}, write $\omega_y=\sum_j\rho_j\star M_{j,y}$ with $\rho_j\succeq0$, $\sum_j\Tr\rho_j=1$, and each $\{M_{j,0},M_{j,1}\}$ a binary POVM. Summing over $y$ gives $\rho=\sum_j\rho_j$, because $M_{j,0}+M_{j,1}=\one$. Set $H_j=M_{j,0}-M_{j,1}=2M_{j,0}-\one$. Since $0\preceq M_{j,0}\preceq\one$, we have $-\one\preceq H_j\preceq\one$ and hence $H_j^2\preceq\one$. Subtracting the two output blocks gives $X=\sum_j\rho_j\star H_j$. Take $Y=\sum_j\rho_jH_j$ and $T=\sum_jH_j\rho_jH_j$. Then
\begin{equation}\label{eq:binary-classical-necessary}
\begin{pmatrix}\rho&Y\\Y^\dagger&T\end{pmatrix}=\sum_j\begin{pmatrix}\one\\H_j\end{pmatrix}\rho_j\begin{pmatrix}\one&H_j\end{pmatrix}\succeq0,\qquad \Tr T=\sum_j\Tr(\rho_jH_j^2)\le\sum_j\Tr\rho_j=1.
\end{equation}
The trace inequality follows from $\Tr[\rho_j(\one-H_j^2)]\ge0$, since both factors are positive semidefinite. The Hermitian part of $Y$ is $\sum_j(\rho_jH_j+H_j\rho_j)/2=X$, so all constraints in \eqref{eq:binary-classical-sdp} hold.

\emph{Extracting a quadratic bound.} Conversely, suppose \eqref{eq:binary-classical-sdp} holds and $\Tr\rho=1$. Its principal block $\rho$ is positive semidefinite. Taking a positive square root of the full block matrix and splitting it into its upper and lower block rows gives a factorization
\begin{equation}\label{eq:binary-classical-factor}
\begin{pmatrix}\rho&Y\\Y^\dagger&T\end{pmatrix}=\begin{pmatrix}R\\S\end{pmatrix}\begin{pmatrix}R^\dagger&S^\dagger\end{pmatrix},\qquad \rho=RR^\dagger,\quad Y=RS^\dagger,\quad T=SS^\dagger.
\end{equation}
Let $Q$ be the orthogonal projector onto $\ker\rho$. Since $(QR)(QR)^\dagger=Q\rho Q=0$, we have $QR=0$, and therefore $QY=QRS^\dagger=0$. Using $2X=Y+Y^\dagger$ gives $QXQ=0$. Thus, in an eigenbasis of $\rho$, $X_{ij}=0$ whenever $p_i=p_j=0$. Define
\begin{equation}\label{eq:binary-classical-jordan-solution}
L_{ij}=\begin{cases}2X_{ij}/(p_i+p_j),&p_i+p_j>0,\\0,&p_i=p_j=0.\end{cases}
\end{equation}
This operator is Hermitian because $X$ is Hermitian and the denominators are real and symmetric in $i,j$. Entrywise, $(\rho\star L)_{ij}=(p_i+p_j)L_{ij}/2=X_{ij}$. This proves $\rho\star L=X$ even for singular $\rho$. Only the kernel-to-kernel block is set to zero; entries connecting $\operatorname{supp}\rho$ and $\ker\rho$ are retained.

Set $c=\Tr(\rho L^2)\ge0$. Cyclicity of the trace and $\rho\star L=X$ give
\begin{equation}\label{eq:binary-classical-c-identity}
\Tr(XL)=\tfrac12\Tr(\rho L^2+L\rho L)=\Tr(\rho L^2)=c.
\end{equation}
Since $L=L^\dagger$ and $X=(Y+Y^\dagger)/2$, we also have $\Tr(XL)=\operatorname{Re}\Tr(LY)$. Furthermore, \eqref{eq:binary-classical-factor} gives $\hsnorm{S}^2=\Tr T$ and $\hsnorm{LR}^2=\Tr(R^\dagger L^2R)=c$. Hilbert--Schmidt Cauchy--Schwarz \cite[Ch.~IV]{Bhatia1997} therefore yields
\begin{equation}\label{eq:binary-classical-quadratic}
c=\operatorname{Re}\Tr(LY)=\operatorname{Re}\Tr(S^\dagger LR)\le\hsnorm{S}\,\hsnorm{LR}=\sqrt{\Tr T}\sqrt{c}.
\end{equation}
If $c>0$, divide by $\sqrt c$ and square to obtain $c\le\Tr T\le1$; if $c=0$, the same bound is immediate. When $\rho$ is faithful, the unique solution is $L=\varphi_\rho^{-1}(X)$ by Proposition~\ref{prop:schur-multiplier}, and
\begin{equation}\label{eq:binary-classical-spectral-calculation}
c=\Tr(XL)=\sum_{i,j}X_{ij}L_{ji}=\sum_{i,j}\frac{2\abs{X_{ij}}^2}{p_i+p_j}.
\end{equation}
Thus SDP feasibility implies \eqref{eq:binary-classical-spectral} in the faithful case. To prove sufficiency of both tests, the remaining argument starts only from a density operator $\rho$ and a Hermitian $L$ satisfying $\rho\star L=X$ and $c=\Tr(\rho L^2)\le1$, and constructs a realizing mixture.

\emph{Constructing an ensemble with a uniform quadratic bound.} The operator $L$ need not itself be a contraction. We will construct a pure-state decomposition $\rho=\sum_j\lambda_jP_{\psi_j}$ satisfying $\norm{L\psi_j}^2=c\le1$ for every $j$. Note an arbitrary pure-state decomposition guarantees only the average identity $\sum_j\lambda_j\norm{L\psi_j}^2=\Tr(\rho L^2)=c$.

Put $s=\operatorname{rank}\rho\ge1$ and consider $A=\sqrt\rho(L^2-c\one)\sqrt\rho$ as a Hermitian operator on $\operatorname{supp}\rho$. Its trace is $\Tr A=\Tr(\rho L^2)-c\Tr\rho=c-c=0$. The Fourier construction below gives an explicit constant diagonal, a special case of the Schur--Horn theorem \cite{Horn1954}; see \cite[Thm.~4.33]{Watrous2018} for a modern treatment. Choose an orthonormal eigenbasis $\{u_\ell\}_{\ell=1}^s$ of $A$, with eigenvalues $\alpha_\ell$, and define
\begin{equation}\label{eq:binary-classical-ensemble}
f_j=\frac1{\sqrt{s}}\sum_{\ell=1}^s e^{2\pi i j\ell/s}u_\ell,\qquad v_j=\sqrt\rho f_j,\qquad \lambda_j=\norm{v_j}^2,\qquad \psi_j=\frac{v_j}{\sqrt{\lambda_j}},\qquad 1\le j\le s.
\end{equation}
The finite geometric-series formula gives
\begin{equation}\label{eq:binary-classical-fourier-orthogonality}
\langle f_j,f_k\rangle=\frac1s\sum_{\ell=1}^s e^{2\pi i(k-j)\ell/s}=\delta_{jk},\qquad 1\le j,k\le s.
\end{equation}
Thus the $s$ vectors $f_j$ form an orthonormal basis of $\operatorname{supp}\rho$. Since $\sqrt\rho$ is invertible on this support, every $\lambda_j$ is strictly positive. If $P_{\operatorname{supp}\rho}$ denotes its orthogonal projector, then
\begin{equation}\label{eq:binary-classical-ensemble-sum}
\sum_{j=1}^s\lambda_jP_{\psi_j}=\sum_{j=1}^s\ketbra{v_j}{v_j}=\sqrt\rho\left(\sum_{j=1}^s\ketbra{f_j}{f_j}\right)\sqrt\rho=\sqrt\rho P_{\operatorname{supp}\rho}\sqrt\rho=\rho.
\end{equation}
Taking traces gives $\sum_j\lambda_j=1$, so these are convex weights. Each $f_j$ has squared overlap $1/s$ with every eigenvector $u_\ell$. It follows that
\begin{equation}\label{eq:binary-classical-constant-expectation}
\begin{aligned}
0=\frac1s\sum_{\ell=1}^s\alpha_\ell&=\langle f_j,A f_j\rangle=\langle v_j,(L^2-c\one)v_j\rangle\\
&=\lambda_j\bigl(\langle\psi_j,L^2\psi_j\rangle-c\bigr).
\end{aligned}
\end{equation}
Dividing by $\lambda_j>0$ proves the promised equality $\norm{L\psi_j}^2=\langle\psi_j,L^2\psi_j\rangle=c$ for each $j$: the Fourier construction has made the value equal to $c$ for every ensemble vector, not just on average. No assumption that $L$ preserves $\operatorname{supp}\rho$ is used here.

\emph{Constructing the binary measurements.} For each $j$, set $a_j=\langle\psi_j,L\psi_j\rangle\in\mathbb{R}$ and $b_j=\norm{L\psi_j-a_j\psi_j}\ge0$. If $b_j>0$, let $\eta_j=(L\psi_j-a_j\psi_j)/b_j$. This is a unit vector orthogonal to $\psi_j$, and $L\psi_j=a_j\psi_j+b_j\eta_j$, with $a_j^2+b_j^2=\norm{L\psi_j}^2=c$. Define $H_j$ to have the following matrix in the ordered orthonormal basis $(\psi_j,\eta_j)$ and to vanish on its orthogonal complement:
\begin{equation}\label{eq:binary-classical-contraction}
H_j\big|_{\operatorname{span}\{\psi_j,\eta_j\}}=\begin{pmatrix}a_j&b_j\\b_j&-a_j\end{pmatrix},\qquad H_j\psi_j=L\psi_j,\qquad \opnorm{H_j}=\sqrt{a_j^2+b_j^2}=\sqrt c\le1.
\end{equation}
The norm identity follows because the square of the displayed matrix is $(a_j^2+b_j^2)\one_2$. When $b_j=0$, use $H_j=a_jP_{\psi_j}$ instead; then $H_j\psi_j=L\psi_j$ and $\opnorm{H_j}=\abs{a_j}=\sqrt c\le1$ still hold. Thus in both cases $H_j$ is Hermitian and $-\one\preceq H_j\preceq\one$. Moreover, $H_j\psi_j=L\psi_j$ gives $H_jP_{\psi_j}=LP_{\psi_j}$; taking adjoints gives $P_{\psi_j}H_j=P_{\psi_j}L$. Hence $P_{\psi_j}\star H_j=P_{\psi_j}\star L$.

It follows that $M_{j,0}=(\one+H_j)/2$ and $M_{j,1}=(\one-H_j)/2$ are positive semidefinite and sum to $\one$, so they form a binary POVM. Finally, using \eqref{eq:binary-classical-ensemble-sum} and $\rho\star L=X$,
\begin{equation}\label{eq:binary-classical-realization}
\begin{aligned}
\sum_{j=1}^s\lambda_jP_{\psi_j}\star H_j&=\left(\sum_{j=1}^s\lambda_jP_{\psi_j}\right)\star L=\rho\star L=X,\\
\sum_{j=1}^s\lambda_jP_{\psi_j}\star M_{j,y}&=\frac12\left(\rho+(-1)^yX\right)=\omega_y,\qquad y\in\{0,1\}.
\end{aligned}
\end{equation}
For each $j$, the two operators $P_{\psi_j}\star M_{j,y}$ are the output blocks of the stote with input $P_{\psi_j}$ and the measurement channel associated with this POVM. By Lemma~\ref{lem:classical-output}, equation~\eqref{eq:binary-classical-realization} expresses $\omega$ as a convex combination of $s=\operatorname{rank}\rho$ such stotes, proving both $\omega\in\base$ and the realization bound.
\end{proof}

\begin{remark}[recognizing and using the binary classical-output criterion]\label{rem:binary-classical-use}
Theorem~\ref{thm:binary-classical-membership} gives an exact membership test for the binary classical-output sector, including operators with indefinite output blocks and mixtures that are not single stotes. Its applicability does not require an output basis to be supplied in advance. Given any Hermitian $\omega$ of unit trace on $\HA\otimes\mathbb{C}^2$, such a basis exists if and only if there is a nonzero traceless Hermitian operator $Z$ on the output satisfying
\begin{equation}\label{eq:binary-classical-commutator}
[\omega,\one_A\otimes Z]=0,\qquad Z=\begin{pmatrix}z&x-iy\\x+iy&-z\end{pmatrix},\qquad (x,y,z)\in\mathbb{R}^3\setminus\{0\}.
\end{equation}
Indeed, a classical-output decomposition commutes with $\one_A\otimes(\ketbra{e_0}{e_0}-\ketbra{e_1}{e_1})$. Conversely, a nonzero $Z$ as above has distinct eigenvalues $\pm\sqrt{x^2+y^2+z^2}$. In an orthonormal eigenbasis $\{e_0,e_1\}$ of $Z$, the commutator equation therefore forces the off-diagonal output blocks of $\omega$ to vanish. This is a direct application of the spectral theorem \cite{HornJohnson2013}.

Equation~\eqref{eq:binary-classical-commutator} is a homogeneous linear system in three real unknowns: equate the real and imaginary parts of its matrix entries to zero. If its kernel is nontrivial, choose any nonzero solution, diagonalize $Z$, and extract
\begin{equation}\label{eq:binary-classical-extract}
\omega_y=(\one_A\otimes\bra{e_y})\,\omega\,(\one_A\otimes\ket{e_y}),\qquad \omega=\sum_{y=0}^1\omega_y\otimes\ketbra{e_y}{e_y}.
\end{equation}
The theorem then tests membership using $\rho=\omega_0+\omega_1$ and $X=\omega_0-\omega_1$, by the SDP \eqref{eq:binary-classical-sdp} or, when $\rho$ is positive definite, the scalar test \eqref{eq:binary-classical-spectral}. The output basis change preserves membership by local unitary invariance (Proposition~\ref{prop:cone-properties}(iii)); an arbitrary global diagonalization does not provide the same reduction. If the linear system has only the zero solution, the theorem is inapplicable, but this does not imply nonmembership.

There is also a one-sided test for arbitrary trace-one Hermitian operators on $\HA\otimes\mathbb{C}^2$: dephase the output in any chosen basis and apply the theorem to the resulting classical-output operator. If that operator fails the membership test, the original operator is outside $\base$, by the post-processing property in Proposition~\ref{prop:cone-properties}(iv). Passing the test after dephasing does not establish membership of the original operator.

This criterion is not restricted to the positive-block case of Corollary~\ref{cor:measurement-reconstruction} or to single stotes. For example, take $\rho=\one_2/2$ and $X=\operatorname{diag}(2/3,0)$. Then $\omega_0=\operatorname{diag}(7/12,1/4)$ and $\omega_1=\operatorname{diag}(-1/12,1/4)$ satisfy \eqref{eq:binary-classical-spectral}, since $\Tr[X\varphi_\rho^{-1}(X)]=8/9$. Thus their classical-output operator belongs to $\base$, although it is not a single stote: its unique reconstructed measurement operators would be $2\omega_0$ and $2\omega_1$, and the latter is not positive.
\end{remark}

\section{Convergent hierarchies and strict cone inclusions}\label{sec:3}

This section gives semidefinite approximations of the full cone and an exact description of its isotropic sector. Its main result is the \emph{outer hierarchy} of Section~\ref{sec:outer-hierarchy} with its explicit error bound: a nested family of SDP-representable cones $\bQ_k \supseteq \bQ$, obtained by specializing constrained symmetric-extension methods \cite{DohertyParriloSpedalieri2004,BertaBorderiFawziScholz2022} to the input--channel representation. Theorem~\ref{thm:outer-hierarchy} gives a self-contained convergence proof for this application; Theorem~\ref{thm:outer-rate} specializes Bose-extension rounding \cite{NavascuesOwariPlenio2009,Chiribella2011} to obtain an explicit $O(1/k)$ stote approximation bound for those same cones. A witness against $\bQ_k$ certifies non-membership in $\bQ$. Section~\ref{sec:inner-hierarchy} supplies the complementary \emph{inner hierarchy} $\bQ^{(m)} \subseteq \bQ$, built from finite nets of pure inputs and convergent at an explicit rate, giving the two-sided family of SDP-representable cones
\begin{equation}\label{eq:hierarchy-chain}
\bQ^{(1)} \subseteq \bQ^{(2)} \subseteq \cdots \subseteq \bQ \subseteq \cdots \subseteq \bQ_2 \subseteq \bQ_1,
\end{equation}
both sides converging with explicit error bounds on their trace-one slices; Section~\ref{sec:sandwich} assembles the two into a two-sided certificate scheme. Section~\ref{sec:outer-physical} interprets the outer certificates as channels with symmetric auxiliary outputs and gives a realization using a correlated ancilla. Section~\ref{sec:sectors} gives a general twirling reduction \cite{Werner1989,HorodeckiHorodecki1999} and computes the isotropic boundary as an application. For equal local dimensions $d\ge2$, the goal is the strict chain $\PPT \subsetneq \bQ \cap \PSD \subsetneq \PSD$, proved using the resulting isotropic witnesses.

\subsection{Outer hierarchy from symmetric extensions}\label{sec:outer-hierarchy}

The outer hierarchy specializes the symmetric-extension method of Doherty--Parrilo--Spedalieri \cite{DohertyParriloSpedalieri2004} and its affine-constrained formulation by Berta, Borderi, Fawzi and Scholz \cite{BertaBorderiFawziScholz2022}. Related approaches include Bose-symmetric hierarchies and rounding for constrained separability \cite[Sections~4--5]{ZeissKossmannFawziBerta2025}, de~Finetti hierarchies for convex bodies \cite{ZeissKossmannSchwonnekPlavala2026}, and hierarchies based on fixed-point conditions \cite{KossmannZeiss2026}. We first express $\bQ$ as a linear image of a constrained separable cone. The main hierarchy $\bQ_k$ uses certificates supported on the Bose-symmetric subspace, together with the channel marginal constraint (3) below. Theorem~\ref{thm:outer-hierarchy} gives this application with a self-contained proof of closedness, nestedness, and convergence; Theorem~\ref{thm:outer-rate} gives an explicit stote approximation bound for these same cones. Bose support is not needed for convergence, but it supplies the particular $O(1/k)$ estimate below and reduces the certificate dimension (Remark~\ref{rem:outer-rate}).

Let $\mathcal{H}_{A'} \cong \HA$ and define the constrained separable cone and contraction
\begin{equation}\label{eq:constrained-sep-cone}
S := \overline{\cone}\{\rho_{A'} \otimes \tilde C_{AB} : \rho \ge 0,\ \tilde C \ge 0,\ \Tr_B[\tilde C] \propto \one_A\}, \qquad \Lambda(\rho \otimes Y) := (\rho_A \otimes \one_B) \star Y,
\end{equation}
extended linearly. Let $F_{A'A}$ be the swap exchanging $A'$ and $A$, defined in the corresponding computational bases by $F_{A'A}(\ket a_{A'}\otimes\ket b_A)=\ket b_{A'}\otimes\ket a_A$. Then
\begin{equation}\label{eq:outer-swap-contraction}
\Lambda(X) = \tfrac12 \Tr_{A'}[(F_{A'A} \otimes \one_B) X + X (F_{A'A} \otimes \one_B)].
\end{equation}
Theorem~\ref{thm:outer-hierarchy}(i) below shows that $\bQ = \Lambda\bigl((\id \otimes \TA \otimes \id)(S)\bigr)$: the cone of stotes is a linear image of a constrained separable cone across the $A'{:}AB$ cut, and it is this product structure that the hierarchy relaxes. The level-$k$ relaxation $\bQ_k$ consists of all $\omega$ admitting $X$ on $\mathcal{H}_{A'}^{\otimes k} \otimes \HA \otimes \HB$ with
\begin{enumerate}
\item[(1)] $X \ge 0$, supported on $\mathrm{Sym}^k(\mathcal{H}_{A'})\otimes\HA\otimes\HB$, and normalized by $\Tr[X] = d_A \Tr[\omega]$;
\item[(2)] $X^{T_I} \ge 0$ for every subset $I$ of $A'$ copies;
\item[(3)] $\Tr_B[X] = \Tr_{AB}[X] \otimes \one_A / d_A$ (channel marginal constraint);
\item[(4)] $\omega = \Lambda\bigl((\id \otimes \TA \otimes \id)(\Tr_{A'_2 \cdots A'_k}[X])\bigr)$.
\end{enumerate}
Here $\mathrm{Sym}^k(\mathcal{H}_{A'})$ is the subspace fixed by every permutation of the $k$ copies. Writing $P_k^{\mathrm{sym}}$ for its orthogonal projector, the support condition means $X=(P_k^{\mathrm{sym}}\otimes\one_{AB})X(P_k^{\mathrm{sym}}\otimes\one_{AB})$. This is stronger than permutation invariance $U_\pi XU_\pi^\dagger=X$: an invariant operator can have support outside the symmetric subspace. The trace normalization follows from (3)--(4), since $\Tr[\Lambda((\id\otimes\TA\otimes\id)X_1)]=\Tr[X]/d_A$, but we include it explicitly to make compactness clear.

To compare with \cite[Theorems~2.3 and~3.1]{BertaBorderiFawziScholz2022}, take $AB$ as the unrepeated subsystem and $A'$ as the repeated subsystem. For a trace-one $\omega$, the normalized certificate $\hat X=X/d_A$ satisfies $\Tr_B\hat X=\hat X_{A'^k}\otimes\one_A/d_A$. This is their affine constraint on the unrepeated subsystem, with linear map $\Tr_B$ and prescribed output $\one_A/d_A$; the repeated subsystem has only the state-normalization constraint. Their constrained de~Finetti theorem therefore applies to permutation-invariant certificates. Applying the continuous linear map in (4) gives convergence to mixtures of stotes. Imposing Bose support or the PPT constraints preserves every pure-input product certificate, so convergence also holds for the stronger relaxations defined above. Thus the general convergence mechanism is already available in that theory; the proof below makes its application to $\bQ$ explicit.

\begin{theorem}[outer hierarchy]\label{thm:outer-hierarchy}
\begin{enumerate}
\item[(i)] $\bQ = \Lambda\bigl((\id \otimes \TA \otimes \id)(S)\bigr)$, and both sides are closed cones with compact bases.
\item[(ii)] Each $\bQ_k$ is a closed SDP-representable convex cone, and the hierarchy is nested and outer: $\bQ \subseteq \bQ_{k+1} \subseteq \bQ_k$ for every $k \ge 1$.
\item[(iii)] The hierarchy converges: $\bigcap_{k \ge 1} \bQ_k = \bQ$.
\end{enumerate}
All three statements remain true if the PPT constraints (2) are dropped from the definition of $\bQ_k$.
\end{theorem}

\begin{proof}
\emph{Step 0: the two cones agree, $\bQ = \Lambda\bigl((\id \otimes \TA \otimes \id)(S)\bigr)$ --- part (i).} Consider a nonzero product generator $\rho \otimes \tilde C$ of $S$ with $\Tr_B[\tilde C] = c\, \one_A$. Positivity gives $r:=\Tr[\rho]>0$ and $c>0$; if $c=0$, then $\Tr[\tilde C]=d_Ac=0$ would force $\tilde C=0$. Set $\hat\rho:=\rho/r$ and $\hat C:=\tilde C/c$. Then $\hat\rho$ is a state, $\hat C$ is a channel Choi matrix, and, with $J:=\hat C^{\TA}\in\cJ$,
\begin{equation}\label{eq:outer-image-generator}
\Lambda\bigl((\id \otimes \TA \otimes \id)(\rho \otimes \tilde C)\bigr)
=rc\,(\hat\rho\otimes\one_B)\star J\in\bQ.
\end{equation}
Here we used the product identity $\Lambda(\rho \otimes Y)=\tfrac12[(\rho_A\otimes\one_B)Y+Y(\rho_A\otimes\one_B)]$, which follows directly from the swap formula for $\Lambda$. Conversely, every generator $(\hat\rho\otimes\one_B)\star J$ of $\bQ$ is the image of $\hat\rho\otimes J^{\TA}$, so the generated cones agree.

It remains to justify the closures in this equality. The trace-one base of $S$ is the convex hull of
\begin{equation}\label{eq:S-base}
\left\{\hat\rho\otimes\frac{\hat C}{d_A}:\ \hat\rho\text{ a state},\ \hat C\succeq0,\ \Tr_B\hat C=\one_A\right\}.
\end{equation}
This set is compact. Every element of its image under $\Lambda\circ(\id\otimes\TA\otimes\id)$ has trace $1/d_A$, so the image is compact and does not contain zero. Its conic hull is therefore closed, by the same compact-base argument as Proposition~\ref{prop:cone-properties}(ii), and its trace-one base is exactly $\base$. Thus the image cone and $\bQ$ are both closed and have compact bases.

\emph{Step 1: representability, nestedness and $\bQ \subseteq \bQ_k$ --- part (ii).} Constraints (1)--(3) are semidefinite and linear conditions on $X$, and (4) is a linear map of $X$; hence $\bQ_k$, the projection onto $\omega$ of the spectrahedral set of pairs $(\omega, X)$, is an SDP-representable convex cone. It is also closed: on the unit-trace slice, (1) restricts $X$ to the compact set of PSD operators with trace $d_A$, and constraints (2)--(3) select a closed subset; its image under (4) is compact. The full cone has this compact unit-trace set as a base and is closed by Proposition~\ref{prop:cone-properties}(ii). For nestedness, let $X$ certify $\omega \in \bQ_{k+1}$ and set $X' := \Tr_{A'_{k+1}}[X]$ on $\mathcal{H}_{A'}^{\otimes k} \otimes \HA \otimes \HB$. Then $X'$ certifies $\omega \in \bQ_k$: (1) partial trace preserves positivity and the trace, and $\mathrm{Sym}^{k+1}(\mathcal{H}_{A'})\subseteq\mathrm{Sym}^k(\mathcal{H}_{A'})\otimes\mathcal{H}_{A'}$ gives the required support of $X'$; (2) for a subset $I$ of the remaining copies, $(X')^{T_I} = \Tr_{A'_{k+1}}[X^{T_I}] \ge 0$; (3) applying $\Tr_{A'_{k+1}}$ to both sides of the channel marginal constraint for $X$ gives $\Tr_B[X'] = \Tr_{AB}[X'] \otimes \one_A / d_A$, since $\Tr_{A'_{k+1}}$ commutes with $\Tr_B$ and with $\Tr_{AB}$; (4) $\Tr_{A'_2 \cdots A'_k}[X'] = \Tr_{A'_2 \cdots A'_{k+1}}[X]$, so the reproduced $\omega$ is unchanged. Hence $\bQ_{k+1} \subseteq \bQ_k$.

For $\bQ \subseteq \bQ_k$, let $\omega = \sum_\alpha \lambda_\alpha (\rho_\alpha \otimes \one) \star J_\alpha$ with $\lambda_\alpha \ge 0$ (finitely many terms by Carath\'eodory). By Lemma~\ref{lem:pure-input}, we may take every $\rho_\alpha$ to be pure; set $\tilde C_\alpha:=J_\alpha^{\TA}$. Then $\Tr[\omega] = \sum_\alpha \lambda_\alpha$. Set $X := \sum_\alpha \lambda_\alpha\, \rho_\alpha^{\otimes k} \otimes \tilde C_\alpha$ on $\mathcal{H}_{A'}^{\otimes k} \otimes \HA \otimes \HB$ (the $A'$ copies carrying copies of $\rho_\alpha$). Then $X$ satisfies each of the four defining constraints of $\bQ_k$:
\begin{itemize}
\item[(1)] $X \ge 0$, supported on the Bose-symmetric subspace because every $\rho_\alpha$ is pure, and $\Tr[X] = \sum_\alpha \lambda_\alpha \cdot 1 \cdot d_A = d_A \Tr[\omega]$;
\item[(2)] partial transposes on $A'$ subsets turn $\rho_\alpha^{\otimes k}$ into mixed tensor powers of $\rho_\alpha$ and $\rho_\alpha^T$, all PSD, so $X^{T_I} \ge 0$;
\item[(3)] $\Tr_B[X] = \sum_\alpha \lambda_\alpha \rho_\alpha^{\otimes k} \otimes \one_A$ while $\Tr_{AB}[X] = d_A \sum_\alpha \lambda_\alpha \rho_\alpha^{\otimes k}$, matching;
\item[(4)] $\Tr_{A'_2 \cdots A'_k}[X] = \sum_\alpha \lambda_\alpha \rho_\alpha \otimes \tilde C_\alpha$, and applying $\id \otimes \TA \otimes \id$ then $\Lambda$ recovers $\omega$ by Step 0.
\end{itemize}
So $\omega \in \bQ_k$.

\emph{Step 2: extraction of an exchangeable limit --- towards part (iii).} Let $\omega \in \bigcap_k \bQ_k$. If $\Tr[\omega] = 0$ then (1) forces $X = 0$ and (4) gives $\omega = 0 \in \bQ$; so assume, after rescaling, $\Tr[\omega] = 1$, and pick a feasible certificate $X^{(k)}$ for each $k$, with $\Tr X^{(k)} = d_A$. For $j \le k$ write $X_j^{(k)} := \Tr_{A'_{j+1} \cdots A'_k}[X^{(k)}]$, a PSD operator of trace $d_A$ on the \emph{fixed} space $\mathcal{H}_{A'}^{\otimes j} \otimes \HA \otimes \HB$.\footnote{For $j = k$ the partial trace is over the empty set of copies and is the identity map, $X_k^{(k)} = X^{(k)}$; this boundary case is included only so that the family is defined for all $j \le k$ --- the subsequential limits below are determined by the tails $k \gg j$.} For fixed $j\geq1$, the sequence $(X_j^{(k)})_{k \ge j}$ lives in the compact set $\{Z \ge 0, \Tr[Z] = d_A\}$; by the diagonal subsequence argument, spelled out as Lemma~\ref{lem:diagonal} in Appendix~\ref{app:C}, choose a common subsequence $k_1 < k_2 < \cdots$ along which $Y_j := \lim_n X_j^{(k_n)}$ exists for every $j$. The family $(Y_j)_{j \ge 1}$ inherits, by closedness of the constraints: positivity and Bose support (hence permutation invariance); the marginal consistency $\Tr_{A'_{j+1}}[Y_{j+1}] = Y_j$ (which holds exactly at every finite stage); the channel marginal constraint (3) at every level $j$ (apply $\Tr_{A'_{j+1} \cdots A'_k}$ to (3) for $X^{(k)}$ and pass to the limit); the PPT conditions (2) on every finite subset; and, from (4), $\omega = \Lambda\bigl((\id \otimes \TA \otimes \id) Y_1\bigr)$ (the level-$k$ certificates all reproduce $\omega$ through their first marginal, and $\Lambda \circ (\id \otimes \TA \otimes \id)$ is continuous).

\emph{Step 3: infinite de Finetti representation.} The diagonal argument in Step 2 gives an infinite consistent family $(Y_j/d_A)_j$ of permutation-invariant states on $\mathcal{H}_{A'}^{\otimes j}\otimes\HA\otimes\HB$, with reference system $AB$. We now use the infinite quantum de~Finetti theorem \cite{Stormer1969,HudsonMoody1976}, stated and proved in \cite[Sections~III--IV]{CavesFuchsSchack2002}, together with its reference-system construction in \cite[Section~V, Eqs.~(31)--(39)]{DohertyParriloSpedalieri2004}. This representation requires only permutation invariance, not Bose support. Applied to the consistent family above, it gives a probability measure $\mu$ on the compact state space of $\mathcal{H}_{A'}$ and a $\mu$-measurable family of states $\tau_\pi$ on $\HA\otimes\HB$. Setting $\sigma_\pi:=d_A\tau_\pi$, we have $\int d\mu(\pi)\Tr[\sigma_\pi]=d_A$ and
\begin{equation}\label{eq:definetti-rep}
Y_j = \int d\mu(\pi)\, \pi^{\otimes j} \otimes \sigma_\pi \qquad \text{for every } j \ge 1.
\end{equation}

The same measure $\mu$ and conditional operators $\sigma_\pi$ represent every level $j$, because the representation is applied to the entire consistent family, not separately to its finite marginals. The equality is exact; no finite de~Finetti approximation is used in this step.

\emph{Step 4: the channel marginal constraint forces the conditional marginal.} Substituting the representation \eqref{eq:definetti-rep} into constraint (3) at level $j$ --- inherited by $Y_j$ in Step 2 --- and collecting both sides under a single integral, the channel marginal constraint becomes the operator identity
\begin{equation}\label{eq:linkage-pointwise}
\begin{aligned}
&\int d\mu(\pi)\ \pi^{\otimes j} \otimes \Delta_\pi = 0 \quad \text{on } \mathcal{H}_{A'}^{\otimes j} \otimes \HA
\qquad \text{for every } j \ge 1,\\
&\qquad\text{where } \Delta_\pi := \Tr_B[\sigma_\pi] - \Tr[\sigma_\pi]\, \frac{\one_A}{d_A}
\end{aligned}
\end{equation}
(note $\Tr[\Delta_\pi] = 0$). Now let $M$ on $\mathcal{H}_{A'}$ and $H$ on $\HA$ be \emph{arbitrary} Hermitian operators, and take the Hilbert--Schmidt inner product of both sides of \eqref{eq:linkage-pointwise} with $M^{\otimes j} \otimes H$: the trace factorizes over the tensor legs, $\ip{\pi^{\otimes j} \otimes \Delta_\pi}{M^{\otimes j} \otimes H} = \ip{\pi}{M}^j \ip{\Delta_\pi}{H}$, and the pairing passes under the integral by linearity, so
\begin{equation}\label{eq:moment-annihilation}
\int d\mu(\pi)\, \ip{\pi}{M}^j \ip{\Delta_\pi}{H} = 0 \qquad \text{for all Hermitian } M, H, \text{ all } j \ge 0
\end{equation}
($j = 0$ is the $A'$-marginalized form of the $j = 1$ identity: take $M = \one$, using $\ip{\pi}{\one} = \Tr[\pi] = 1$). We now upgrade the averaged identities \eqref{eq:moment-annihilation} to a pointwise statement. Write $\mathcal{S} := \cS(\mathcal{H}_{A'})$ for the state space of $\mathcal{H}_{A'}$ --- a compact metrizable subset of the Hermitian operators --- and $C(\mathcal{S})$ for the real-valued continuous functions on it, with the uniform norm.

\emph{A signed measure.} For fixed Hermitian $H$ on $\HA$, the function $\pi \mapsto \ip{\Delta_\pi}{H}$ is measurable and $\mu$-integrable ($\abs{\ip{\Delta_\pi}{H}} \le \opnorm{H} \trnorm{\Delta_\pi} \le 2 \opnorm{H} \Tr[\sigma_\pi]$, and $\int \Tr[\sigma_\pi]\, d\mu = d_A$), so $d\nu_H := \ip{\Delta_\pi}{H}\, d\mu$ is a finite signed measure on $\mathcal{S}$, and \eqref{eq:moment-annihilation} says precisely that $\nu_H$ annihilates the linear span of the \emph{moment functions},
\begin{equation}\label{eq:moment-span}
\mathcal{P} := \operatorname{span}\bigl\{\, \pi \mapsto \ip{\pi}{M}^j \ : \ M \text{ Hermitian on } \mathcal{H}_{A'},\ j \ge 0 \,\bigr\} \ \subset\ C(\mathcal{S}).
\end{equation}

\emph{$\mathcal{P}$ is an algebra, by polarization.} Let $\mathcal{A} \subseteq C(\mathcal{S})$ be the unital algebra generated by the linear functionals $\ell_M : \pi \mapsto \ip{\pi}{M}$ --- the linear span of the constants together with all finite products $\ell_{M_1} \cdots \ell_{M_r}$. Clearly $\mathcal{P} \subseteq \mathcal{A}$; the reverse inclusion is the polarization identity for the symmetric $r$-linear form $(x_1, \dots, x_r) \mapsto x_1 \cdots x_r$,
\begin{equation}\label{eq:polarization}
x_1 x_2 \cdots x_r = \frac{1}{2^r\, r!} \sum_{\varepsilon \in \{\pm 1\}^r} \varepsilon_1 \cdots \varepsilon_r\, (\varepsilon_1 x_1 + \cdots + \varepsilon_r x_r)^r
\end{equation}
(for $r = 2$, \eqref{eq:polarization} reads $x_1 x_2 = \tfrac14\bigl[(x_1 + x_2)^2 - (x_1 - x_2)^2\bigr]$), applied with $x_i = \ell_{M_i}(\pi)$: by linearity of the pairing, $\varepsilon_1 x_1 + \cdots + \varepsilon_r x_r = \ip{\pi}{\textstyle\sum_i \varepsilon_i M_i}$ with $\sum_i \varepsilon_i M_i$ again Hermitian, so every product $\ell_{M_1} \cdots \ell_{M_r}$ is a finite linear combination of $r$-th powers $\ell_M^r \in \mathcal{P}$. Hence $\mathcal{P} = \mathcal{A}$, and $\nu_H$ annihilates the whole algebra $\mathcal{A}$.

\emph{Stone--Weierstrass.} $\mathcal{A}$ contains the constants (the case $j = 0$) and separates the points of $\mathcal{S}$: for states $\pi \ne \pi'$, the Hermitian choice $M = \pi - \pi'$ gives $\ell_M(\pi) - \ell_M(\pi') = \hsnorm{\pi - \pi'}^2 > 0$, by nondegeneracy of the Hilbert--Schmidt pairing. These are the hypotheses of the real Stone--Weierstrass theorem \cite[Thm.~7.32]{Rudin1976}, so $\mathcal{A}$ is uniformly dense in $C(\mathcal{S})$. A finite signed measure annihilating a uniformly dense subspace annihilates every $g \in C(\mathcal{S})$ ($\abs{\int g\, d\nu_H} = \abs{\int (g - f)\, d\nu_H} \le \norm{g - f}_\infty\, \abs{\nu_H}(\mathcal{S})$ for $f \in \mathcal{A}$ approximating $g$), so $\nu_H = 0$ as a measure, and its density with respect to $\mu$ vanishes: $\ip{\Delta_\pi}{H} = 0$ for $\mu$-almost every $\pi$, with the exceptional null set depending on $H$. This is the statement ``for every $H$, for almost every $\pi$'', whereas what is needed below is the exchanged order ``for almost every $\pi$, for every $H$'' --- and the exchange is not automatic, since the union of the exceptional null sets over the uncountably many Hermitian $H$ need not be null; finite dimensionality resolves this. Since the Hermitian operators on $\HA$ form a real vector space of finite dimension $d_A^2$, running over a basis $H_1, \dots, H_{d_A^2}$ and taking the finite (hence still null) union of the exceptional sets gives $\ip{\Delta_\pi}{H} = 0$ for \emph{all} $H$ simultaneously, $\mu$-almost everywhere; by nondegeneracy of the pairing once more, $\Delta_\pi = 0$ for $\mu$-almost every $\pi$:
\begin{equation}\label{eq:linkage-as}
\Tr_B[\sigma_\pi] = \Tr[\sigma_\pi]\, \frac{\one_A}{d_A} \qquad \text{$\mu$-a.s.}
\end{equation}

\emph{Step 5: conclusion.} By Step 4, $\mu$-almost every integrand $\pi \otimes \sigma_\pi$ of $Y_1 = \int d\mu\, \pi \otimes \sigma_\pi$ is a generator of the constrained separable cone $S$ (its $\tilde C$-factor $\sigma_\pi$ satisfies $\Tr_B[\sigma_\pi] \propto \one_A$ exactly). $S$ is a closed convex cone, and the integral is a limit of convex combinations of its generators (for every $W$ in the dual cone $S^*$, $\ip{W}{Y_1} = \int d\mu(\pi)\, \ip{W}{\pi \otimes \sigma_\pi} \ge 0$, the integral converging absolutely since $\pi \mapsto \Tr[\sigma_\pi]$ is $\mu$-integrable; hence $Y_1 \in S^{**} = S$ by the bipolar theorem for closed convex cones \cite[Thm.~14.1]{Rockafellar1970}), so $Y_1 \in S$. Now recall from Step 2 where $Y_1$ came from: it is the limit of the first marginals $X_1^{(k_n)} = \Tr_{A'_2 \cdots A'_{k_n}}[X^{(k_n)}]$ of the certificates of the fixed element $\omega \in \bigcap_k \bQ_k$, and each certificate reproduces $\omega$ through its first marginal via constraint (4), $\omega = \Lambda\bigl((\id \otimes \TA \otimes \id)\, X_1^{(k_n)}\bigr)$; since $\Lambda \circ (\id \otimes \TA \otimes \id)$ is continuous, passing to the limit gives $\omega = \Lambda\bigl((\id \otimes \TA \otimes \id)\, Y_1\bigr)$, which by $Y_1 \in S$ and Step 0 lies in $\Lambda\bigl((\id \otimes \TA \otimes \id)(S)\bigr) = \bQ$. This proves $\bigcap_k \bQ_k \subseteq \bQ$; together with Step 1, equality, which is part (iii). The PPT conditions are preserved in Step 2 but are not used to establish membership in $S$ or $\bQ$. All three parts therefore hold without them. Adding them can only tighten each outer approximation.
\end{proof}

\begin{remark}[consistency in infinite de Finetti arguments]\label{rem:definetti-consistency}
Existence of a symmetric extension at every finite level does not by itself make separately chosen extensions mutually consistent. In the completeness proof of \cite[Section~V, between Eqs.~(29) and~(30)]{DohertyParriloSpedalieri2004}, the assumption that every candidate at a fixed level fails to extend to some larger level is used to infer a single larger level at which all candidates fail. This quantifier exchange is not justified as written: the failure level may depend on the candidate. Compactness and diagonal extraction, as in Step~2 above and Lemma~\ref{lem:diagonal}, supply the missing argument by producing a consistent limiting family before applying the infinite de~Finetti theorem. The gap therefore concerns that proof step, not the validity of the completeness result. Example~\ref{ex:dicke-extension-consistency} in Appendix~\ref{app:consistency-example} gives a construction within the PPT-free hierarchy: each chosen two-copy candidate extends only to a finite level, these levels are unbounded, and the candidates converge to an infinitely extendible limit. The paragraph following the example shows that independently chosen extensions can also be incompatible when all PPT constraints are imposed.

Finite de~Finetti estimates avoid this issue altogether. For ordinary separability, \cite[Section~II.4, Theorem~II.7$'$]{ChristandlKonigMitchisonRenner2007} gives a separable approximation from each finite symmetric extension; \cite[Cor.~5]{NavascuesOwariPlenio2009} gives explicit bounds for Bose extensions. For affine-constrained problems, \cite[Theorem~2.3]{BertaBorderiFawziScholz2022} constructs approximations satisfying the constraints term by term, with error tending to zero, and its Theorem~3.1 deduces convergence of the optimization hierarchy. For a fixed marginal admitting extensions at all levels, these approximants converge to that marginal, which belongs to the relevant closed convex set. No compatibility between extensions at different levels is needed in this finite-level approach.
\end{remark}

Bose support and permutation invariance coincide at level $1$; the distinction matters at higher levels. The following theorem uses exactly the cones $\bQ_k$ just defined. Its proof applies the coherent-state measurement to all $k$ copies and prepares one pure state from its outcome. The resulting one-body identity is the standard Bose-extension rounding formula \cite{NavascuesOwariPlenio2009,Chiribella2011}. We give the identity and the channel marginal constraint argument explicitly.

\begin{theorem}[quantitative outer convergence]\label{thm:outer-rate}
For every $k\ge1$ and every $\omega\in\bQ_k$ with $\Tr[\omega]=1$, there is $\tilde\omega\in\base$ such that
\begin{equation}\label{eq:outer-rate}
\trnorm{\omega-\tilde\omega}\le\frac{2d_A^2(d_A-1)}{k+d_A-1}.
\end{equation}
Consequently, for the trace-norm Hausdorff distance $d_H$ defined in Section~\ref{sec:1.4}, and hence also for its Hilbert--Schmidt analogue,
\begin{equation}\label{eq:outer-rate-hausdorff}
d_H\bigl(\bQ_k\cap\{\Tr=1\},\,\base\bigr)\le\frac{2d_A^2(d_A-1)}{k+d_A-1}.
\end{equation}
Both statements hold with or without the PPT constraints (2).
\end{theorem}

\begin{proof}
\emph{Step 1: a normalized measurement on all copies.} Put $d=d_A$, $R=AB$, and $\hat X=X/d$ for a certificate $X$ of $\omega$. Thus $\hat X$ is a state supported on $\mathrm{Sym}^k(\mathcal{H}_{A'})\otimes R$. Write $D_j=\binom{j+d-1}{d-1}$ for the dimension of the symmetric subspace, and let $d\varphi$ be normalized Haar measure on pure states $\varphi=\ketbra{\varphi}{\varphi}$. Schur's lemma gives the coherent-state resolution \cite{Chiribella2011}
\begin{equation}\label{eq:outer-rate-resolution}
D_j\int d\varphi\,\varphi^{\otimes j}=P_j^{\mathrm{sym}}.
\end{equation}
Define the positive conditional operators on $R$ by
\begin{equation}\label{eq:outer-rate-tau}
\tau_\varphi
 :=D_k\Tr_{A'^k}\bigl[(\varphi^{\otimes k}\otimes\one_R)\hat X\bigr]\succeq0,
\qquad t_\varphi:=\Tr[\tau_\varphi].
\end{equation}
Positivity follows by writing the partial trace with $\varphi^{\otimes k}$ on both sides of $\hat X$. Equation~\eqref{eq:outer-rate-resolution} gives $\int\tau_\varphi\,d\varphi=\hat X_R$ and $\int t_\varphi\,d\varphi=1$.

\emph{Step 2: exact preservation of the channel marginal constraint.} Constraint (3) is $\hat X_{A'^k A}=\hat X_{A'^k}\otimes\one_A/d$. Since the measurement acts only on the copies,
\begin{equation}\label{eq:outer-rate-linkage}
\Tr_B[\tau_\varphi]
 =D_k\Tr\bigl[\varphi^{\otimes k}\hat X_{A'^k}\bigr]\frac{\one_A}{d}
 =t_\varphi\frac{\one_A}{d}.
\end{equation}
For $t_\varphi>0$, the operator $\tilde C_\varphi=d\tau_\varphi/t_\varphi$ is therefore a channel Choi matrix. If $t_\varphi=0$, positivity implies $\tau_\varphi=0$. It follows that
\begin{equation}\label{eq:outer-rate-rounded}
Z:=\int d\varphi\,\varphi\otimes\tau_\varphi
\end{equation}
is a trace-one element of $S$, and $dZ$ is a convex combination of pure-input/channel product generators.

\emph{Step 3: the exact one-body identity.} We first specify the tensor factors so that every term in the identity acts on the same space. Let $A'_0$ be an additional copy of $\mathcal{H}_{A'}$, separate from the $k$ copies supporting $\hat X$. For $i=1,\ldots,k$, let $V_i:\mathcal{H}_{A'_i}\to\mathcal{H}_{A'_0}$ be the unitary that maps each computational basis vector $\ket{a}_i$ to $\ket{a}_0$. Write $\hat X_{A'_iR}$ for the marginal of $\hat X$ retaining only copy $i$ and $R$, and define
\begin{equation}\label{eq:outer-rate-output-copy}
\hat X_{0R}:=(V_1\otimes\one_R)\hat X_{A'_1R}(V_1^\dagger\otimes\one_R),
\qquad Z_{0R}:=\int d\varphi\,\varphi_0\otimes\tau_\varphi.
\end{equation}
Here $\varphi_0=\ketbra{\varphi}{\varphi}$ on $A'_0$, with the same coordinates as $\varphi$ in \eqref{eq:outer-rate-rounded}. Thus $Z_{0R}$ is $Z$ written on the new copy: if the $A'$ factor of $Z$ is labelled $A'_1$, then $Z_{0R}=(V_1\otimes\one_R)Z(V_1^\dagger\otimes\one_R)$. These unitary maps only specify how to transfer matrices between isomorphic copies; they do not equate distinct physical subsystems. We will prove
\begin{equation}\label{eq:outer-rate-definetti}
Z_{0R}=\frac{k\hat X_{0R}+\one_0\otimes\hat X_R}{k+d}.
\end{equation}
Both summands on the right, and the operator on the left, act on $\mathcal{H}_{A'_0}\otimes R$; $\one_0$ denotes the identity on $\mathcal{H}_{A'_0}$.

\emph{From the integral to the symmetric projector.} Substitute \eqref{eq:outer-rate-tau} into the definition of $Z_{0R}$. The partial trace below removes the original copies $1,\ldots,k$ but leaves copy $0$ and $R$. By linearity, we can move the integral inside this partial trace:
\begin{equation}\label{eq:outer-rate-projector}
\begin{aligned}
Z_{0R}
 &=D_k\int d\varphi\,\varphi_0\otimes
   \Tr_{1,\ldots,k}\bigl[(\varphi^{\otimes k}\otimes\one_R)\hat X\bigr]\\
 &=D_k\Tr_{1,\ldots,k}\left[
   \left(\left(\int d\varphi\,\varphi_0\otimes\varphi^{\otimes k}\right)\otimes\one_R\right)
   (\one_0\otimes\hat X)\right]\\
 &=\frac{D_k}{D_{k+1}}\Tr_{1,\ldots,k}
   \bigl[(P_{k+1}^{\mathrm{sym}}\otimes\one_R)(\one_0\otimes\hat X)\bigr].
\end{aligned}
\end{equation}
The last equality is \eqref{eq:outer-rate-resolution} with $j=k+1$, applied to copies $0,1,\ldots,k$. In particular, the factor $D_k$ from the conditional operator remains, while the integral contributes $1/D_{k+1}$.

\emph{Expanding the projector.} Let $F_{0i}$ swap copies $0$ and $i$ and act as the identity on the other copies. The symmetric projector is the average of all permutation operators \cite[Prop.~7.1, eq.~(7.8)]{Watrous2018}. Each permutation of $0,\ldots,k$ has a unique decomposition as a transposition $(0\,i)$, or the identity, followed by a permutation fixing $0$: choose $i$ to be the image of $0$. Averaging first over permutations fixing $0$ therefore gives
\begin{equation}\label{eq:outer-rate-cosets}
P_{k+1}^{\mathrm{sym}}
 =\frac1{k+1}\left(\one+\sum_{i=1}^kF_{0i}\right)
   (\one_0\otimes P_k^{\mathrm{sym}}).
\end{equation}
Bose support means $(P_k^{\mathrm{sym}}\otimes\one_R)\hat X=\hat X$, so the final projector in \eqref{eq:outer-rate-cosets} can be removed when that expression is substituted into \eqref{eq:outer-rate-projector}. To obtain the $\one_0\otimes\hat X_R$ term in \eqref{eq:outer-rate-definetti}, consider the identity summand in \eqref{eq:outer-rate-cosets}: its full contribution to $Z_{0R}$ is
\begin{equation}\label{eq:outer-rate-identity-term}
\frac{D_k}{(k+1)D_{k+1}}\Tr_{1,\ldots,k}[\one_0\otimes\hat X]
 =\frac{D_k}{(k+1)D_{k+1}}(\one_0\otimes\hat X_R)
 =\frac{\one_0\otimes\hat X_R}{k+d}.
\end{equation}
Here the partial trace leaves copy $0$ unchanged, and we used $D_{k+1}/D_k=(k+d)/(k+1)$.

\emph{Evaluating a swap term.} First trace out all copies other than $0$ and $i$. Since $F_{0i}$ acts on neither those copies nor $R$, the remaining calculation is $\Tr_i[(F_{0i}\otimes\one_R)(\one_0\otimes\hat X_{A'_iR})]$. To evaluate it, write $\hat X_{A'_iR}=\sum_{a,b}\ketbra{a}{b}_i\otimes M_{ab}$, with $M_{ab}$ operators on $R$. Using independent indices for $F_{0i}=\sum_{p,q}\ketbra{p}{q}_0\otimes\ketbra{q}{p}_i$ gives
\begin{equation}\label{eq:outer-rate-swap-marginal}
\begin{aligned}
&\Tr_i\bigl[(F_{0i}\otimes\one_R)(\one_0\otimes\hat X_{A'_iR})\bigr]\\
&\quad=\sum_{p,q,a,b}\Tr\bigl[\ketbra{q}{p}_i\ketbra{a}{b}_i\bigr]\,\ketbra{p}{q}_0\otimes M_{ab}\\
&\quad=\sum_{p,q,a,b}\delta_{pa}\Tr\bigl[\ketbra{q}{b}_i\bigr]\,\ketbra{p}{q}_0\otimes M_{ab}\\
&\quad=\sum_{p,q,a,b}\delta_{pa}\delta_{qb}\,\ketbra{p}{q}_0\otimes M_{ab}\\
&\quad=\sum_{a,b}\ketbra{a}{b}_0\otimes M_{ab}\\
&\quad=(V_i\otimes\one_R)\hat X_{A'_iR}(V_i^\dagger\otimes\one_R).
\end{aligned}
\end{equation}
The first equality expands both operators and traces only the factor on copy $i$, leaving the factors on $0$ and $R$ unchanged. The next two equalities use $\ketbra{q}{p}_i\ketbra{a}{b}_i=\delta_{pa}\ketbra{q}{b}_i$ and $\Tr[\ketbra{q}{b}_i]=\delta_{qb}$. Summing over $p,q$ then sets $p=a$ and $q=b$, and the last equality uses $V_i\ket{a}_i=\ket{a}_0$. Thus the swap term transfers the marginal's matrix entries from copy $i$ to copy $0$, without transposing them. Permutation invariance of $\hat X$ makes all $k$ transported marginals equal to $\hat X_{0R}$. Combining the identity term and the $k$ swap terms yields
\begin{equation}\label{eq:outer-rate-combined}
Z_{0R}
 =\frac{D_k}{(k+1)D_{k+1}}
   \bigl(\one_0\otimes\hat X_R+k\hat X_{0R}\bigr).
\end{equation}
The dimension formula gives $D_{k+1}/D_k=(k+d)/(k+1)$, so the prefactor is $1/(k+d)$. This proves \eqref{eq:outer-rate-definetti} with every term on $\mathcal{H}_{A'_0}\otimes R$.

\emph{The trace-norm estimate.} To obtain the bound of \cite[Cor.~5]{NavascuesOwariPlenio2009}, put $L:=\one_0\otimes\hat X_R-Z_{0R}=\int(\one_0-\varphi_0)\otimes\tau_\varphi\,d\varphi$. Since $\varphi_0$ is a rank-one projector and $\tau_\varphi\succeq0$, each integrand is positive, so $L\succeq0$. Moreover, $\Tr L=\Tr[\one_0]\Tr[\hat X_R]-\Tr[Z_{0R}]=d\cdot1-1=d-1$. Substituting $\one_0\otimes\hat X_R=Z_{0R}+L$ into \eqref{eq:outer-rate-definetti} gives the following rearrangement:
\begin{equation}\label{eq:outer-rate-rearrangement}
\begin{aligned}
(k+d)Z_{0R}&=k\hat X_{0R}+Z_{0R}+L,\\
(k+d-1)Z_{0R}&=k\hat X_{0R}+L,\\
(k+d-1)(\hat X_{0R}-Z_{0R})&=(k+d-1)\hat X_{0R}-(k\hat X_{0R}+L)\\
&=(d-1)\hat X_{0R}-L.
\end{aligned}
\end{equation}
\begin{samepage}
Finally, conjugating by $V_1^\dagger\otimes\one_R$ transfers both $\hat X_{0R}$ and $Z_{0R}$ back to the space used for $\hat X_{A'_1R}$ and $Z$. Unitary conjugation preserves trace norm. As $k+d-1>0$, we obtain
\begin{equation}\label{eq:outer-rate-state-bound}
\begin{aligned}
\trnorm{\hat X_{A'_1R}-Z}
 &=\trnorm{\hat X_{0R}-Z_{0R}}\\
 &=\frac{\trnorm{(d-1)\hat X_{0R}-L}}{k+d-1}\\
 &\le\frac{(d-1)\trnorm{\hat X_{0R}}+\trnorm{L}}{k+d-1}\\
 &=\frac{(d-1)\Tr[\hat X_{0R}]+\Tr L}{k+d-1}\\
 &=\frac{(d-1)\cdot1+(d-1)}{k+d-1}
 =\frac{2(d-1)}{k+d-1}.
\end{aligned}
\end{equation}
\end{samepage}
The inequality uses the triangle inequality and homogeneity of the norm, with $d-1\ge0$. The following equality uses positivity of $\hat X_{0R}$ and $L$, for which the trace norm equals the trace. The final line substitutes $\Tr[\hat X_{0R}]=1$ and $\Tr L=d-1$.

\emph{Step 4: transfer to the stote.} Define
\begin{equation}\label{eq:outer-rate-tildeomega}
\tilde\omega:=\Lambda\bigl((\id\otimes\TA\otimes\id)(dZ)\bigr)
 =\int d\varphi\,t_\varphi\,(\varphi\otimes\one_B)\star\tilde C_\varphi^{\TA}.
\end{equation}
For every $\varphi$ with $t_\varphi>0$, Step 2 shows that $\tilde C_\varphi$ is a channel Choi matrix. Thus $(\varphi\otimes\one_B)\star\tilde C_\varphi^{\TA}$ is a trace-one stote. Outcomes with $t_\varphi=0$ contribute zero. Since $t_\varphi\ge0$ and $\int t_\varphi\,d\varphi=1$, equation~\eqref{eq:outer-rate-tildeomega} is an average of elements of the compact convex set $\base$, and hence $\tilde\omega\in\base$.

We now estimate its distance from $\omega$. In the remainder of this step, write $A'=A'_1$ and set
\begin{equation}\label{eq:outer-rate-transfer-definitions}
\begin{aligned}
\Theta&:=\id_{A'}\otimes\TA\otimes\id_B,
\qquad \Delta:=\hat X_{A'_1R}-Z,\\
\Lambda(Y)&:=\tfrac12\Tr_{A'}[FY+YF],
\qquad F:=F_{A'A}\otimes\one_B.
\end{aligned}
\end{equation}
Here $F_{A'A}$ is the swap defined by $F_{A'A}(\ket a_{A'}\otimes\ket b_A)=\ket b_{A'}\otimes\ket a_A$; thus $F$ exchanges $A'$ and $A$ and leaves $B$ unchanged. Both $\Delta$ and $\Theta(\Delta)$ are Hermitian operators on $\mathcal H_{A'}\otimes\HA\otimes\HB$. Constraint (4) for the certificate $X=d\hat X$ gives $\omega=d\Lambda(\Theta(\hat X_{A'_1R}))$, whereas \eqref{eq:outer-rate-tildeomega} gives $\tilde\omega=d\Lambda(\Theta(Z))$. By linearity,
\begin{equation}\label{eq:outer-rate-transfer-difference}
\omega-\tilde\omega=d\Lambda(\Theta(\Delta)).
\end{equation}
We bound the two maps $\Lambda$ and $\Theta$ separately.

\emph{The contraction map $\Lambda$.} The swap $F$ in \eqref{eq:outer-rate-transfer-definitions} satisfies $F=F^\dagger=F^{-1}$. For a Hermitian operator $Y$ on $A'AB$, put $H:=(FY+YF)/2$. This is Hermitian, and the defining swap formula gives $\Lambda(Y)=\Tr_{A'}H$. The partial trace is trace-norm contractive \cite[eq.~(1.183)]{Watrous2018}. In the Hermitian case needed here, this follows directly by writing $H=H_+-H_-$ for its positive and negative parts: their partial traces are positive, so
\begin{equation}\label{eq:outer-rate-partial-trace}
\begin{aligned}
\trnorm{\Tr_{A'}H}
 &\le\trnorm{\Tr_{A'}H_+}+\trnorm{\Tr_{A'}H_-}\\
 &=\Tr[\Tr_{A'}H_+]+\Tr[\Tr_{A'}H_-]\\
 &=\Tr H_++\Tr H_-=\trnorm{H}.
\end{aligned}
\end{equation}
Applying \eqref{eq:outer-rate-partial-trace}, then the triangle inequality, and finally invariance of the trace norm under multiplication by a unitary, gives
\begin{equation}\label{eq:outer-rate-lambda-bound}
\begin{aligned}
\trnorm{\Lambda(Y)}
 &=\trnorm{\Tr_{A'}H}\le\trnorm{H}\\
 &=\tfrac12\trnorm{FY+YF}\\
 &\le\tfrac12\bigl(\trnorm{FY}+\trnorm{YF}\bigr)
 =\trnorm{Y}.
\end{aligned}
\end{equation}

\emph{The partial transpose $\Theta$.} Transposing the $d$-dimensional factor $A$ satisfies the following bound for every Hermitian $Y$ on $A'AB$ \cite[Example~3.42 and Thm.~3.46]{Watrous2018}:
\begin{equation}\label{eq:outer-rate-transpose-bound}
\trnorm{\Theta(Y)}\le d\,\trnorm{Y}.
\end{equation}
To prove it, first take a unit vector $z$ and its Schmidt decomposition across $A{:}A'B$. For this calculation, order the tensor factors as $A\otimes(A'B)$; this unitary reordering does not change trace norms. In this ordering,
\begin{equation}\label{eq:outer-rate-schmidt}
\ket z=\sum_{i=1}^r s_i\ket{u_i}\otimes\ket{v_i},
\qquad r\le d,\quad s_i>0,\quad\sum_{i=1}^r s_i^2=1,
\end{equation}
where both families of vectors are orthonormal. Let $\ket{\bar u_i}$ be the vector obtained by complex-conjugating the coordinates of $\ket{u_i}$ in the fixed computational basis of $A$ used to define $\TA$. Partial transpose then gives
\begin{equation}\label{eq:outer-rate-transpose-action}
\begin{aligned}
\Theta(\ketbra{z}{z})
 &=\sum_{i,j}s_is_j\ketbra{\bar u_j}{\bar u_i}\otimes\ketbra{v_i}{v_j},\\
\Theta(\ketbra{z}{z})(\ket{\bar u_i}\otimes\ket{v_j})
 &=s_is_j\ket{\bar u_j}\otimes\ket{v_i}.
\end{aligned}
\end{equation}
Thus $\ket{\bar u_i}\otimes\ket{v_i}$ is an eigenvector with eigenvalue $s_i^2$. For $i<j$, the span of $\ket{\bar u_i}\otimes\ket{v_j}$ and $\ket{\bar u_j}\otimes\ket{v_i}$ has matrix $\bigl(\begin{smallmatrix}0&s_is_j\\s_is_j&0\end{smallmatrix}\bigr)$, whose eigenvalues are $s_is_j$ and $-s_is_j$. The operator is zero on the remaining subspace. Summing the absolute eigenvalues gives the pure-state formula of \cite[Prop.~8]{VidalWerner2002}, followed by Cauchy--Schwarz:
\begin{equation}\label{eq:outer-rate-pure-transpose}
\begin{aligned}
\trnorm{\Theta(\ketbra{z}{z})}
 &=\sum_i s_i^2+2\sum_{i<j}s_is_j
 =\left(\sum_i s_i\right)^2\\
 &\le r\sum_i s_i^2=r\le d.
\end{aligned}
\end{equation}
For general Hermitian $Y$, take a spectral decomposition $Y=\sum_j\lambda_j\ketbra{z_j}{z_j}$ with real $\lambda_j$ and orthonormal $z_j$. Linearity, the triangle inequality, and \eqref{eq:outer-rate-pure-transpose} give
\begin{equation}\label{eq:outer-rate-spectral-transpose}
\begin{aligned}
\trnorm{\Theta(Y)}
 &=\trnorm{\sum_j\lambda_j\Theta(\ketbra{z_j}{z_j})}\\
 &\le\sum_j|\lambda_j|\,\trnorm{\Theta(\ketbra{z_j}{z_j})}\\
 &\le d\sum_j|\lambda_j|=d\,\trnorm{Y},
\end{aligned}
\end{equation}
which proves \eqref{eq:outer-rate-transpose-bound}.

\emph{Combining the bounds.} Starting from \eqref{eq:outer-rate-transfer-difference}, apply \eqref{eq:outer-rate-lambda-bound} to the Hermitian operator $\Theta(\Delta)$, then \eqref{eq:outer-rate-transpose-bound} to $\Delta$, and finally the approximation bound \eqref{eq:outer-rate-state-bound}:
\begin{equation}\label{eq:outer-rate-final}
\begin{aligned}
\trnorm{\omega-\tilde\omega}
 &=d\,\trnorm{\Lambda(\Theta(\Delta))}\\
 &\le d\,\trnorm{\Theta(\Delta)}\\
 &\le d^2\trnorm{\Delta}
 =d^2\trnorm{\hat X_{A'_1R}-Z}\\
 &\le\frac{2d^2(d-1)}{k+d-1}.
\end{aligned}
\end{equation}
Thus one factor of $d$ comes from the normalization $X=d\hat X$, and the other comes from partial transpose; $\Lambda$ contributes no additional factor.
Since $\base\subseteq\bQ_k\cap\{\Tr=1\}$, the nested-set formula \eqref{eq:hausdorff-nested} turns the uniform bound just proved into \eqref{eq:outer-rate-hausdorff}. No PPT constraint was used.
\end{proof}

\begin{remark}[the role of symmetry and the computational scope]\label{rem:outer-rate}
Bose support is unnecessary for qualitative convergence. If it is replaced by permutation invariance, let the resulting cones be $\bQ_k^{\mathrm{perm}}$, with the same choice of PPT constraints. The proof of Theorem~\ref{thm:outer-hierarchy} still applies: the extraction, de~Finetti representation, and moment argument use only permutation invariance. Hence
\begin{equation}\label{eq:outer-permutation-variant}
\bQ\subseteq\bQ_{k+1}\subseteq\bQ_k\subseteq\bQ_k^{\mathrm{perm}},
\qquad \bigcap_k\bQ_k^{\mathrm{perm}}=\bQ.
\end{equation}
The constrained de~Finetti estimate of \cite[Theorem~2.3]{BertaBorderiFawziScholz2022}, applied with the subsystem identification above and followed by the linear map in (4), also gives $d_H(\bQ_k^{\mathrm{perm}}\cap\{\Tr=1\},\base)=O_{d_A,d_B}(k^{-1/2})$. This holds with or without the PPT constraints. Theorem~\ref{thm:outer-rate} instead uses Bose support through \eqref{eq:outer-rate-resolution} and \eqref{eq:outer-rate-cosets} to obtain an explicit $O(1/k)$ bound independent of $d_B$. No strict improvement of the actual approximation error at every finite level is asserted. Recent de~Finetti bounds with improved dimension dependence \cite{JeronimoWuXu2026definetti} concern unconstrained approximating product states; they do not by themselves establish the channel marginal constraint required here.

For the PPT-free hierarchy, the PSD variable can be stored on $\mathrm{Sym}^k(\mathcal{H}_{A'})\otimes\HA\otimes\HB$, of dimension $D_kd_Ad_B=O_{d_A}(k^{d_A-1}d_B)$. The channel marginal constraint can likewise be imposed on the compressed space. Thus, for fixed $d_A$, choosing $k\ge\max\{1,2d_A^2(d_A-1)/\varepsilon\}$ gives an outer approximation of $\base$ with trace-norm error at most $\varepsilon$ by an SDP whose matrix dimensions are polynomial in $d_B$ and $1/\varepsilon$. This size statement refers to the PPT-free formulation and does not count unreduced partial-transpose matrices as if they were already compressed.

Both quantitative estimates provide finite-accuracy guarantees and a finite sufficient level for excluding any trace-one operator a known positive distance from $\base$, with different dimension dependence and error scaling. Neither asserts finite termination of the two-sided exact-membership procedure at every boundary point, nor does polynomial SDP size alone establish a bit-complexity bound for exact feasibility.
\end{remark}

\subsection{Physical interpretation of the outer hierarchy certificates}\label{sec:outer-physical}

The extension in the DPS hierarchy describes one system sharing the same bipartite marginal with several interchangeable systems \cite[Sec.~II]{DohertyParriloSpedalieri2004}. Here the analogous extension concerns the positive input--channel certificate, not the generally non-positive operator $\omega$. We explain how condition (3) gives a channel with symmetric auxiliary outputs, how that channel can be implemented using a correlated ancilla, and what the convergence theorem says about this implementation.

\paragraph{A channel with input-independent auxiliary outputs.}
Fix $\Tr\omega=1$ and a level-$k$ certificate $X$, so $\Tr X=d_A$, and write $R=A'_1\cdots A'_k$. Introduce a physical input $A_{\rm in}\cong A$. The factor $A$ already present in $X$ will be the Choi reference, not a second input or an output. Define
\begin{equation}\label{eq:physical-linkage}
\tau_R:=\frac{\Tr_{AB}X}{d_A},
\qquad
\Tr_B X=\tau_R\otimes\one_A,
\qquad
\Tr_{RB}X=\one_A.
\end{equation}
The first operator is a state, the second identity is condition (3), and the third follows by tracing out $R$. Thus positivity and \eqref{eq:physical-linkage} make $X$ the unnormalized Choi matrix of a channel $\cE_k:A_{\rm in}\to RB$ \cite{Choi1975,Watrous2018}. More precisely, after reordering the factors of $X$ as $ARB$,
\begin{equation}\label{eq:physical-choi}
\frac{X}{d_A}=(\id_A\otimes\cE_k)(\ketbra{\Phi}{\Phi}),
\qquad
\ket{\Phi}_{AA_{\rm in}}=\frac1{\sqrt{d_A}}\sum_{a=1}^{d_A}\ket a_A\otimes\ket a_{A_{\rm in}}.
\end{equation}
The reduced Choi matrix in \eqref{eq:physical-linkage} is that of a replacement channel. Consequently,
\begin{equation}\label{eq:physical-no-signalling}
\Tr_B\cE_k(\eta)=\tau_R\,\Tr\eta.
\end{equation}
In particular, the joint auxiliary output is $\tau_R$ for every input state $\eta$. This is stronger than requiring only its individual $A'_i$ marginals to be input-independent. It does not require $R$ and $B$ to be uncorrelated: a channel can, for example, ignore its input and prepare correlated outputs. Bose support makes the $A'_i$ interchangeable and restricts their joint state to the symmetric subspace. The PPT requirements (2) are further constraints on the Choi matrix, not physical partial-transpose operations. These auxiliary systems are neither successive times nor copies of an unknown input.

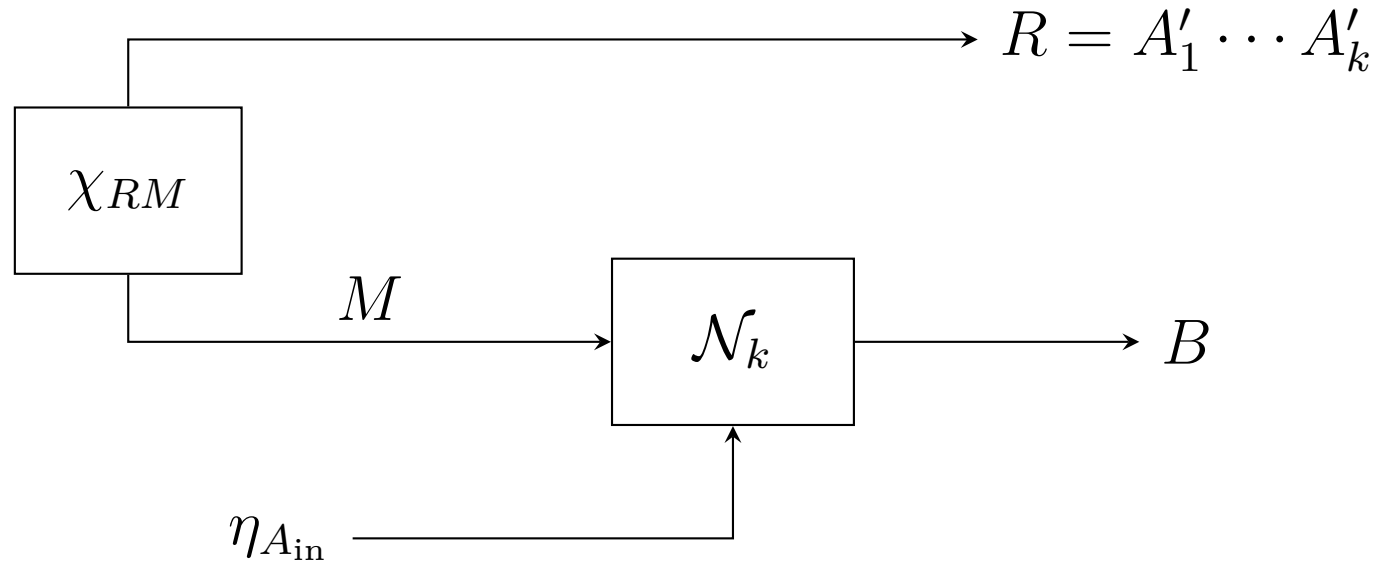
\begin{figure}[htbp]
\centering
\begin{tikzpicture}[>=stealth]
\node[draw,minimum width=1.5cm,minimum height=1.1cm] (prep) at (0,0) {$\chi_{RM}$};
\node[draw,minimum width=1.6cm,minimum height=1.1cm] (channel) at (4,-1) {$\mathcal N_k$};
\node (aux) at (7,1) {$R=A'_1\cdots A'_k$};
\node (output) at (7,-1) {$B$};
\node (input) at (1,-2.3) {$\eta_{A_{\rm in}}$};
\draw[->] (prep.north) -- (0,1) -- (aux.west);
\draw[->] (prep.south) -- (0,-1) -- node[above] {$M$} (channel.west);
\draw[->] (input.east) -- (4,-2.3) -- (channel.south);
\draw[->] (channel.east) -- (output.west);
\end{tikzpicture}
\caption{Schematic circuit for a realization of the certificate channel with a correlated ancilla. The fixed preparation supplies the retained auxiliary output $R$ and an ancilla $M$; the channel acts jointly on the ancilla and the physical input. The reconstruction of $\omega$ is not a further physical operation in this diagram.}\label{fig:physical-memory}
\end{figure}

\paragraph{A realization with a correlated ancilla.}
Equation~\eqref{eq:physical-no-signalling} permits a realization of the form
\begin{equation}\label{eq:physical-memory}
\cE_k(\eta)=(\id_R\otimes\mathcal N_k)(\chi_{RM}\otimes\eta),
\qquad
\mathcal N_k:MA_{\rm in}\longrightarrow B,
\end{equation}
where $M$ is an ancilla that may be correlated with $R$ in the fixed state $\chi_{RM}$, with $\Tr_M\chi_{RM}=\tau_R$, and $\mathcal N_k$ is a channel. One first prepares $\chi_{RM}$, retains $R$, and feeds $M$ together with the input into $\mathcal N_k$ (Figure~\ref{fig:physical-memory}). The input never acts on $R$, but correlations between $R$ and $M$ can produce correlations between $R$ and $B$. This is a special case of the semilocalizable realization for channels with one-way no-signalling \cite{EggelingSchlingemannWerner2002}: the party retaining $R$ has no input.

For completeness, choose $\chi_{RM}=\ketbra{\chi}{\chi}$ to be a minimal purification of $\tau_R$, and purify the normalized Choi state $X/d_A$ on $RABF$. Both this purification and $\ket\chi_{RM}\otimes\ket\Phi_{AA_{\rm in}}$ purify $\tau_R\otimes\one_A/d_A$. Uniqueness of purification \cite[Thm.~2.12]{Watrous2018} therefore gives an isometry $V:MA_{\rm in}\to BF$, enlarging $F$ if necessary, that maps the latter purification to the former while leaving $RA$ unchanged. The channel $\mathcal N_k(\zeta)=\Tr_F[V\zeta V^\dagger]$ then gives \eqref{eq:physical-memory}, because the two channels have the same Choi matrix.

This construction needs only an ancilla of dimension
\begin{equation}\label{eq:physical-memory-dimension}
\dim M=\operatorname{rank}\tau_R
\leq\dim\mathrm{Sym}^k(\mathcal H_{A'})
=\binom{k+d_A-1}{d_A-1}.
\end{equation}
The equality is achieved by a minimal purification of $\tau_R$, and the inequality follows from Bose support. For $d_A=2$, an ancilla of dimension at most $k+1$ suffices. With permutation invariance alone, the immediate bound is instead $\dim M\leq d_A^k$. These are sufficient bounds, not claims about the smallest possible ancilla dimension when mixed preparations $\chi_{RM}$ are allowed. They also do not bound the environment used to implement $\mathcal N_k$. The hierarchy optimizes over $X$, without a separate ancilla-dimension parameter; its level counts symmetric auxiliary outputs, not ancilla dimension.

\paragraph{Shared classical randomness in the limit.}
For $\omega$ belonging to every outer level, the compactness argument in Theorem~\ref{thm:outer-hierarchy} selects a consistent limiting family $Y_j$; independently chosen finite-level certificates need not themselves form such a family. Equations~\eqref{eq:definetti-rep} and \eqref{eq:linkage-as} give
\begin{equation}\label{eq:physical-limit-choi}
Y_j=\int \pi^{\otimes j}\otimes C_\pi\,d\nu(\pi),
\qquad
C_\pi\succeq0,\quad \Tr_B C_\pi=\one_A,
\end{equation}
where $d\nu(\pi)=\Tr(\sigma_\pi)\,d\mu(\pi)/d_A$ is a probability measure and $C_\pi=d_A\sigma_\pi/\Tr(\sigma_\pi)$ whenever $\Tr(\sigma_\pi)>0$; zero-weight terms are omitted. The constraints on $C_\pi$ hold for $\nu$-almost every $\pi$. Write $\cE_k$ for the channel associated with the certificate $X^{(k)}$ chosen in Step 2 of Theorem~\ref{thm:outer-hierarchy}, and let $(k_n)_n$ be the common subsequence selected there. For each fixed $j\geq1$, the channel with Choi matrix $Y_j$ is
\begin{equation}\label{eq:physical-reduced-channel-limit}
\cE_j^{\mathrm{lim}}(\eta)
=\lim_{n\to\infty}\Tr_{A'_{j+1}\cdots A'_{k_n}}\bigl[\cE_{k_n}(\eta)\bigr].
\end{equation}
Here only the tail with $k_n\geq j$ is used. The index $j$ counts the retained auxiliary outputs and remains fixed; the extension level $k_n$ tends to infinity. The limit follows from convergence of the corresponding Choi matrices on the fixed space $A\otimes(A')^{\otimes j}\otimes B$. By \eqref{eq:physical-limit-choi}, this limiting channel has the form
\begin{equation}\label{eq:physical-limit-channel}
\cE_j^{\mathrm{lim}}(\eta)=\int \pi^{\otimes j}\otimes\mathcal N_\pi(\eta)\,d\nu(\pi),
\end{equation}
with $\mathcal N_\pi$ the channel whose Choi matrix is $C_\pi$. A shared classical label $\pi$ now specifies both the auxiliary preparation and the channel applied to the input. The channels $\mathcal N_\pi$ remain arbitrary quantum channels; they need not destroy entanglement with an input reference. It is the correlation between the auxiliary preparation and the choice of channel that admits a classical description. This is the constrained counterpart of the separable-state interpretation of the DPS limit \cite{DohertyParriloSpedalieri2004,BertaBorderiFawziScholz2022}.

Finally, applying the reconstruction map to $Y_1$ gives
\begin{equation}\label{eq:physical-limit-stote}
\omega=\int(\pi_A\otimes\one_B)\star C_\pi^{T_A}\,d\nu(\pi)\in\base.
\end{equation}
Here $\pi_A$ is $\pi$ transferred from $A'$ to $A$ using the fixed basis identification. This is a convex mixture of stotes, not necessarily a single stote. The contraction and partial transpose in condition (4) are a mathematical reconstruction, not a quantum channel, and $\omega$ need not be positive. At finite level, Theorem~\ref{thm:outer-rate} supplies the quantitative version: measuring the auxiliary copies and preparing a state from the outcome gives a state--channel mixture with the stated error after reconstruction. These statements interpret the certificates without asserting that $\omega$ itself is an ordinary output state.

\subsection{Finite-input inner hierarchy}\label{sec:inner-hierarchy}

The inner construction discretizes only the pure-input variable. Lemma~\ref{lem:pure-input} shows that pure inputs generate the full convex hull. Definition~\ref{def:finite-inner-cone} associates an inner cone with any set of $N$ pure-input directions, retaining the full channel set at each direction. Theorem~\ref{thm:inner-rate} bounds its Hausdorff error and constructs nested $N$-direction sets with an explicit bound in terms of $N$. Definition~\ref{def:inner-relaxation} uses those nested sets to define the hierarchy, and Corollary~\ref{prop:inner-validity} establishes its closedness, SDP representation, inner inclusion, monotonicity, and convergence. Remark~\ref{rem:covering-numbers} prescribes the number of directions for a desired error.

For a unit vector $\ket\psi$, write $P_\psi:=\ketbra{\psi}{\psi}$ for its pure-state projector. Input directions are the corresponding rays in $\mathbb{CP}^{d_A-1}$; expressions indexed by a direction are independent of the phase of its unit representative. For a fixed pure input $P_\psi$, define the channel slice
\begin{equation}\label{eq:channel-slice}
B_\psi := \{(P_\psi \otimes \one_B) \star J : J \in \cJ\} = \{(P_\psi \otimes \one_B) \star \tilde{C}^{\TA} : \tilde{C} \succeq 0,\ \Tr_B[\tilde{C}] = \one_A\}.
\end{equation}
This is the linear image of the Choi spectrahedron, so it is convex and SDP-representable (a spectrahedral shadow) \cite{BlekhermanParriloThomas2013}. Thus only the input direction needs to be discretized; the channel optimization remains exact.

\begin{definition}[finite-input inner cone]\label{def:finite-inner-cone}
Fix an integer $N\ge1$. For a set $\Psi=\{\psi_1,\ldots,\psi_N\}\subset\mathbb{CP}^{d_A-1}$ of $N$ pure-input directions, define
\begin{equation}\label{eq:finite-inner-cone}
\bQ(\Psi) := \left\{ \sum_{l=1}^{N} (P_{\psi_l} \otimes \one_B) \star \tilde{C}_l^{\TA} \ : \ \tilde{C}_l \succeq 0,\ \Tr_B[\tilde{C}_l] \propto \one_A \right\} = \sum_{l=1}^{N} \cone(B_{\psi_l}).
\end{equation}
\end{definition}

Its unit-trace slice is
\begin{equation}\label{eq:finite-inner-base}
\base(\Psi):=\bQ(\Psi)\cap\{\Tr=1\}=\conv\left(\bigcup_{\psi\in\Psi}B_\psi\right)\subseteq\base.
\end{equation}
Indeed, write $\Tr_B[\tilde C_l]=t_l\one_A$. Positivity implies $t_l\ge0$, and $t_l=0$ forces $\tilde C_l=0$. For $t_l>0$, the matrix $\tilde C_l/t_l$ is a channel Choi matrix, and the corresponding summand in \eqref{eq:finite-inner-cone} has trace $t_l$. Unit trace therefore means $\sum_l t_l=1$. Conversely, convexity of each $B_{\psi_l}$ allows all terms with the same input to be combined into a single channel block. Each $B_\psi$ is compact, being a continuous image of the compact channel set, so $\base(\Psi)$ is compact and convex.

For the approximation estimates, equip $\mathbb{CP}^{d_A-1}$ with the Fubini--Study distance \cite[Ch.~4]{BengtssonZyczkowski2017},
\begin{equation}\label{eq:FS-distance}
\theta_{\mathrm{FS}}(\psi, \psi') := \arccos \abs{\braket{\psi}{\psi'}} \in [0, \pi/2],
\end{equation}
the angle between the rays of $\ket{\psi}$ and $\ket{\psi'}$. The \emph{covering radius} of a finite set $\Psi \subset \mathbb{CP}^{d_A-1}$ is the standard notion from metric geometry \cite[\S 4.2]{Vershynin2018}: the largest distance from a point of the space to the set,
\begin{equation}\label{eq:covering-radius}
\delta(\Psi) := \max_{\psi \in \mathbb{CP}^{d_A-1}}\ \min_{\psi_l \in \Psi}\ \theta_{\mathrm{FS}}(\psi, \psi_l),
\end{equation}
equivalently the smallest $\delta$ such that the closed Fubini--Study balls of radius $\delta$ centered at the points of $\Psi$ cover $\mathbb{CP}^{d_A-1}$.

\begin{theorem}[quantitative finite-input approximation]\label{thm:inner-rate}
Let $d_A\ge2$, and let $d_H$ be the trace-norm Hausdorff distance defined in Section~\ref{sec:1.4}. For any nonempty finite input set $\Psi$,
\begin{equation}\label{eq:inner-rate}
d_H\big(\base(\Psi),\,\base\big)\le 2(d_A-1)\sin^2\delta(\Psi).
\end{equation}
There exist nested sets $\Psi_N\subset\Psi_{N+1}$ with $|\Psi_N|=N$ for every integer $N\ge1$ such that
\begin{equation}\label{eq:rate-in-N}
d_H\big(\base(\Psi_N),\,\base\big)\le 2\pi^2(d_A-1)\,N^{-1/(d_A-1)}.
\end{equation}
\end{theorem}

\begin{proof}
\emph{Step 1: an input bound independent of the output dimension.} Write $d=d_A$, let $A_0$ be a copy of $A$, and let $F$ swap $A$ and $A_0$, so $F(\ket a_A\otimes\ket b_{A_0})=\ket b_A\otimes\ket a_{A_0}$. For a channel $\cE:A_0\to B$, its Jamio{\l}kowski matrix is $J=(\id_A\otimes\cE)(F)$. For unit vectors $\ket\psi,\ket\phi$, set $\Delta=P_\psi-P_\phi$ and $\theta=\theta_{\mathrm{FS}}(\psi,\phi)$. Since the channel acts only on the second factor, multiplication by $\Delta$ on $A$ can be taken inside the map on either side: for every operator $Y$ on $AA_0$,
\begin{equation}\label{eq:input-channel-intertwining}
\begin{aligned}
(\Delta\otimes\one_B)(\id_A\otimes\cE)(Y)
&=(\id_A\otimes\cE)\bigl((\Delta\otimes\one_{A_0})Y\bigr),\\
(\id_A\otimes\cE)(Y)(\Delta\otimes\one_B)
&=(\id_A\otimes\cE)\bigl(Y(\Delta\otimes\one_{A_0})\bigr).
\end{aligned}
\end{equation}
For a simple tensor $Y=X\otimes Z$, both sides of the first identity equal $\Delta X\otimes\cE(Z)$, and both sides of the second equal $X\Delta\otimes\cE(Z)$. Linearity extends these identities to every $Y$. Define $T_\Delta:=(\Delta\otimes\one_{A_0})\star F$. By bilinearity and \eqref{eq:input-channel-intertwining} with $Y=F$,
\begin{equation}\label{eq:input-difference}
\begin{aligned}
&(P_\psi\otimes\one_B)\star J-(P_\phi\otimes\one_B)\star J\\
&\quad=(\Delta\otimes\one_B)\star J\\
&\quad=(\id_A\otimes\cE)\!\left(\tfrac12\bigl[(\Delta\otimes\one_{A_0})F+F(\Delta\otimes\one_{A_0})\bigr]\right)\\
&\quad=(\id_A\otimes\cE)(T_\Delta).
\end{aligned}
\end{equation}
To compute the norm of $T_\Delta$, use $F(\Delta\otimes\one_{A_0})=(\one_A\otimes\Delta)F$, where $\Delta$ on $A_0$ denotes its copy under the fixed identification with $A$. Thus
\begin{equation}\label{eq:input-swap-form}
\begin{aligned}
T_\Delta
&=\tfrac12\bigl[(\Delta\otimes\one_{A_0})F+F(\Delta\otimes\one_{A_0})\bigr]\\
&=\tfrac12(\Delta\otimes\one_{A_0}+\one_A\otimes\Delta)F.
\end{aligned}
\end{equation}
The eigenvalues of $\Delta$ are $s,-s$ and $d-2$ zeros, where $s=\sin\theta$: on the span of the two input vectors, $\Delta$ is traceless and has determinant $-\sin^2\theta$. If the rays coincide, $\Delta=0$ and the same formulas hold with $s=0$. Since $F$ is unitary, writing these eigenvalues as $\lambda_1,\ldots,\lambda_d$ gives
\begin{equation}\label{eq:input-swap-norm}
\trnorm{T_\Delta}
=\tfrac12\sum_{i,j=1}^d\abs{\lambda_i+\lambda_j}
=\tfrac12\bigl(4s+4(d-2)s\bigr)
=2(d-1)\sin\theta.
\end{equation}
Channels contract the trace norm of Hermitian operators \cite[\S 3.3]{Watrous2018}. Indeed, for a Hermitian $H=H_+-H_-$, positivity and trace preservation give $\trnorm{\Phi(H)}\le\Tr\Phi(H_+)+\Tr\Phi(H_-)=\trnorm{H}$. Applying this to the channel $\id_A\otimes\cE$ and the Hermitian $T_\Delta$ yields
\begin{equation}\label{eq:input-lipschitz}
\trnorm{(P_\psi\otimes\one_B)\star J-(P_\phi\otimes\one_B)\star J}
\le 2(d-1)\sin\theta,
\end{equation}
uniformly over $J\in\cJ$.

\emph{Step 2: a quadratic bound through support functions.} We combine \eqref{eq:input-lipschitz} with the support-function characterization of Hausdorff distance to bound the distance between $\base$ and $\base(\Psi)$. Schneider \cite[Lemma~1.8.14, p.~66]{Schneider2014} states that the Euclidean Hausdorff distance between compact convex sets equals the supremum of the absolute difference of their support functions over unit directions. For the trace norm, the corresponding formula uses the dual operator-norm unit ball. Writing $h_\Psi(K):=\max_{\omega\in\base(\Psi)}\Tr(K\omega)$, the inclusion $\base(\Psi)\subseteq\base$ removes the absolute value and gives
\begin{equation}\label{eq:inner-hausdorff-support}
d_H(\base(\Psi),\base)
=\max_{K=K^\dagger,\,\opnorm{K}\le1}\bigl(h(K)-h_\Psi(K)\bigr).
\end{equation}
We justify this trace-norm version at the end of this step. It therefore suffices to bound the support-function difference uniformly over Hermitian costs $K$ with $\opnorm{K}\le1$. Fix such a $K$. For each $J\in\cJ$, define the Hermitian operator $A_J(K):=\Tr_B(J\star K)$ on $A$. Cyclicity of the trace gives
\begin{equation}\label{eq:inner-input-objective}
\Tr\bigl[K((P_\psi\otimes\one_B)\star J)\bigr]
=\Tr\bigl[(P_\psi\otimes\one_B)(J\star K)\bigr]
=\bra\psi A_J(K)\ket\psi.
\end{equation}
Apply \eqref{eq:input-lipschitz} to orthogonal unit eigenvectors $u_+,u_-$ for the largest and smallest eigenvalues of $A_J(K)$. Trace--operator norm duality then bounds its spectral diameter:
\begin{equation}\label{eq:inner-spectral-diameter}
\begin{aligned}
\lambda_{\max}(A_J(K))-\lambda_{\min}(A_J(K))
&=\Tr\bigl[K\bigl(((P_{u_+}-P_{u_-})\otimes\one_B)\star J\bigr)\bigr]\\
&\le\opnorm{K}\,\trnorm{((P_{u_+}-P_{u_-})\otimes\one_B)\star J}
\le2(d-1).
\end{aligned}
\end{equation}
By Lemma~\ref{lem:pure-input}, $h(K)$ is the maximum over pure-input generators. Thus \eqref{eq:inner-input-objective} gives
\begin{equation}\label{eq:inner-support-eigenvalue}
h(K)=\max_{J\in\cJ}\max_{\langle\psi|\psi\rangle=1}\bra\psi A_J(K)\ket\psi
=\max_{J\in\cJ}\lambda_{\max}(A_J(K)).
\end{equation}
Compactness of the pure-input and channel sets gives an attaining pair $(\psi,J)$. Fix such a pair. Then
\begin{equation}\label{eq:inner-optimal-pair}
h(K)=\bra\psi A_J(K)\ket\psi=\lambda_{\max}(A_J(K)),
\qquad A_J(K)\ket\psi=h(K)\ket\psi.
\end{equation}
Indeed, for this fixed $J$, $\ket\psi$ must be a top eigenvector of $A_J(K)$, since otherwise another input would increase the objective. Choose a nearest direction $\phi\in\Psi$, and let $\theta:=\theta_{\mathrm{FS}}(\psi,\phi)\le\delta(\Psi)$ be the Fubini--Study angle between $\psi$ and $\phi$. After choosing its phase, write $\ket\phi=\cos\theta\ket\psi+\sin\theta\ket v$ with $\ket v$ unit and orthogonal to $\ket\psi$. When $\theta=0$, any such $v$ may be used. Because $\psi$ is an eigenvector, the cross terms vanish, and
\begin{equation}\label{eq:inner-rayleigh-loss}
\bra\phi A_J(K)\ket\phi
=\cos^2\theta\,\lambda_{\max}(A_J(K))
+\sin^2\theta\,\bra v A_J(K)\ket v.
\end{equation}
The generator with input $\phi$ and the same channel $J$ belongs to $\base(\Psi)$, so $h_\Psi(K)\ge\bra\phi A_J(K)\ket\phi$. Using \eqref{eq:inner-optimal-pair} for $h(K)$ and then \eqref{eq:inner-rayleigh-loss}, we obtain
\begin{equation}\label{eq:finite-input-approximation}
\begin{aligned}
0\le h(K)-h_\Psi(K)
&\le\lambda_{\max}(A_J(K))-\bra\phi A_J(K)\ket\phi\\
&=\sin^2\theta\bigl(\lambda_{\max}(A_J(K))-\bra v A_J(K)\ket v\bigr)\\
&\le2(d-1)\sin^2\theta
\le2(d-1)\sin^2\delta(\Psi).
\end{aligned}
\end{equation}
To justify \eqref{eq:inner-hausdorff-support}, let $D_1$ be the closed trace-norm unit ball of Hermitian operators. By \eqref{eq:hausdorff-nested}, $d_H(\base(\Psi),\base)\le r$ is equivalent to $\base\subseteq\base(\Psi)+rD_1$. The support function of the latter set is $h_\Psi(K)+r\opnorm{K}$, by trace--operator norm duality. The separating-hyperplane theorem \cite[\S 11]{Rockafellar1970} makes this inclusion equivalent to $h(K)\le h_\Psi(K)+r\opnorm{K}$ for every Hermitian $K$, which proves \eqref{eq:inner-hausdorff-support}. Combining it with \eqref{eq:finite-input-approximation} proves \eqref{eq:inner-rate}. The quadratic dependence on the covering radius comes from the Rayleigh quotient at a maximizing input and the convexity of the bases; Step 1 alone gives only a linear bound for replacing an individual generator by one with a nearby input.

\emph{Step 3: nested input sets with prescribed cardinality.} Assume $d_A\ge2$. Choose any first direction $\psi_1$ and apply the farthest-point rule \cite{Gonzalez1985} on $\mathbb{CP}^{d_A-1}$:
\begin{equation}\label{eq:farthest-point-rule}
\psi_{n+1}\in\operatorname*{arg\,max}_{\psi\in\mathbb{CP}^{d_A-1}}\ \min_{1\le l\le n}\theta_{\mathrm{FS}}(\psi,\psi_l).
\end{equation}
The maximum exists by compactness and continuity. For any starting direction, we obtain an $N$-point set of input directions $\Psi_N:=\{\psi_1,\ldots,\psi_N\}$ and covering radius $\delta_N:=\delta(\Psi_N)$. A finite set does not cover the positive-dimensional pure-state space with radius zero, so each new direction is distinct and $|\Psi_N|=N$. The covering radii are nonincreasing in $N$. It remains to show that $\delta_N\leq\pi N^{-1/(2(d_A-1))}$.

For $1\le i<j\le N$, the choice of $\psi_j$ implies
\begin{equation}\label{eq:farthest-point-separation}
\theta_{\mathrm{FS}}(\psi_i,\psi_j)\ge\min_{1\le l<j}\theta_{\mathrm{FS}}(\psi_l,\psi_j)=\delta_{j-1}\ge\delta_N.
\end{equation}
The radius-$\delta_N$ balls cover the space by definition of $\delta_N$, and these covering balls may overlap. To count the points, use instead the open balls of radius $\delta_N/2$ about the same centres. These smaller balls are disjoint: a point in two of them would, by the triangle inequality, put their centres at distance less than $\delta_N$. This is the usual packing argument for bounding covering numbers \cite[\S 4.2]{Vershynin2018}.

We compute the normalized Haar volume of a Fubini--Study ball of radius $r$ from the angular volume element in \cite[\S 4.7, Eqs.~(4.86)--(4.88), p.~130]{BengtssonZyczkowski2017}, and then apply the formula with $r=\delta_N/2$ to the disjoint balls above. Put $n=d_A-1$ and let $\mu$ be normalized Haar measure on $\mathbb{CP}^n$. For any centre represented by a unit vector $\psi$, choose a unitary $U$ with $U\psi=e_n$. Both the Fubini--Study metric and $\mu$ are unitarily invariant \cite[\S\S 4.5, 4.7]{BengtssonZyczkowski2017}. Hence, writing $B_{\mathrm{FS}}(\psi,r)$ for the ball centred at the ray of $\psi$,
\begin{equation}\label{eq:FS-ball-centre-invariance}
U\bigl(B_{\mathrm{FS}}(\psi,r)\bigr)=B_{\mathrm{FS}}(e_n,r),\qquad
\mu\bigl(B_{\mathrm{FS}}(\psi,r)\bigr)=\mu\bigl(B_{\mathrm{FS}}(e_n,r)\bigr)=:v(r).
\end{equation}
This is a change of coordinates adapted to each individual centre; it does not require a single unitary sending all the centres to $e_n$.

The reference writes a unit vector as $Z=(n_0,n_1e^{i\nu_1},\ldots,n_ne^{i\nu_n})$, with $n_n=\cos\vartheta_n$ and $0\le\vartheta_i\le\pi/2$ \cite[Eqs.~(4.68)--(4.70), pp.~122--123]{BengtssonZyczkowski2017}. Thus $\abs{\braket{e_n}{Z}}=\cos\vartheta_n$, so the Fubini--Study distance from $e_n$ is precisely $\vartheta_n$. The unnormalized volume element underlying Eq.~(4.88) of that reference is
\begin{equation}\label{eq:FS-angular-volume-element}
\mathrm{d}V=\prod_{i=1}^{n}\cos\vartheta_i\,\sin^{2i-1}\vartheta_i\,\mathrm{d}\vartheta_i\,\mathrm{d}\nu_i,
\qquad 0\le\vartheta_i\le\frac{\pi}{2},\quad 0\le\nu_i<2\pi.
\end{equation}
Although the components of $Z$ depend on several angles, this volume element factorizes and the angular ranges are independent. For $1\le i\le n$ and $0\le R\le\pi/2$, define and evaluate the individual factors:
\begin{equation}\label{eq:FS-angular-factor}
\begin{aligned}
I_i(R)&:=\int_0^R\int_0^{2\pi}\cos\vartheta_i\,\sin^{2i-1}\vartheta_i\,\mathrm{d}\nu_i\,\mathrm{d}\vartheta_i\\
&=2\pi\int_0^R\cos t\,\sin^{2i-1}t\,\mathrm{d}t
=\frac{\pi}{i}\left[\sin^{2i}t\right]_0^R=\frac{\pi}{i}\sin^{2i}R.
\end{aligned}
\end{equation}
The ball condition restricts only $\vartheta_n\le r$; all other angles and phases still range over their full domains. Consequently, all factors with $i<n$ are integrated and then cancel against the normalization:
\begin{equation}\label{eq:FS-ball-volume}
\begin{aligned}
v(r)&=\frac{\left[\prod_{i=1}^{n-1}I_i(\pi/2)\right]I_n(r)}{\prod_{i=1}^{n}I_i(\pi/2)}
=\frac{I_n(r)}{I_n(\pi/2)}\\
&=2n\int_0^r\cos\theta\,\sin^{2n-1}\theta\,\mathrm{d}\theta
=\left[\sin^{2n}\theta\right]_0^r=\sin^{2(d_A-1)}r.
\end{aligned}
\end{equation}
The $N$ disjoint smaller balls have total normalized volume at most one, so
\begin{equation}\label{eq:net-packing-volume}
N\,\sin^{2(d_A-1)}(\delta_N/2)=N\,v(\delta_N/2)\le1.
\end{equation}
Since $0<\delta_N\le\pi/2$ and $\sin(\delta_N/2)\ge\delta_N/\pi$, this yields
\begin{align}
N\left(\frac{\delta_N}{\pi}\right)^{2(d_A-1)}&\le1,\label{eq:net-size-bound}\\
\delta_N&\le\pi\,N^{-1/(2(d_A-1))}.\label{eq:net-tradeoff}
\end{align}
Combining \eqref{eq:net-tradeoff} with \eqref{eq:inner-rate} and $\sin\delta_N\le\delta_N$ proves \eqref{eq:rate-in-N} for these nested input sets.
\end{proof}

\begin{definition}[inner hierarchy]\label{def:inner-relaxation}
For $d_A\ge2$, fix one nested sequence $(\Psi_N)_{N\ge1}$ constructed in Step 3 of the proof of Theorem~\ref{thm:inner-rate}. Let $(N_m)_{m\ge1}$ be any nondecreasing sequence of positive integers tending to infinity. Define the level-$m$ inner cone and its input covering radius by
\begin{equation}\label{eq:inner-relaxation-def}
\bQ^{(m)}:=\bQ(\Psi_{N_m}),\qquad\delta_m:=\delta(\Psi_{N_m}).
\end{equation}
\end{definition}

\begin{corollary}[validity, representability, monotonicity, convergence]\label{prop:inner-validity}
Each $\bQ^{(m)}$ is: (i) a closed SDP-representable cone (a spectrahedral shadow), with an explicit representation using $N_m$ PSD blocks of size $d_A d_B\times d_A d_B$ once the input directions are fixed; (ii) inner, $\bQ^{(m)}\subseteq\bQ$; and (iii) monotone, $\bQ^{(m)}\subseteq\bQ^{(m+1)}$. Moreover, (iv) the hierarchy exhausts $\bQ$ in the sense that
\begin{equation}\label{eq:inner-exhaustion}
\overline{\bigcup_m\bQ^{(m)}}=\bQ.
\end{equation}
\end{corollary}

\begin{proof}
(i) follows from \eqref{eq:finite-inner-cone}: stack the block-diagonal PSD variables $\tilde C_1,\ldots,\tilde C_{N_m}$, impose the linear constraints $\Tr_B[\tilde C_l]\propto\one_A$, and apply the displayed linear map. This is a spectrahedral shadow \cite{BlekhermanParriloThomas2013}. Closedness follows separately from the compact trace-one base \eqref{eq:finite-inner-base}, by the argument of Proposition~\ref{prop:cone-properties}(ii). (ii) follows from \eqref{eq:finite-inner-base} by taking cones. For (iii), $N_m\le N_{m+1}$ and the nested construction imply $\Psi_{N_m}\subseteq\Psi_{N_{m+1}}$; the additional channel blocks can be set to zero.

For (iv), Theorem~\ref{thm:inner-rate} and $N_m\to\infty$ give
\begin{equation}\label{eq:hierarchy-inner-rate}
d_H\big(\bQ^{(m)}\cap\{\Tr=1\},\,\base\big)\le 2\pi^2(d_A-1)\,N_m^{-1/(d_A-1)}\longrightarrow0.
\end{equation}
For any nonzero $X\in\bQ$, write $X=t\omega$ with $t=\Tr[X]>0$ and $\omega\in\base$. Choose $\omega_m\in\bQ^{(m)}\cap\{\Tr=1\}$ converging to $\omega$. Then $t\omega_m\in\bQ^{(m)}$ and $\trnorm{X-t\omega_m}=t\trnorm{\omega-\omega_m}\to0$. The zero element belongs to every level, so $\bQ\subseteq\overline{\bigcup_m\bQ^{(m)}}$. The reverse inclusion follows from (ii) and closedness of $\bQ$.
\end{proof}

\begin{remark}[size for a prescribed error]\label{rem:covering-numbers}
To prescribe Hausdorff errors $\eta_m>0$ with $\eta_m\downarrow0$, choose in Definition~\ref{def:inner-relaxation}
\begin{equation}\label{eq:prescribed-inner-size}
N_m:=\left\lceil\left(\frac{2\pi^2(d_A-1)}{\eta_m}\right)^{d_A-1}\right\rceil.
\end{equation}
These cardinalities are nondecreasing and tend to infinity. Equation~\eqref{eq:hierarchy-inner-rate} then gives
\begin{equation}\label{eq:prescribed-inner-error}
d_H\big(\bQ^{(m)}\cap\{\Tr=1\},\,\base\big)\le 2\pi^2(d_A-1)\,N_m^{-1/(d_A-1)}\le\eta_m.
\end{equation}
This is a sufficient number of PSD blocks, not a claim of minimality. For fixed $d_A$, \eqref{eq:prescribed-inner-size} grows as $O_{d_A}(\eta_m^{-(d_A-1)})$, independently of $d_B$; each block still has size $d_Ad_B$. Exact farthest-point selection is not, by itself, an efficient finite-precision algorithm.
\end{remark}

\subsection{Membership certificates and optimization error bounds}\label{sec:sandwich}

Retain $d_A\ge2$ from Definition~\ref{def:inner-relaxation}. The two hierarchies give certificates of membership and nonmembership at finite levels, together with error bounds for optimization. These conclusions do not imply that the hierarchy decides exact membership at every boundary point after finitely many levels.

\begin{remark}[a two-sided certificate]\label{rem:two-sided}
At every finite $(m,k)$, both sides of $\bQ^{(m)}\subseteq\bQ\subseteq\bQ_k$ have semidefinite representations. A feasible inner decomposition $\omega=\sum_{l=1}^{N_m}(P_{\psi_l}\otimes\one)\star\tilde C_l^{\TA}$, using the fixed directions $\{\psi_1,\ldots,\psi_{N_m}\}=\Psi_{N_m}$ and blocks satisfying $\tilde C_l\succeq0$ and $\Tr_B\tilde C_l\propto\one_A$, certifies $\omega\in\bQ^{(m)}\subseteq\bQ$. A witness $K\in\bQ_k^*$ with $\ip{\omega}{K}<0$ certifies $\omega\notin\bQ$. These are exact statements about feasible certificates; floating-point solver output requires additional feasibility and error checks before it gives a rigorous certificate.

For Hermitian $K$, let $h_k^{\mathrm{out}}(K)$ and $h_m^{\mathrm{in}}(K)$ denote the support functions of the trace-one slices of $\bQ_k$ and $\bQ^{(m)}$. By Theorem~\ref{thm:outer-rate} and \eqref{eq:hierarchy-inner-rate}, valid upper bounds on their trace-norm Hausdorff distances from $\base$ are
\begin{equation}\label{eq:two-sided-error-bounds}
\varepsilon_k^{\mathrm{out}}:=\frac{2d_A^2(d_A-1)}{k+d_A-1},
\qquad
\varepsilon_m^{\mathrm{in}}:=2\pi^2(d_A-1)\,N_m^{-1/(d_A-1)}.
\end{equation}
With the choice of $N_m$ in Remark~\ref{rem:covering-numbers}, the inner bound satisfies $\varepsilon_m^{\mathrm{in}}\le\eta_m$. Trace--operator norm duality gives
\begin{equation}\label{eq:support-sandwich-gap}
h_m^{\mathrm{in}}(K)\le h(K)\le h_k^{\mathrm{out}}(K),\qquad 0\le h_k^{\mathrm{out}}(K)-h_m^{\mathrm{in}}(K)\le\opnorm{K}\bigl(\varepsilon_k^{\mathrm{out}}+\varepsilon_m^{\mathrm{in}}\bigr).
\end{equation}
Indeed, each point of the outer base is close to a point of $\base$, and each point of $\base$ is close to a point of the inner base. The change in the pairing with $K$ is bounded by the corresponding trace-norm distance times $\opnorm{K}$. Validated SDP upper and lower bounds, including solver errors, therefore give a stopping criterion for the support value in the chosen direction. A small gap in one direction alone is not a membership certificate for an arbitrary candidate.
\end{remark}

\begin{remark}[exact boundary membership]\label{rem:boundary-membership}
If $\Tr_B\omega=P_\psi$ is pure, then $\omega\in\base$ if and only if $\omega\in B_\psi$. Indeed, in any convex decomposition of $\omega$ into stotes, the input marginals average to $P_\psi$, so extremality forces every positively weighted component to have that same marginal. Convexity of $B_\psi$ then combines their channels into one channel. The reverse inclusion holds by definition.

For $d_A\ge2$, the inner hierarchy need not contain every element of $\bQ$ at a finite level. Choose a unit vector $\ket\psi$ whose ray lies outside the countable set $\bigcup_m\Psi_{N_m}$ and any output state $\sigma$. The product state $P_\psi\otimes\sigma$ lies in $\base$. If it belonged to some $\bQ^{(m)}$, the preceding extremality argument would force every positively weighted input direction in its finite-input decomposition to equal $\psi$, contrary to $\psi\notin\Psi_{N_m}$. This example is consistent with Hausdorff convergence: arbitrarily accurate approximants exist, but exact membership is never certified by these inner cones. In contrast, every point outside $\bQ$ is excluded by some finite outer level, since $\bigcap_k\bQ_k=\bQ$.
\end{remark}

\subsection{Strict cone inclusions through isotropic states}\label{sec:sectors}

The goal of this section is Corollary~\ref{cor:strict-chain}: the strict inclusions $\PPT \subsetneq \bQ \cap \PSD \subsetneq \PSD$, with explicit witnesses for both. We first give a general symmetry reduction for compact groups of local unitaries, using the twirling method applied to Werner and isotropic states \cite{Werner1989,HorodeckiHorodecki1999}. We then apply it to the $U\otimes\bar U$-invariant sector, whose trace-one commutant is one-dimensional. The resulting isotropic boundary, stated as Theorem~\ref{thm:isotropic-boundary}, supplies the witnesses needed for the strict inclusions. Positive single stotes with negative partial transpose (NPT) were already exhibited by Song et al.\ \cite[Result~3]{SongEtAl2024}, so the existence of NPT members of $\bQ$ is not new. The result here determines the full isotropic section of the convex cone, including the exact endpoints, and shows that the maximally entangled state remains excluded even when convex mixtures of stotes are allowed.

\begin{proposition}[twirling reduction]\label{prop:twirling-reduction}
Let $\Gamma$ be a compact group represented by local unitaries $W_g=U_g\otimes V_g$, and let
\begin{equation}\label{eq:general-twirl}
\mathcal{T}_\Gamma(X):=\int_\Gamma W_gXW_g^\dagger\,dg
\end{equation}
be its normalized Haar twirl. For any set $S$ of Hermitian operators, $\mathcal{T}_\Gamma(S)$ denotes its image under this map:
\begin{equation}\label{eq:twirling-set-image}
\mathcal{T}_\Gamma(S):=\{\mathcal{T}_\Gamma(X):X\in S\}.
\end{equation}
For every $\Gamma$-invariant Hermitian operator $\omega$,
\begin{equation}\label{eq:twirling-membership}
\omega\in\bQ\quad\Longleftrightarrow\quad
\omega\in\mathcal{T}_\Gamma(\bQ)=\cone\bigl(\mathcal{T}_\Gamma(\base)\bigr).
\end{equation}
The image $\mathcal{T}_\Gamma(\base)$ is compact and convex, and coincides with the invariant part of $\base$:
\begin{equation}\label{eq:twirling-base}
\mathcal{T}_\Gamma(\base)
=\{\omega\in\base:W_g\omega W_g^\dagger=\omega\ \text{for all }g\in\Gamma\}
\subseteq\base.
\end{equation}
For every Hermitian $W$ satisfying $W_gWW_g^\dagger=W$ for all $g\in\Gamma$, its support function satisfies
\begin{equation}\label{eq:invariant-support-function}
\max_{\omega\in\mathcal{T}_\Gamma(\base)}\ip{\omega}{W}=h(W),
\end{equation}
where $h$ is the support function of $\base$ defined in \eqref{eq:support-function}.
\end{proposition}

\begin{proof}
First fix any $X\in\bQ$, without assuming that $X$ is invariant. Local-unitary covariance \eqref{eq:lu-covariance} sends each generator to another generator; by linearity, it also preserves their nonnegative linear combinations. Thus $W_gXW_g^\dagger\in\bQ$ for every $g\in\Gamma$, as stated in Proposition~\ref{prop:cone-properties}(iii). Since $\bQ$ is a convex cone, every finite weighted average satisfies
\begin{equation}\label{eq:twirling-finite-average}
\sum_{i=1}^N p_iW_{g_i}XW_{g_i}^\dagger\in\bQ,
\qquad p_i\ge0,\quad\sum_{i=1}^N p_i=1.
\end{equation}
The normalized Haar integral defining $\mathcal{T}_\Gamma(X)$ is a norm limit of such averages: one approximates the continuous map $g\mapsto W_gXW_g^\dagger$ by simple functions, using the Haar measures of the partition sets as the weights $p_i$. Closedness of $\bQ$ (Proposition~\ref{prop:cone-properties}(ii)) then gives $\mathcal{T}_\Gamma(X)\in\bQ$. This proves cone preservation for every $X\in\bQ$, not only for invariant elements. In particular, $\mathcal{T}_\Gamma(\base)\subseteq\base$.

We now prove both directions of \eqref{eq:twirling-membership}. If $\omega\in\bQ$ is invariant, then every conjugate $W_g\omega W_g^\dagger$ equals $\omega$, so $\mathcal{T}_\Gamma(\omega)=\omega$. Hence $\omega$ is the twirl of an element of $\bQ$, namely itself, and belongs to $\mathcal{T}_\Gamma(\bQ)$. Conversely, if $\omega\in\mathcal{T}_\Gamma(\bQ)$, there exists $X\in\bQ$ such that $\omega=\mathcal{T}_\Gamma(X)$. Each conjugate of this $X$ lies in $\bQ$, and the averaging argument above therefore gives $\omega\in\bQ$. This direction does not require $X$ to be invariant. Finally, linearity and $\bQ=\cone(\base)$ give $\mathcal{T}_\Gamma(\bQ)=\cone(\mathcal{T}_\Gamma(\base))$.

Every twirled operator is invariant, by invariance of Haar measure. Together with preservation of $\base$, this gives one inclusion in \eqref{eq:twirling-base}. Conversely, an invariant $\omega\in\base$ satisfies $\omega=\mathcal{T}_\Gamma(\omega)$, so it belongs to $\mathcal{T}_\Gamma(\base)$. This proves the equality in \eqref{eq:twirling-base}. The set is compact by continuity of the twirl and convex by its linearity.

For invariant $W$, self-adjointness of the Haar twirl for the Hilbert--Schmidt pairing gives
\begin{equation}\label{eq:twirled-support}
\max_{Q\in\base}\ip{\mathcal{T}_\Gamma(Q)}{W}
=\max_{Q\in\base}\ip{Q}{\mathcal{T}_\Gamma(W)}
=\max_{Q\in\base}\ip{Q}{W}=h(W).
\end{equation}
\end{proof}

By the characterization of a compact convex set through its supporting halfspaces \cite[Thm.~13.1]{Rockafellar1970}, a $\Gamma$-invariant trace-one Hermitian operator $\omega$ belongs to $\bQ$ if and only if $\ip{\omega}{W}\le h(W)$ for every $\Gamma$-invariant Hermitian $W$. Thus Proposition~\ref{prop:twirling-reduction} restricts the directions needed for the membership test to the invariant subspace, but does not compute the support values in those directions. In particular, it does not guarantee that finitely many supporting inequalities suffice to describe the invariant section exactly.

\begin{remark}[relation to unitarily invariant transport costs]
Proposition~\ref{prop:twirling-reduction} generalizes the symmetry-reduction principle underlying \cite[Prop.~5.1]{HoogstederCalsamigliaWinter2025} from simultaneous unitaries $U\otimes U$ to arbitrary compact groups represented by local unitaries. That result combines symmetry with nonnegativity of costs and zero cost for the identity channel to classify the admissible invariant cost matrices. Here the conclusion concerns invariant sections of $\bQ$ and their support functions; by duality, it also reduces the positivity test for invariant cost matrices. It does not impose the zero-identity-cost condition or provide a classification of those matrices.
\end{remark}

Take $d_A = d_B = d\ge2$ and let $\ket{\Phi^+} = \tfrac{1}{\sqrt d} \sum_i \ket{ii}$ be the maximally entangled vector. The \emph{singlet fraction} used here is the overlap of an operator $\omega$ on $\mathbb{C}^d \otimes \mathbb{C}^d$ with this fixed maximally entangled state,
\begin{equation}\label{eq:singlet-fraction}
F(\omega) := \bra{\Phi^+}\,\omega\,\ket{\Phi^+},
\end{equation}
a quantity studied for density operators in entanglement purification \cite{BennettDiVincenzoSmolinWootters1996} and teleportation \cite{HorodeckiHorodeckiHorodecki1999}. Unlike the fully entangled fraction, which maximizes the overlap over maximally entangled vectors, $F$ is the linear functional $\ip{\omega}{\Phi^+}$ of the fixed projector $\Phi^+ = \ketbra{\Phi^+}{\Phi^+}$. Its role here is that of the single affine coordinate on the $U \otimes \bar U$-invariant sector. The vector $\ket{\Phi^+}$ is invariant under $U \otimes \bar U$ for every unitary $U$, and the operators commuting with all $U \otimes \bar U$ form the two-dimensional span of $\one$ and $\Phi^+$, whose trace-one section is the \emph{isotropic} family \cite{HorodeckiHorodecki1999}
\begin{equation}\label{eq:isotropic-family}
\omega_t = t\,\Phi^+ + (1-t)\,\frac{\one}{d^2}, \qquad F(\omega_t) = t + \frac{1-t}{d^2}, \qquad -\frac{1}{d^2-1}\le t\le1,
\end{equation}
so an invariant trace-one operator is determined by its singlet fraction alone; the stated range of $t$ is exactly the range for which $\omega_t\succeq0$. Apply Proposition~\ref{prop:twirling-reduction} to the group $U\otimes\bar U$. Its twirl maps every trace-one Hermitian operator onto the isotropic line and preserves $F$. For fixed $d$, define the extremal singlet fractions over the full trace-one base by
\begin{equation}\label{eq:singlet-extrema-definition}
\begin{aligned}
F_{\max}(d)&:=\max_{\omega\in\base}F(\omega)=h(\Phi^+),\\
F_{\min}(d)&:=\min_{\omega\in\base}F(\omega)=-h(-\Phi^+).
\end{aligned}
\end{equation}
Both extrema are attained because $\base$ is compact and $F$ is continuous. The invariant section of $\base$ is therefore an interval whose endpoints in the coordinate $F$ are $F_{\min}(d)$ and $F_{\max}(d)$. The following theorem computes these two values. Comparing the upper endpoint with the PPT threshold $F=1/d$ \cite{HorodeckiHorodecki1999,HorodeckiHorodeckiHorodecki1999} and with $F(\Phi^+)=1$ will then prove the strict inclusions.

\begin{theorem}[extremal temporal singlet fractions and the isotropic boundary]\label{thm:isotropic-boundary}
For equal local dimensions $d\ge2$, the maximum singlet fraction over $\base$ is
\begin{equation}\label{eq:Fmax}
F_{\max}(d) =
\begin{cases}
\dfrac{1}{d}\Bigl(1 + \dfrac{d-1}{4}\Bigr), & 2 \le d \le 5,\\[2ex]
\dfrac{\sqrt{d-1}}{d}, & d \ge 6,
\end{cases}
\end{equation}
so $F_{\max}(2) = 5/8$, $F_{\max}(3) = 1/2$. The minimum over $\base$ is $F_{\min}(d)=-\sqrt{d-1}/d$. On the entire trace-one isotropic line, allowing every real $t$, one has $\omega_t\in\bQ$ if and only if $F_{\min}(d)\le F(\omega_t)\le F_{\max}(d)$. For the physical range $-1/(d^2-1)\le t\le1$, the lower inequality is automatic, so membership is equivalent to
\begin{equation}\label{eq:tstar}
t\le t^*(d):=\frac{d^2F_{\max}(d)-1}{d^2-1}.
\end{equation}
For qubits $t^*=1/2$, strictly between the PPT/separability threshold $1/3$ and $1$.
\end{theorem}

\begin{proof}
\emph{Step 1: reduction to a fixed pure input.} Use the basis $\{\ket{j}\}_{j=0}^{d-1}$ defining $\ket{\Phi^+}$ and the partial transpose $T_A$. By Lemma~\ref{lem:pure-input}, $\base$ is the convex hull of pure-input generators. Since $F$ is linear, its maximum and minimum over $\base$ equal its maximum and minimum over those generators. The maximally entangled vector is $U\otimes\bar U$-invariant: the transfer identity \cite[Eq.~(1.132)]{Watrous2018} $(X\otimes\one)\ket{\Phi^+}=(\one\otimes X^T)\ket{\Phi^+}$ gives $(U\otimes\bar U)\ket{\Phi^+}=(\one\otimes\bar U U^T)\ket{\Phi^+}=\ket{\Phi^+}$, since $\bar U U^T=\overline{UU^\dagger}=\one$.

For any pure input $\ket{\psi}$, choose a unitary $U$ with $U\ket{\psi}=\ket{0}$. Given $Q=(P_\psi\otimes\one)\star J$, local-unitary covariance as in Proposition~\ref{prop:cone-properties}(iii) gives
\begin{equation}\label{eq:singlet-input-reduction}
Q':=(U\otimes\bar U)Q(U\otimes\bar U)^\dagger
=(P_0\otimes\one)\star J',
\qquad J':=(U\otimes\bar U)J(U\otimes\bar U)^\dagger\in\cJ,
\end{equation}
where $P_0:=\ketbra{0}{0}$. The invariance just proved gives $F(Q')=F(Q)$. Thus every value on a pure-input generator is attained with input $P_0$, and restricting to this input changes neither extremum. The generators $Q$ themselves need not be invariant.

\emph{Step 2: the objective in Choi coordinates.} By the Choi characterization of channels recalled in Section~\ref{sec:1.4}, the allowed $J$ are exactly $J=\tilde C^{\TA}$ with $\tilde C\succeq0$ and $\Tr_B[\tilde C]=\one_A$. Write $Q(\tilde C):=(P_0\otimes\one)\star\tilde C^{\TA}$ and define $f(\tilde C):=F(Q(\tilde C))$; this is not the overlap $F(\tilde C)$ of the Choi matrix itself. The two terms of the Jordan product give complex-conjugate expectation values, and $\bra{\Phi^+}(P_0\otimes\one)=d^{-1/2}\bra{00}$, so
\begin{equation}\label{eq:singlet-objective}
\begin{aligned}
f(\tilde C)
&=\operatorname{Re}\bra{\Phi^+}(P_0\otimes\one)J\ket{\Phi^+}\\
&=\frac{1}{d}\operatorname{Re}\sum_{j=0}^{d-1}\bra{00}J\ket{jj}\\
&=\frac{1}{d}\operatorname{Re}\sum_{j=0}^{d-1}\bra{j0}\tilde C\ket{0j},
\end{aligned}
\end{equation}
using $\bra{ab}X^{\TA}\ket{a'b'}=\bra{a'b}X\ket{ab'}$ in the last step. Set $a:=\bra{00}\tilde C\ket{00}\ge0$. We must therefore maximize and minimize $f(\tilde C)=d^{-1}\bigl(a+\sum_{j=1}^{d-1}\operatorname{Re}\bra{j0}\tilde C\ket{0j}\bigr)$ over the stated Choi constraints.

\emph{Step 3: upper bound.} For $1\le j\le d-1$, set $b_j:=\bra{0j}\tilde C\ket{0j}\ge0$ and $c_j:=\bra{j0}\tilde C\ket{j0}\ge0$. The determinant of the positive $2\times2$ compression to $\operatorname{span}\{\ket{0j},\ket{j0}\}$ is nonnegative, giving $\abs{\bra{j0}\tilde C\ket{0j}}\le\sqrt{b_jc_j}$. The diagonal entries of the marginal constraint give
\begin{equation}\label{eq:singlet-diagonal-constraints}
a+\sum_{j=1}^{d-1}b_j=1,
\qquad
c_j\le\sum_{\ell=0}^{d-1}\bra{j\ell}\tilde C\ket{j\ell}=1.
\end{equation}
In particular, $0\le a\le1$. Bounding each real part by its absolute value and then using Cauchy--Schwarz for the $d-1$ numbers $\sqrt{b_j}$ gives
\begin{equation}\label{eq:singlet-upper}
\begin{aligned}
f(\tilde C)
&\le\frac{1}{d}\left(a+\sum_{j=1}^{d-1}\sqrt{b_jc_j}\right)
\le\frac{1}{d}\left(a+\sum_{j=1}^{d-1}\sqrt{b_j}\right)\\
&\le\frac{1}{d}\left(a+\sqrt{(d-1)\sum_{j=1}^{d-1}b_j}\right)
=\frac{a+\sqrt{(d-1)(1-a)}}{d}
=:\frac{g(a)}{d}.
\end{aligned}
\end{equation}
The function $g$ is continuous and concave on $[0,1]$, with $g'(a)=1-\tfrac12\sqrt{(d-1)/(1-a)}$ for $0\le a<1$. For $2\le d\le4$, the stationary point $a=1-\tfrac{d-1}{4}$ lies in the interior and gives the maximum $1+\tfrac{d-1}{4}$. At $d=5$, the maximum is at the endpoint $a=0$ and has the same value. For $d\ge6$, $g'(0)<0$ and concavity imply that $g$ is decreasing, so the maximum is $g(0)=\sqrt{d-1}$. This proves the upper bounds in \eqref{eq:Fmax}.

\emph{Step 4: attainment.} Fix $d\ge2$, let $a^*$ be the maximizing value of $a$ found in Step 3, set $b_j := (1-a^*)/(d-1)$ for $j \ge 1$, and define
\begin{equation}\label{eq:singlet-attain}
\tilde C^+ := a^* \ketbra{00}{00} + \sum_{j=1}^{d-1} \ketbra{w_j}{w_j}, \qquad \ket{w_j} := \sqrt{b_j}\, \ket{0j} + \ket{j0}.
\end{equation}
This matrix is positive as a sum of positive rank-one terms. To check the full marginal constraint, expand each summand. Its cross terms are proportional to $\ketbra{0}{j}_A\otimes\ketbra{j}{0}_B$ and its adjoint, whose partial traces vanish because $\braket{0}{j}=0$. Consequently,
\begin{align}
\Tr_B[\ketbra{w_j}{w_j}]&=b_j\ketbra{0}{0}+\ketbra{j}{j},\label{eq:singlet-attain-marginal-term}\\
\Tr_B[\tilde C^+]&=\left(a^*+\sum_{j=1}^{d-1}b_j\right)\ketbra{0}{0}+\sum_{j=1}^{d-1}\ketbra{j}{j}
=\one_A.\label{eq:singlet-attain-marginal}
\end{align}
Thus $\tilde C^+$ is a feasible Choi matrix, not merely a matrix satisfying the diagonal constraints used in Step 3. For every $j\ge1$, only the $j$th summand contributes to $\bra{j0}\tilde C^+\ket{0j}$, giving $\bra{j0}\tilde C^+\ket{0j}=\sqrt{b_j}$. Therefore
\begin{equation}\label{eq:singlet-attain-value}
f(\tilde C^+)=\frac{1}{d}\left(a^*+(d-1)\sqrt{\frac{1-a^*}{d-1}}\right)=\frac{g(a^*)}{d}=F_{\max}(d).
\end{equation}
Note that, as expected, this construction saturates every inequality in Step 3: the relevant off-diagonal entries are positive real numbers of magnitude $\sqrt{b_jc_j}$, each $c_j=1$, and all $b_j$ are equal. We denote the attaining generator by $Q^+:=Q(\tilde C^+)$.

\emph{Step 5: the lower endpoint.} Step 1 also reduced the minimum to the same feasible Choi matrices. With the variables of Step 3, positivity gives $\operatorname{Re}\bra{j0}\tilde C\ket{0j}\ge-\sqrt{b_jc_j}$. Using $c_j\le1$ and the same Cauchy--Schwarz bound as before yields
\begin{equation}\label{eq:singlet-lower}
\begin{aligned}
f(\tilde C)
&\ge\frac{1}{d}\left(a-\sum_{j=1}^{d-1}\sqrt{b_jc_j}\right)
\ge\frac{1}{d}\left(a-\sum_{j=1}^{d-1}\sqrt{b_j}\right)\\
&\ge\frac{a-\sqrt{(d-1)(1-a)}}{d}
\ge-\frac{\sqrt{d-1}}{d}.
\end{aligned}
\end{equation}
The last inequality uses $a\ge0$ and $\sqrt{1-a}\le1$. To attain it for all $d\ge2$, take $a=0$, $b_j=1/(d-1)$, and reverse the sign of the second term in each vector from Step 4:
\begin{equation}\label{eq:singlet-lower-attain}
\tilde C^-:=\sum_{j=1}^{d-1}\ketbra{w_j^-}{w_j^-},
\qquad
\ket{w_j^-}:=\frac{\ket{0j}}{\sqrt{d-1}}-\ket{j0}.
\end{equation}
Again $\tilde C^-\succeq0$. The sign change only affects the cross terms, which vanish under the partial trace, so \eqref{eq:singlet-attain-marginal-term} gives $\Tr_B[\tilde C^-]=\ketbra{0}{0}+\sum_{j=1}^{d-1}\ketbra{j}{j}=\one_A$. Its relevant off-diagonal entries are all $-1/\sqrt{d-1}$, and hence
\begin{equation}\label{eq:singlet-lower-value}
f(\tilde C^-)=-\frac{d-1}{d\sqrt{d-1}}=-\frac{\sqrt{d-1}}{d}.
\end{equation}
The generator $Q^-:=Q(\tilde C^-)$ attains the lower bound, proving the formula for $F_{\min}(d)$.

\emph{Step 6: the full isotropic line.} Since $\Tr\omega_t=1$, membership of $\omega_t$ in $\bQ$ is equivalent to membership in $\base$. Necessity of the bounds on $F(\omega_t)$ follows from the two extrema just proved. For sufficiency, take $\Gamma$ to be the group represented by $U\otimes\bar U$. Since $Q^\pm\in\base$, the image-set definition \eqref{eq:twirling-set-image} and the inclusion proved in \eqref{eq:twirling-base} give
\begin{equation}\label{eq:isotropic-twirled-attainment}
\mathcal{T}_\Gamma(Q^\pm)\in\mathcal{T}_\Gamma(\base)\subseteq\base.
\end{equation}
These twirls are isotropic by \eqref{eq:twirling-base}. By Step 1 and linearity of $F$, their singlet fractions remain $F_{\max}(d)$ and $F_{\min}(d)$, respectively. Since $F$ is an affine coordinate on the isotropic line \eqref{eq:isotropic-family}, their convex combinations give every isotropic operator with singlet fraction between these extrema. Convexity of $\base$ therefore proves sufficiency.

For physical isotropic states, $F(\omega_t)\ge0>F_{\min}(d)$, so only the upper bound remains. Substituting $F(\omega_t)=t+(1-t)/d^2$ gives \eqref{eq:tstar}; for $d=2$, this yields $t^*=1/2$.
\end{proof}

We can now prove the strict inclusions that are the goal of this section. The isotropic calculation provides explicit witnesses; the conclusion concerns the full cones. Lorenz, Brunner and Benedetti \cite[Appendix~A.2]{LorenzBrunnerBenedetti2026} gave a numerical example of a positive, non-PPT single stote in dimensions $3\times2$. Here the isotropic calculation gives exact witnesses for both strict inclusions for every equal local dimension $d\ge2$.

\begin{corollary}[strict cone inclusions]\label{cor:strict-chain}
For all $d \ge 2$, $\PPT \subsetneq \bQ \cap \PSD \subsetneq \PSD$. Both inclusions have explicit witnesses: every isotropic state with $t \in (\tfrac{1}{d+1}, t^*(d)]$ is NPT and belongs to $\bQ$, while $\Phi^+$ is positive and does not belong to $\bQ$.
\end{corollary}

\begin{proof}
The inclusions hold by Theorem~\ref{thm:spectral-test} ($\PPT \subseteq \bQ$, and $\PPT \subseteq \PSD$ by definition) and by $\bQ \cap \PSD \subseteq \PSD$. For the first strictness, an isotropic state is PPT iff it is separable iff $t \le \tfrac{1}{d+1}$ \cite{HorodeckiHorodecki1999,HorodeckiHorodeckiHorodecki1999}. Directly, $(\Phi^+)^{\TA}$ is $1/d$ times the swap, so for $t\ge0$ the minimum eigenvalue of $\omega_t^{\TA}$ is $\tfrac{1-t}{d^2}-\tfrac{t}{d}$ and changes sign at this value of $t$. Also $F(\omega_{1/(d+1)})=1/d<F_{\max}(d)$ for every $d\ge2$: this follows from the first branch of \eqref{eq:Fmax} for $d\le5$ and the second branch for $d\ge6$. Since $F(\omega_t)$ is strictly increasing in $t$, $t^*(d)>1/(d+1)$, and every $\omega_t$ in the stated interval is NPT and belongs to $\bQ$ by Theorem~\ref{thm:isotropic-boundary}. For the second strictness, $\Phi^+=\omega_1$ is PSD and has $F(\Phi^+)=1>F_{\max}(d)$ (indeed $F_{\max}(d)\le5/8$), so it does not belong to $\bQ$.
\end{proof}

\section{Conclusion}\label{sec:4}

This paper studies the convex-cone characterization problem for states over time posed in \cite{HoogstederCalsamigliaWinter2025}. The known single-stote spectral criterion and PPT inclusion \cite{SongParzygnat2025}, recalled in Theorem~\ref{thm:spectral-test}, provide useful structural input. Theorem~\ref{thm:binary-classical-membership} gives an exact SDP description of the binary classical-output sector and constructs a realizing mixture for every member, including those with indefinite blocks.

The outer hierarchy applies constrained symmetric-extension theory \cite{BertaBorderiFawziScholz2022} to the input--channel representation, using Bose-symmetric certificates with a channel marginal constraint (Theorem~\ref{thm:outer-hierarchy}). Its existence and qualitative convergence are applications of that theory. The quantitative specialization in Theorem~\ref{thm:outer-rate} bounds the trace-norm Hausdorff distance between its trace-one slices and $\base$ by $2d_A^2(d_A-1)/(k+d_A-1)$, independently of $d_B$. This gives an explicit sufficient level for a prescribed error. The proof applies Bose-extension rounding \cite{NavascuesOwariPlenio2009,Chiribella2011}, verifies that it preserves the channel marginal constraint pointwise, and bounds the error after the stote map. For permutation-invariant extensions without Bose support, constrained de~Finetti theory already gives an $O(k^{-1/2})$ bound at fixed dimensions (Remark~\ref{rem:outer-rate}).

The inner hierarchy instead restricts the pure inputs to finitely many directions while retaining every channel at each direction. Its closed SDP-representable cones $\bQ^{(m)}$ increase within $\bQ$ and have dense union in $\bQ$ (Corollary~\ref{prop:inner-validity}). Theorem~\ref{thm:inner-rate} quantifies the approximation in terms of both the covering radius $\delta_m$ and the number $N_m$ of input directions. For $d_A\ge2$ and the nested direction sets constructed there,
\begin{equation}\label{eq:conclusion-inner-bounds}
d_H\bigl(\bQ^{(m)}\cap\{\Tr=1\},\base\bigr)
\le 2(d_A-1)\sin^2\delta_m
\le 2\pi^2(d_A-1)N_m^{-1/(d_A-1)}.
\end{equation}
Thus a target trace-norm Hausdorff error $\eta>0$ is guaranteed by $N_m\ge\lceil(2\pi^2(d_A-1)/\eta)^{d_A-1}\rceil$, a sufficient cardinality rather than a claim of optimality (Remark~\ref{rem:covering-numbers}). The quadratic covering-radius bound follows by optimizing the input Rayleigh quotient for each support direction. For fixed $d_A$, the sufficient number of blocks grows as $O_{d_A}(\eta^{-(d_A-1)})$ and is independent of $d_B$, although each block has size $d_Ad_B$.

Together the hierarchies give the certified inclusions $\bQ^{(m)}\subseteq\bQ\subseteq\bQ_k$ and the support-function error bounds of Remark~\ref{rem:two-sided}. Inner feasibility certifies membership, while exclusion from an outer cone certifies nonmembership. Every nonmember is excluded at a finite outer level. Exact inner membership need not occur at any finite level for boundary points, as Remark~\ref{rem:boundary-membership} shows.

The exact isotropic computation gives $F_{\max}(d)$ and $\PPT\subsetneq\bQ\cap\PSD\subsetneq\PSD$ for equal local dimensions $d\ge2$; Proposition~\ref{prop:twirling-reduction} states what twirling provides for other invariant sectors.

Further directions include sharper dimension dependence in the convergence bounds and improved rates for the broader permutation-invariant hierarchy. For Bose extensions, PPT-assisted rounding already gives an $O(k^{-2})$ trace-norm bound for separability at fixed dimension \cite[Thm.~3 and Cor.~5]{NavascuesOwariPlenio2009}; the corresponding refinement preserving the channel marginal constraint is not developed here. Other questions include a complete description for $d_A=d_B=2$, where extreme channels are classified \cite{RuskaiSzarekWerner2002}; extending the binary classical-output membership criterion of Theorem~\ref{thm:binary-classical-membership} to three or more outcomes, allowing indefinite output blocks; and exact boundaries for further invariant sectors, such as the $U\otimes U$ Werner sector \cite{Werner1989}.

\section*{Acknowledgements}

Thanks to Ian George for pointing out \cite{BertaBorderiFawziScholz2022} and other symmetric extension results and discussing the interpretation of the outer hierarchy.

M.H.R. is supported by the National Research Foundation, Prime Minister's Office, Singapore, under its Campus for Research Excellence and Technological Enterprise (CREATE) program, Quantum Security and Resilience for Emerging Technologies Enterprise (QUASAR-CREATE).

\section*{LLM use disclaimer}

LLM assistance has been used to derive the results and to assist in writing the manuscript. The author takes full responsibility for the statements in the manuscript, including any errors.

\appendix
\section{Diagonal subsequence extraction}\label{app:C}

Step 2 of Theorem~\ref{thm:outer-hierarchy} requires one subsequence of certificate levels along which every fixed marginal converges. We record the standard diagonal subsequence construction in the form needed here (cf.\ \cite[Thms.~7.23 and 7.25]{Rudin1976}).

\begin{lemma}[diagonal subsequence]\label{lem:diagonal}
For each $j \ge 1$ let $M_j$ be a sequentially compact metric space and let $(x_j^{(k)})_{k \ge j}$ be a sequence in $M_j$. Then there is a common increasing sequence $k_1 < k_2 < \cdots$, with $k_n\ge n$, such that, for every fixed $j$, the tail $(x_j^{(k_n)})_{n\ge j}$ converges in $M_j$.
\end{lemma}

In Step 2, $M_j = \{Z \succeq 0,\ \Tr[Z] = d_A\}$ on the fixed space $\mathcal{H}_{A'}^{\otimes j} \otimes \HA \otimes \HB$ --- a closed bounded subset of the finite-dimensional real vector space of Hermitian operators, hence compact by Heine--Borel, and sequentially compact by Bolzano--Weierstrass \cite[Thms.~2.41 and 3.6]{Rudin1976} --- and $x_j^{(k)} = X_j^{(k)}$.

\begin{proof}
\emph{Nested refinements.} By sequential compactness of $M_1$, choose an infinite index set $K_1 \subseteq \{1, 2, \dots\}$ along which $x_1^{(k)}$ converges. Inductively, given the infinite set $K_{j-1}$, choose an infinite $K_j \subseteq K_{j-1} \cap \{j, j+1, \dots\}$ along which $x_j^{(k)}$ converges (the sequence $(x_j^{(k)})_{k \ge j}$ being defined only from index $j$ on); since $K_j \subseteq K_{j-1} \subseteq \cdots \subseteq K_1$, along $K_j$ the sequences $x_1^{(k)}, \dots, x_j^{(k)}$ \emph{all} converge (a subsequence of a convergent sequence converges, to the same limit). Displaying each $K_j$ as a row, in increasing order,
\begin{equation}\label{eq:diag-array}
\begin{array}{lccccc}
K_1: & k^{(1)}_1 & k^{(1)}_2 & k^{(1)}_3 & k^{(1)}_4 & \cdots\\[2pt]
K_2: & k^{(2)}_1 & k^{(2)}_2 & k^{(2)}_3 & k^{(2)}_4 & \cdots\\[2pt]
K_3: & k^{(3)}_1 & k^{(3)}_2 & k^{(3)}_3 & k^{(3)}_4 & \cdots\\[2pt]
& \vdots
\end{array}
\end{equation}
the sequence indexed by row $K_j$ gives convergence for the first $j$ marginals, but no fixed row need give convergence for all marginals. Taking the intersection does not solve this problem: nested infinite sets can have empty intersection. For example, $K_j=\{j,j+1,\ldots\}$ is infinite for every $j$, but $\bigcap_{j\ge1}K_j=\varnothing$.

\emph{The diagonal.} Set $k_n := k^{(n)}_n$, the $n$-th element of the $n$-th row --- the entries running down the diagonal of the array. The sequence is increasing: $k^{(n)}_n \ge k^{(n-1)}_n > k^{(n-1)}_{n-1}$, since $K_n \subseteq K_{n-1}$ implies that the $n$-th smallest element of $K_n$ is at least the $n$-th smallest element of $K_{n-1}$, and rows are listed in increasing order. It works for every $j$ by a tail argument: for $n \ge j$,
\begin{equation}\label{eq:diag-tail}
k_n = k^{(n)}_n \in K_n \subseteq K_j,
\end{equation}
so $(x_j^{(k_n)})_{n\ge j}$ is well-defined and is a subsequence of the convergent sequence indexed by $K_j$. Also $k_n\ge n$, since every element of $K_n$ is at least $n$. This proves the assertion for every fixed $j$.
\end{proof}

\begin{remark}\label{rem:diagonal}
For each fixed $j$, the diagonal sequence eventually lies in $K_j$; the starting index of this tail may depend on $j$. This is all that simultaneous convergence of the fixed marginals requires.
\end{remark}

\section{Why finite-level extensions require a consistency argument}\label{app:consistency-example}

\begin{example}[finite consistent extensions with no uniform failure level]\label{ex:dicke-extension-consistency}
The following construction makes the issue in Remark~\ref{rem:definetti-consistency} explicit within the PPT-free outer hierarchy of Theorem~\ref{thm:outer-hierarchy}. Each chosen two-copy candidate belongs to a consistent family up to a finite level, but cannot extend beyond that level. These levels are unbounded, and the candidates converge to an infinitely extendible limit.

For $0\le r\le j$, write
\begin{equation}\label{eq:dicke-vector}
\ket{D_{j,r}}:=\binom jr^{-1/2}\sum_{\substack{x\in\{0,1\}^j\\ |x|=r}}\ket{x},
\qquad P_{j,r}:=\ketbra{D_{j,r}}{D_{j,r}},
\end{equation}
where $|x|$ is the number of ones in $x$. These are the symmetric Dicke states \cite{Dicke1954}; see also \cite[Section~1]{NepomechieRavaniniRaveh2024}. Fix an integer $n\ge1$ and take the balanced state $P_{2n,n}$. Its $j$-copy marginals are
\begin{equation}\label{eq:dicke-marginals}
\rho_j^{(2n)}:=\Tr_{j+1,\ldots,2n}[P_{2n,n}]
=\sum_{r=0}^{j}\frac{\binom jr\binom{2n-j}{n-r}}{\binom{2n}{n}}P_{j,r},
\qquad 1\le j\le 2n,
\end{equation}
with binomial coefficients outside their usual range taken to be zero. This reduced-state formula appears in \cite{PopkovSalerno2005}; a modern derivation is the spin-$1/2$ case of the Schmidt decomposition in \cite[Appendix~C, Eqs.~(C1)--(C2)]{NepomechieRavaniniRaveh2024}. For fixed $n$, these marginals are consistent: $\Tr_{j+1}\rho_{j+1}^{(2n)}=\rho_j^{(2n)}$ for $j<2n$. The balanced Dicke family and its two-copy marginal were also used in \cite[Section~III, Eqs.~(30)--(31)]{NavascuesOwariPlenio2009}. In particular,
\begin{equation}\label{eq:dicke-pair}
\begin{aligned}
\rho_1^{(2n)}&=\one_2/2,\\
\rho_2^{(2n)}&=\frac{n-1}{2(2n-1)}\bigl(\ketbra{00}{00}+\ketbra{11}{11}\bigr)
+\frac{n}{2n-1}\ketbra{\psi^+}{\psi^+},
\qquad \ket{\psi^+}:=\frac{\ket{01}+\ket{10}}{\sqrt2}.
\end{aligned}
\end{equation}

To obtain certificates in our hierarchy, set $d_A=2$, fix a state $\sigma_B$, and put
\begin{equation}\label{eq:dicke-certificates}
X^{(2n)}:=P_{2n,n}\otimes\one_A\otimes\sigma_B,
\qquad X_j^{(2n)}=\rho_j^{(2n)}\otimes\one_A\otimes\sigma_B.
\end{equation}
These operators are positive, have Bose support on the $A'$ copies, and have trace $2$. They satisfy the channel marginal constraint because $\Tr_BX^{(2n)}=P_{2n,n}\otimes\one_A$ and $\Tr_{AB}X^{(2n)}=2P_{2n,n}$. By \eqref{eq:constrained-sep-cone} and $\rho_1^{(2n)}=\one_2/2$, condition~(4) reproduces the same operator $\omega=\one_A/2\otimes\sigma_B$ for every $n$.

Nevertheless, the two-copy candidate $\rho_2^{(2n)}$ has no permutation-invariant extension to $m>2n$ copies. To see this, define $Z=\ketbra{0}{0}-\ketbra{1}{1}$ and set $Z_i:=\one_2^{\otimes(i-1)}\otimes Z\otimes\one_2^{\otimes(m-i)}$ on the common $m$-copy space. Equation~\eqref{eq:dicke-pair} gives
\begin{equation}\label{eq:dicke-pair-correlation}
\Tr[\rho_2^{(2n)}(Z\otimes Z)]=-\frac1{2n-1}.
\end{equation}
If a permutation-invariant $m$-copy state $\tau_m$ had this two-copy marginal, every pair would have the same correlation. Since $Z_i^2=\one$, positivity would require
\begin{equation}\label{eq:dicke-extension-obstruction}
0\le\Tr\left[\tau_m\left(\sum_{i=1}^m Z_i\right)^2\right]
=m+2\sum_{i<\ell}\Tr[\tau_m Z_iZ_\ell]
=m-\frac{m(m-1)}{2n-1}
=\frac{m(2n-m)}{2n-1}.
\end{equation}
For $m>2n$ this is negative. Thus $\rho_2^{(2n)}$ extends to $2n$ copies but not to $2n+1$; likewise, a level-$m$ certificate with $m>2n$ extending $X_2^{(2n)}$ would give such an extension after tracing out $AB$ and dividing by $2$, so it cannot exist. Each chosen candidate therefore fails at some level, but no single level excludes all of them: for any prescribed $m$, choose an integer $n$ with $2n\ge m$.

The failure of each selected two-copy candidate to extend indefinitely does not contradict completeness of the hierarchy. In the completeness proof, it is enough to obtain one consistent infinite certificate family for the fixed $\omega$. Such a family can arise by taking limits of the finite certificates: its two-copy marginal is then an additional candidate, distinct from every selected $X_2^{(2n)}$, that admits extensions to all levels. For every fixed $j$ and $r$, the coefficient in \eqref{eq:dicke-marginals} tends to $2^{-j}\binom jr$ as $n\to\infty$, by expanding the binomial coefficients into factorial products. Hence
\begin{equation}\label{eq:dicke-consistent-limit}
\rho_j^{(2n)}\longrightarrow\rho_j^{(\infty)}
:=\sum_{r=0}^j2^{-j}\binom jr P_{j,r}
=\int_0^{2\pi}(\ketbra{\phi_\theta}{\phi_\theta})^{\otimes j}\,\frac{d\theta}{2\pi},
\qquad \ket{\phi_\theta}:=\frac{\ket0+e^{i\theta}\ket1}{\sqrt2}.
\end{equation}
To verify the integral identity, group the computational-basis strings in $\ket{\phi_\theta}^{\otimes j}$ by their number $r$ of ones. Every such string has amplitude $2^{-j/2}e^{ir\theta}$, and there are $\binom jr$ of them, so
\begin{equation}\label{eq:dicke-phase-expansion}
\ket{\phi_\theta}^{\otimes j}
=2^{-j/2}\sum_{r=0}^{j}\sqrt{\binom jr}\,e^{ir\theta}\ket{D_{j,r}}.
\end{equation}
Taking the projector and integrating gives
\begin{equation}\label{eq:dicke-phase-average}
\begin{aligned}
\int_0^{2\pi}(\ketbra{\phi_\theta}{\phi_\theta})^{\otimes j}\,\frac{d\theta}{2\pi}
&=2^{-j}\sum_{r,s=0}^{j}\sqrt{\binom jr\binom js}\,\ketbra{D_{j,r}}{D_{j,s}}
\int_0^{2\pi}e^{i(r-s)\theta}\,\frac{d\theta}{2\pi}\\
&=2^{-j}\sum_{r,s=0}^{j}\sqrt{\binom jr\binom js}\,\delta_{rs}\ketbra{D_{j,r}}{D_{j,s}}\\
&=\sum_{r=0}^{j}2^{-j}\binom jr P_{j,r}=\rho_j^{(\infty)}.
\end{aligned}
\end{equation}
Here the phase integral is $1$ when $r=s$ and $0$ otherwise: for the nonzero integer $r-s$, its value is $(e^{2\pi i(r-s)}-1)/(2\pi i(r-s))=0$.

This family is consistent for all $j$, since tracing one factor of each tensor power leaves the preceding tensor power. Thus $Y_j:=\rho_j^{(\infty)}\otimes\one_A\otimes\sigma_B$ is a consistent infinite certificate family for the same $\omega$. The set of chosen finite-$n$ candidates is not closed: its limit $Y_2$ is an additional candidate that never fails. One cannot infer a uniform failure level from the finite candidates without accounting for such limits. Here every fixed marginal converges explicitly; the compactness and diagonal argument in Step~2 of Theorem~\ref{thm:outer-hierarchy} establishes a consistent limiting family for arbitrary choices of certificates.

These certificates do not satisfy the PPT constraint~(2): $(\rho_2^{(2n)})^{T_1}$ has eigenvalue $-1/[2(2n-1)]$ on $(\ket{00}-\ket{11})/\sqrt2$, so the finite certificates are not PPT.
\end{example}

Separately chosen extensions can also be incompatible when all PPT constraints are imposed. For $k\ge1$, let $Z_k=(P_0^{\otimes k}+P_1^{\otimes k})/2$ and $W_k=(P_+^{\otimes k}+P_-^{\otimes k})/2$, where $P_v=\ketbra{v}{v}$ and $\ket{\pm}=(\ket0\pm\ket1)/\sqrt2$. Both are Bose-supported PPT states with one-qubit marginal $\one_2/2$, since they are mixtures of pure tensor powers. Choose $R_k=Z_k$ for even $k$ and $R_k=W_k$ for odd $k$. These extensions are incompatible: $\Tr_3R_3=W_2\ne Z_2=R_2$, since $\bra{00}W_2\ket{00}=1/4$ whereas $\bra{00}Z_2\ket{00}=1/2$. Tensoring each $R_k$ with $\one_A\otimes\sigma_B$ gives certificates satisfying all four constraints for the same $\omega=\one_A/2\otimes\sigma_B$, as above. Restricting to even levels gives the consistent limiting family $(Z_j)_j$ on the repeated systems; odd levels give $(W_j)_j$. Thus PPT constraints do not ensure consistency of arbitrary choices, and a consistent limiting family need not be unique.


\end{document}